\documentclass[twocolumn,secnumarabic,amssymb, nobibnotes, aps, prl, floatfix]{revtex4-2}
\usepackage{graphicx}
\usepackage{bm}
\usepackage[dvipsnames]{xcolor}
\usepackage[normalem]{ulem}
\usepackage{textcomp}
\usepackage{lineno,hyperref}
\usepackage{hyperref}
\usepackage{xspace}
\usepackage{siunitx}
\usepackage{times}
\usepackage{amsmath}
\usepackage{amssymb}
\usepackage{svg}
\usepackage{upgreek}
\usepackage{braket}

\newcommand{\FGT}{Fe$_3$GeTe$_2$\xspace}
\newcommand{\EF}{$E_{\rm F}$\xspace}
\newcommand{\NbSe}{NbSe$_2$\xspace}

\begin{document}
\title{Giant orbital Zeeman effects in a magnetic topological van der Waals interphase}

	\author{
		Tobias Wichmann,$^{1,2,3}$
	    Mirco Sastges,$^{4,5,6}$
		Keda Jin,$^{1,2,7}$
        Jose Martinez-Castro,$^{1,2,7}$        
	    Tom G. Saunderson,$^{4,5}$	        
        Dongwook Go,$^{5,8}$
        Honey Boban,$^9$
        Samir Lounis,$^{10}$		
        Lukasz Plucinski,$^9$
        Markus Ternes,$^{1,2,7}$
		Yuriy Mokrousov,$^{4,5}$
		F. Stefan Tautz$^{1,2,3}$
		and Felix L\"upke$^{1,2,11,\ast}$
  \\
		\vspace{0.25cm}
		\normalsize{$^{1}$\textit{Peter Gr\"unberg Institute (PGI-3), Forschungszentrum J\"ulich, 52425 J\"ulich, Germany}}\\
		\normalsize{$^{2}$\textit{J\"ulich Aachen Research Alliance, Fundamentals of Future Information Technology, 52425 J\"ulich, Germany}}\\
		\normalsize{$^{3}$\textit{Institut f\"ur Experimentalphysik IV A, RWTH Aachen University, 52074 Aachen, Germany}}\\
		\normalsize{$^{4}$\textit{Institute of Physics, Johannes Gutenberg University, 55099 Mainz, Germany}}\\
        \normalsize{$^{5}$\textit{Peter Gr\"unberg Institute (PGI-1), Forschungszentrum J\"ulich, 52425 J\"ulich, Germany}}\\
        \normalsize{$^{6}$\textit{Department of Physics, RWTH Aachen University, 52056 Aachen, Germany}}\\ 
		\normalsize{$^{7}$\textit{Institut f\"ur Experimentalphysik II B, RWTH Aachen University, 52074 Aachen, Germany}}\\
        \normalsize{$^{8}$\textit{Department of Physics, Korea University, Seoul 02841, Republic of Korea}}\\        
        \normalsize{$^{9}$\textit{Peter Gr\"unberg Institute (PGI-6), Forschungszentrum J\"ulich, 52425 J\"ulich, Germany}}\\
        \normalsize{$^{10}$\textit{Institute of Physics and Halle-Berlin-Regensburg Cluster of Excellence CCE, Martin-Luther-University Halle-Wittenberg, 06099 Halle, Germany}}\\        
        \normalsize{$^{11}$\textit{Institute of Physics II, Universit\"at zu K\"oln, Z\"ulpicher Straße 77, 50937 Cologne, Germany}}\\        
        \vspace{0.25cm}
     \normalsize{$^\ast$E-mail: f.luepke@fz-juelich.de}\\
	}

\begin{abstract}
Van der Waals (vdW) heterostructures allow the engineering of magnetic and electronic properties via interfacial orbital hybridisation and charge transfer. The resulting electric fields can give rise to Rashba spin-orbit coupling and Dzyaloshinskii–Moriya interactions, which may stabilise novel two-dimensional magnetic phases.
While such emergent magnetic ‘interphases’ may offer a promising platform for spin-based electronics, direct spectroscopic evidence for them is still lacking.
Here, we report electric-field tunable Zeeman effects with Landé $g$-factors up to $\approx230$ at the interface of graphene and the vdW ferromagnet \FGT. 
They arise from an interplay of local-moment canting and the itinerant orbital moments of \FGT, producing a giant asymmetric level splitting.
Exploiting the inelastic phonon gap of graphene, scanning tunnelling spectroscopy allows us to access the buried interface properties, and quantify Faraday-like screening.
Our findings provide new insights and opportunities for electrically controlled interface magnetism, e.g. for spintronics {and orbitronics} devices.
\end{abstract}
\maketitle

{The interplay of topological band structures and magnetism provides a rich playground for the realization of exotic materials \cite{armitage2018weyl, wan2011topological, tokura2019magnetic}. This combination of properties may either be achieved by doping magnetism into topological matter, or by employing materials in which intrinsic magnetic order coexists with non-trivial band topologies. 
An example for the latter case is \FGT (FGT), a metallic vdW ferromagnet with a Curie temperature of $T_\mathrm{c}\approx220\rm\,K$ in the bulk \cite{chen2013magnetic, jang2020origin}. 
{While FGT's magnetic properties at temperatures $T\gtrsim T_\mathrm{c}$ originate from localized Fe moments, for $T\ll T_\mathrm{c}$ itinerant Stoner-like magnetism, carried by coherent delocalized Fe states, dominates \cite{zhu2016electronic, zhang2018emergence, kim2022fe3gete2}.
In addition to its ferromagnetic properties,} the band structure of FGT exhibits a topological nodal line gap that originates from a crossing of {hybridized Fe-I and Fe-II} spin-majority bands at its $\mathbf{k}=\rm K/K$' points (Ref.  \onlinecite{kim2018large} and Fig.~\ref{Fig1}a-c). 
The band hybridization near this topological gap results in large Berry curvatures $\boldsymbol{\Omega}(\mathbf{k})$ \cite{kim2018large} and in turn orbital magnetic moments $\mathbf{m}(\mathbf{k}) \sim - \boldsymbol{\Omega}(\mathbf{k})$ \cite{hanke2016role} which dominate the spin moments {at the edges of these bands} (Fig.~\ref{Fig1}d).
{The interplay of strongly correlated itinerant magnetism and Fe local moments results in a large magnetic anisotropy energy \cite{zhao2021electric, zhao2021kondo, srivastava2024unusual} that allows to control the magnetic state of FGT not only by external magnetic fields but also by charge-addressing stimuli, such as electric fields, currents and doping \cite{deng2018gate,wang2019current, park2019controlling, zhao2021electric}. 
This opens up the possibility to engineer magnetic properties at interfaces via the electric fields that arise when different materials are brought into contact, or even by external gates.}

\begin{figure*}[ht!]
  \centering
  \includegraphics[width=\textwidth]{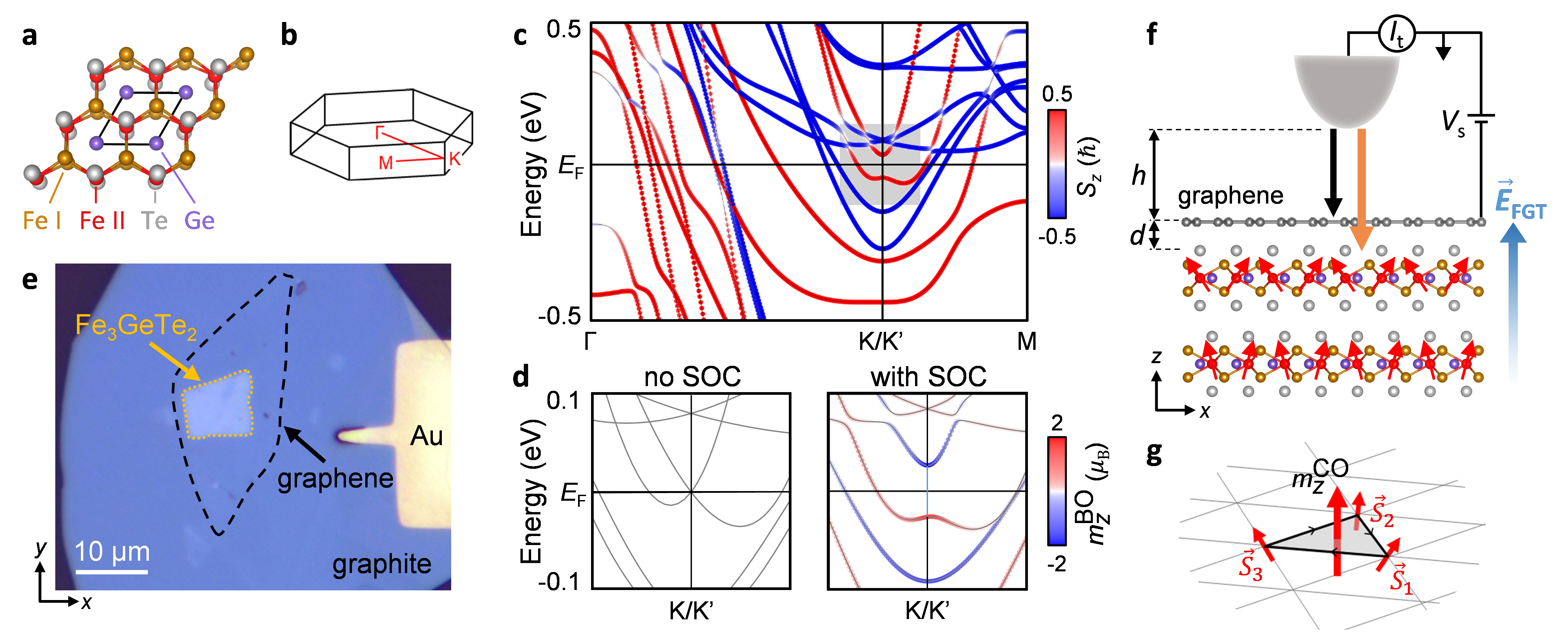}
  \caption{\label{Fig1} \textbf{{Orbital moments in \FGT and scanning tunnelling microscopy experiment of the graphene/\FGT heterostructure.}}
  \textbf{(a)} Top view of the monolayer \FGT unit cell, {highlighting the two inequivalent Fe sites}. \textbf{(b)} First Brillouin zone of \FGT with indicated symmetry points and lines. 
  \textbf{(c)} Spin-resolved band structure of bulk \FGT, calculated by density functional theory including spin-orbit coupling, with color-coded majority (red) and minority (blue) bands. Near $\Gamma$ only dispersive bands cross the Fermi energy \EF, while at K/K' several band extrema are located close to \EF.
  \textbf{(d)} Zoom into the shaded region in  panel c, calculated without (left) and with (right) spin-orbit coupling (SOC). The latter calculation shows out-of-plane orbital magnetic moments $m_{z}^\mathrm{BO}$ near the K/K' points. 
  \textbf{(e)} Optical micrograph of a graphene/\FGT heterostructure. Outlines of the respective flakes are indicated.
  \textbf{(f)} Schematic side view of the heterostructure and the scanning tunnelling microscopy setup. There are two parallel tunnelling channels, from the tip to graphene (black arrow) and to \FGT (orange arrow).
  The tip-sample distance $h$ and graphene-\FGT distance $d$ are indicated.  
  $V_{\rm s}$ is the sample bias and $I_{\rm t}$ the tunnelling current.
  At the graphene/\FGT vdW interface, a charge transfer dipole gives rise to an internal electric field $\vec{ {E}}_\mathrm{FGT}$, which results in a canting of the Fe spins at the interface (indicated by the red arrows at the Fe-II sites), due {to the Rashba effect and} Dzyaloshinskii–Moriya interaction (see main text for details).
  \textbf{(g)} {Schematic of non-collinear spin arrangement in the Fe-I sublattice, where a canting of localized spins $\vec S_i$ results in a conical spin structure.
  The ensuing scalar spin chirality gives rise to a chiral orbital moment $m_z^\mathrm{CO}\sim \vec S_1\cdot(\vec S_2\times \vec S_3)$ \cite{shindou2001orbital, tatara2003quantum, bulaevskii2008electronic, dos2016chirality}. }
  The crystal structures in panels a and f were created using VESTA \cite{momma2011vesta}.
  }
\end{figure*}

Here, we report a giant asymmetric Zeeman effect at the interface between monolayer graphene and FGT, which is {moreover} tunable by varying the interface dipole via electrostatic gating. 
A detailed analysis combining scanning tunnelling microscopy/spectroscopy (STM/STS) and theoretical calculations reveals that this unusual Zeeman effect is the result of two coupled orbital Zeeman effects:
On the one hand, the band hybridization near the topological band gap results in localized out-of-plane band orbital moments (BOM) $m_z^\mathrm{BO}$ near the K/K' points.
On the other hand, {a canting of Fe local moments} that has previously been reported in transport studies of graphene/FGT interfaces \cite{zhao2023strong, srivastava2024unusual} 
gives rise to chiral orbital moments (COM) $m_z^\mathrm{CO}$, which result from the winding of the {itinerant} charge carrier wave functions as they traverse the non-collinear texture {of localized moments} at the vdW interface
and which can directly couple to external magnetic fields with large $g$-factors \cite{hanke2016role, yin2018giant, lux2018engineering, grytsiuk2020topological}.
This complex compound Zeeman effect signals the emergence of subtle correlations between magnetic and topological electronic properties, which may open up the opportunity to engineer the response of the heterostructure to magnetic and electric fields, e.g., for chiral orbitronics applications. 
}

\section{Results}
\subsection{The graphene/FGT heterostructure}
\begin{figure*}[!t]
  \centering
  \includegraphics[width=0.6\textwidth]{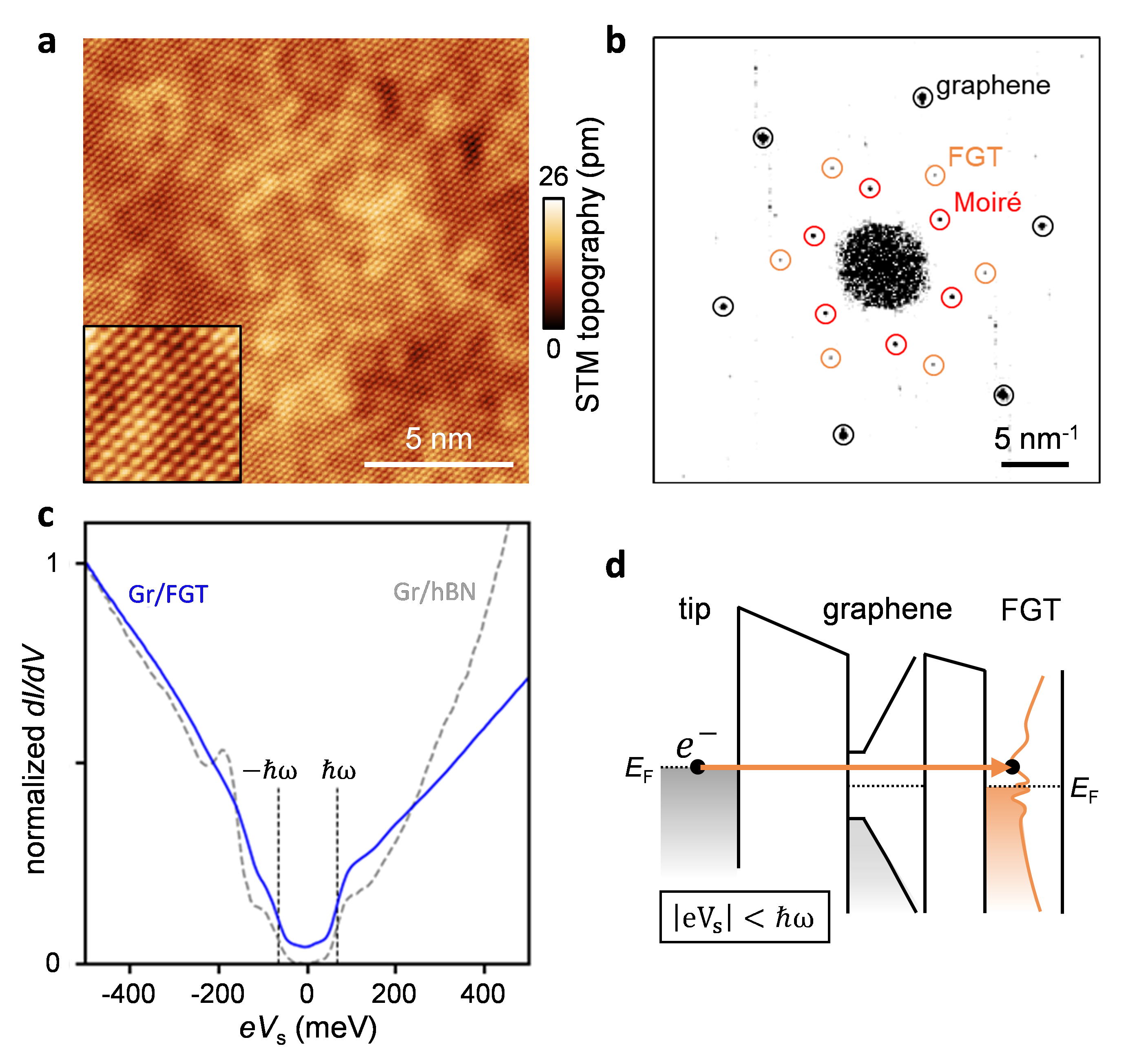}
  \caption{\label{Fig2} \textbf{Tunnelling through the inelastic gap of graphene.}
  \textbf{(a)} STM topography of the graphene/\FGT heterostructure surface ($V_\mathrm{s}=-50\,\rm mV$, $I_\mathrm{t}=70\,\rm pA$), simultaneously revealing atomic resolution of the graphene lattice and a longer-range modulation  stemming from the partial occupancy of the Fe-II sites in \FGT. Inset: Zoom into the graphene lattice (scan size $2.5\rm\,nm$).
  \textbf{(b)} Fourier transform of the topography in panel a. The hexagonal patterns of graphene (black circles), \FGT (orange circles), and their moiré pattern (red circles) are clearly visible.
  \textbf{(c)} Spatially averaged tunnelling conductance spectrum of the heterostructure (blue solid line, $V_\mathrm{s}=500\,\rm mV$, $I_\mathrm{t}=200\,\rm pA$).  
  Inset: Second derivative spectrum.
  Below the inelastic threshold, i.e., for $|eV_\mathrm{s}|<\hbar \omega$ where $\omega$ is the lowest available phonon frequency of graphene, an inelastic tunnelling gap is observed (vertical dotted lines), while outside this range the linear dispersion of the graphene Dirac cone is visible. Within the inelastic gap, a significant conductance is measured, in contrast to the findings for graphene/hBN heterostructures (grey dashed line, data from Ref.~\cite{decker2011local}).
  The graphene is close to charge neutrality in both cases.
  For better comparison, both data sets are normalized at $eV_{\rm s}=-500\rm\,meV$.
  \textbf{(d)} Schematic diagram of the tunnelling processes. With its gapless spectrum, \FGT dominates the tunnelling current within the inelastic gap of the graphene.
}
\end{figure*}

An optical micrograph of a graphene/FGT heterostructure, where a $28\rm\,nm$ thick FGT flake is encapsulated between monolayer graphene on the top and an additional graphite flake underneath, is shown in Fig.~\ref{Fig1}e. For details of the sample fabrication see the Methods section. 
{At this thickness and the experimental temperature of $\sim6\rm\,K$, an isolated FGT flake would be in a single-domain ferromagnetic state \cite{fei2018two, deng2018gate}, however, the symmetry breaking and electric field at the interface to the graphene give rise to a Dzyaloshinskii–Moriya interaction, which cants the {magnetic moments} at the interface away from surface normal \cite{srivastava2024unusual}, as indicated schematically in the sketch of the heterostructure and the tunnelling experiment in Fig.~\ref{Fig1}f.
}
Constant-current STM topography images of the graphene/FGT heterostructure (Fig.~\ref{Fig2}a) show a superposition of the atomic lattice of graphene and a long-range modulation that is a characteristic feature of FGT: It stems from random disorder in the partially occupied Fe-II sites \cite{zhao2021kondo}.
In the corresponding Fourier transform (Fig.~\ref{Fig2}b), we observe sharp spots due to the graphene ($a_{\rm gr}=246\rm\,pm$) and the FGT lattices ($a_{\rm FGT}=399\rm\,pm$), as well as their Moir\'e pattern ($a=524\rm\,pm$), indicating an atomically clean vdW interface. The relative orientation of both lattices is approximately $17^{\circ}$.

\begin{figure*}[!t]
  \centering
  \includegraphics[width=0.75\textwidth]{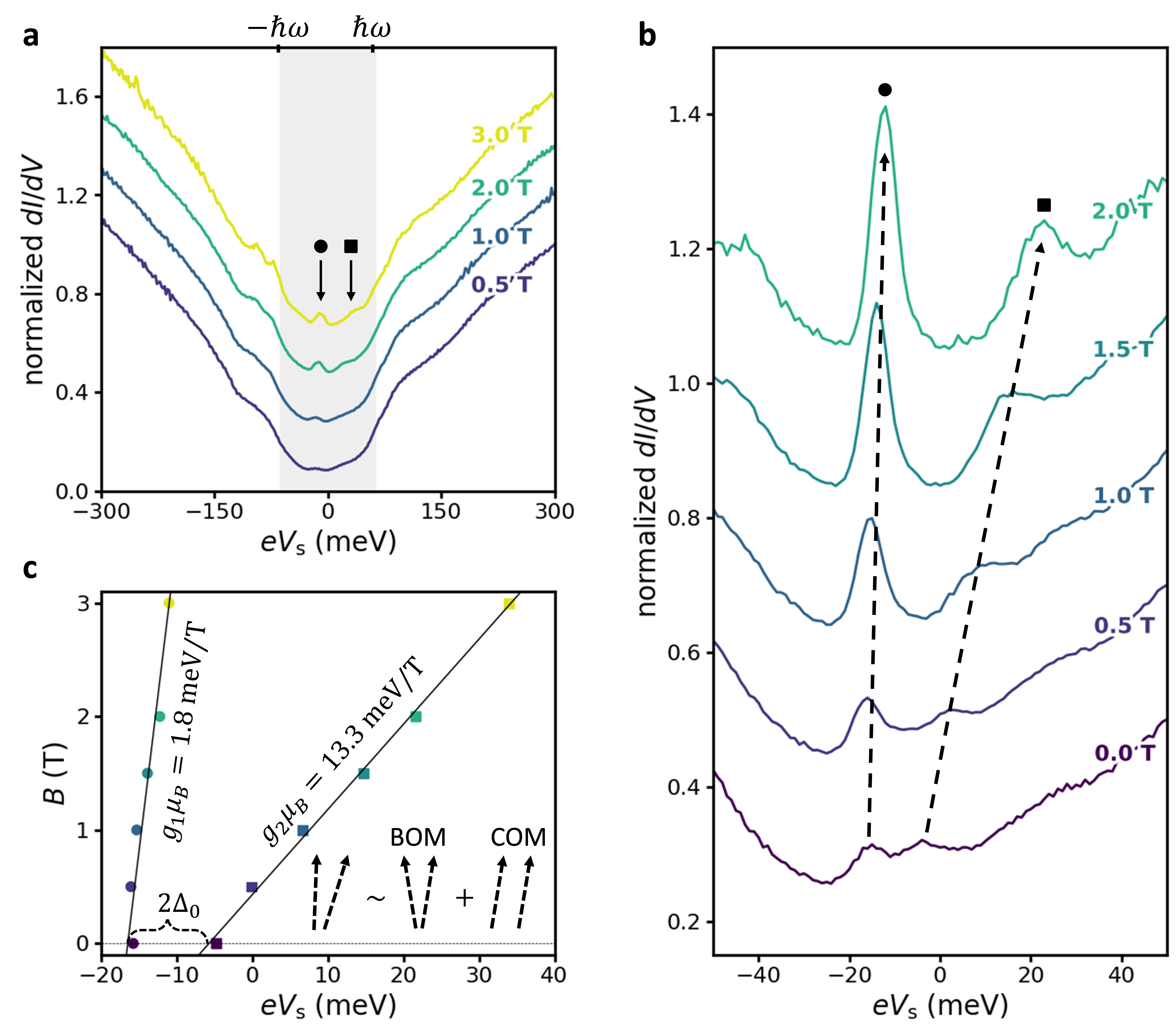}
  \caption{\label{Fig3} \textbf{Giant orbital Zeeman effects at the graphene/\FGT interface.}
  \textbf{(a)} Tunnelling spectra (differential conductance in arbitrary units) for different out-of-plane  $B$ fields (normal to surface).  For clarity, curves are offset by 0.2 arb.~units each.
  In the inelastic gap of graphene (shaded region), two peaks emerge with increasing $B$ field, as indicated by arrows and symbols. Tip stabilization parameters: $V_{\rm s}=300\rm\,mV$ and $I_{\rm t}=1\rm\,nA$.
  \textbf{(b)} Zoom into the inelastic-gap region, revealing both a continuous increase of the two peaks' intensities and shifts to higher energy as a function of applied $B$ field. Curves are offset by 0.2 arb. units each. Dashed arrows are guides to the eye. Tip stabilization parameters: $V_{\rm s}=50\rm\,mV$ and $I_{\rm t}=0.5\rm\,nA$.
  \textbf{(c)} Symbols: Peak positions (horizontal axis) as function of applied $B$ field (vertical axis).
  The solid lines are linear fits, yielding  effective Landé $g$-factors $g_1\approx30$ (circles) and $g_2\approx230$ (squares).
  Inset: Schematic decomposition of the observed peak shifts into a splitting and a shifting contribution. The former arises from  the band orbital moments (BOM) induced by the topological nodal line gap, while the latter has its origin in the chiral orbital moments (COM) induced by spin canting. For more details, see main text.
   }
\end{figure*}
\begin{figure*}[!t]
  \centering
  \includegraphics[width=0.75\textwidth]{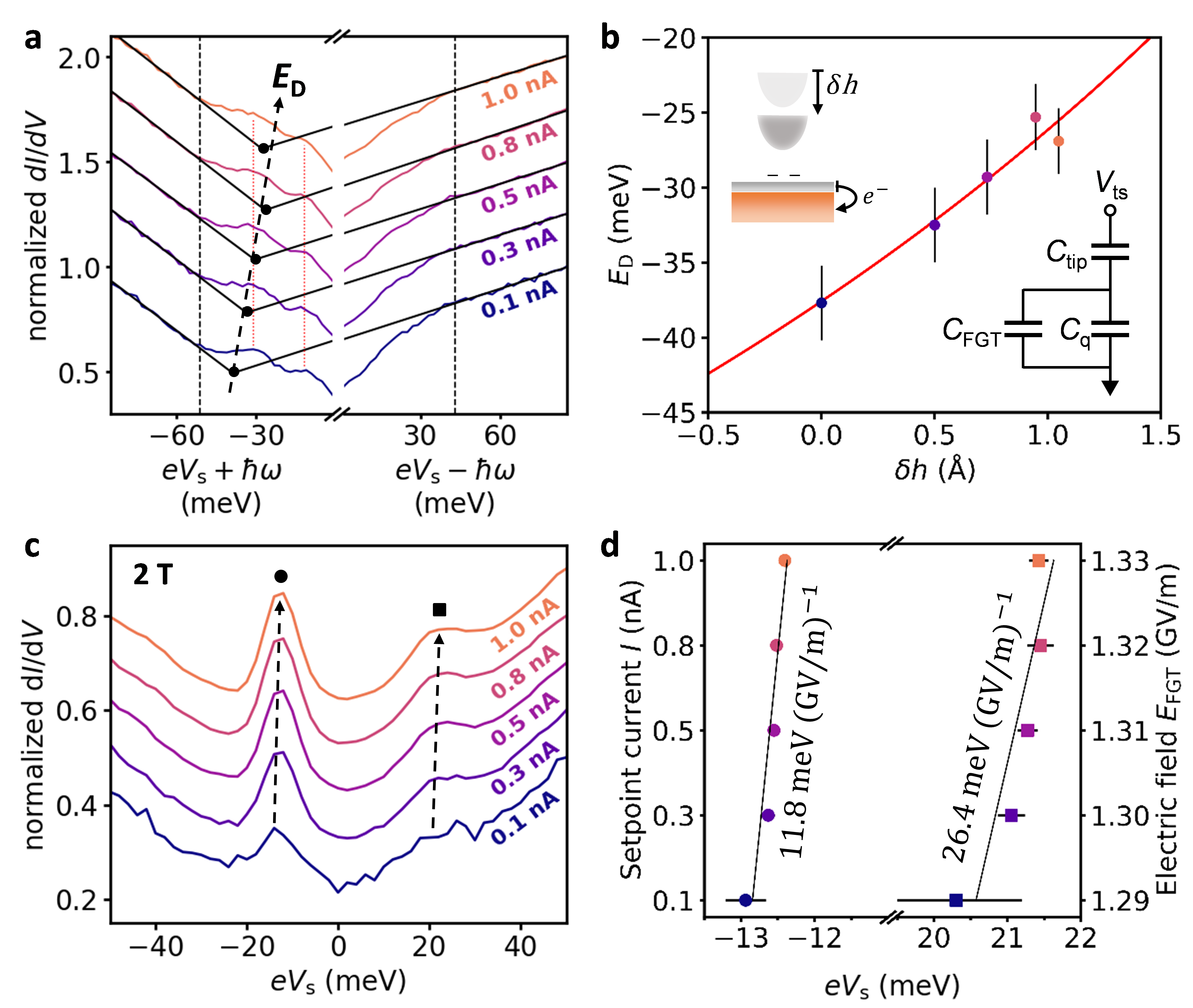}
  \caption{\label{Fig4} \textbf{Tuning the orbital Zeeman effects at the graphene/\FGT interface by electric fields.}
  \textbf{ (a)} Differential conductance spectra (in arbitrary units) for different tunnelling setpoints. The inelastic gap is cut out, and the curves for different setpoints are offset vertically for clarity. The Dirac point energy $E_{\rm D}$ (black dots) is extracted from the crossing points of line fits to the linear graphene spectrum above/below the vertical dashed lines, revealing a shift to higher values as the setpoint is increased. The dashed arrow is a guide to the eye. Remnant features due to \FGT near the edge of the inelastic gap do not shift with setpoint, as indicated by red dotted lines. The tip stabilization voltage was $V_\mathrm{s}=150\rm\,mV$ for all setpoints.
  \textbf{(b)} Dirac point energies $E_{\rm D}$ (filled circles, extracted from panel (a)) plotted as function of the change in tip-sample distance $\delta h$ (see upper inset). Error bars are propagated from the fits of the Dirac spectrum in panel (a).
  The red curve is a fit to the data points, using the capacitor model  in the lower inset (Eqs.~\ref{eq:ngr} and \ref{eq:E_D} in the Methods section).
  \textbf{(c)} Differential conductance spectra (in arbitrary units) of the \FGT-derived peaks in the inelastic gap of graphene for different tunnelling setpoints (curves are vertically offset for clarity. All spectra are recorded at a fixed external field of $B=2\rm\,T$.
  \textbf{ (d)} Energy positions of the peaks in panel c as a function of setpoint current.
  Error bars correspond to the peak fit uncertainties (see Extended Data Fig.~\ref{Fig-SI-peak-extraction2} for details).
 }
\end{figure*}

\subsection{The inelastic tunnelling gap of graphene}
Tunnelling spectra recorded on the heterostructure reveals the characteristic V-shaped density of states of the graphene Dirac cone (Fig.~\ref{Fig2}c), as well as an inelastic tunnelling gap at the Fermi energy \EF ($eV_{\rm s}=0$). 
This gap results from a suppression of the tunnelling probability due to a momentum mismatch between the tunnelling tip and the finite parallel momenta of the graphene Dirac states at the K/K' points {(see Refs.~\onlinecite{zhang2008giant, wehling2008phonon, decker2011local} and the Methods section for a more detailed discussion)}. 
However, above a threshold energy, i.e., for $|eV_{\rm s}|\geq \hbar \omega=(65\pm 2)\rm\,meV$, phonon-assisted tunnelling channels become available, which significantly enhance the tunnelling probability. 
Our observed onset energy of the inelastic tunnelling channels corresponds to the lowest-energy phonon mode {at the graphene K point} \cite{natterer2015strong}, in good agreement with the phonon thresholds reported for graphene on different substrates \cite{zhang2008giant, decker2011local, qiu2021visualizing}. 
This agreement indicates that the graphene is neither significantly strained nor strongly doped. 
However, in contrast to graphene on insulating substrates, where the $dI/dV$ signal almost vanishes below the phonon threshold \cite{zhang2008giant, decker2011local, natterer2015strong, sun2023determining}, we observe a significant tunnelling conductance for $|eV_{\rm s}|<\hbar\omega$. 
The reason is that, while tunnelling to the graphene is suppressed in this energy range, charge carriers can still tunnel between the tip and the metallic FGT underneath the graphene, which allows us to directly access the electronic properties at the graphene/FGT interface (Fig.~\ref{Fig2}d {and the Methods section}).\\

\subsection{Giant asymmetric Zeeman effect}
{A careful analysis of the tunnelling spectrum inside the inelastic phonon gap of graphene reveals two peaks, which strengthen on application of a magnetic field (Fig.~\ref{Fig3}a)}.  At the same time, the linear Dirac spectrum of graphene outside the inelastic tunnelling gap remains mostly unaffected.
Strikingly, the two peaks within the inelastic gap show significant energy shifts as function of the magnetic field.
Linear fits to the data result in 
$g_1\mu_\mathrm{B}=(1.8\pm0.3)\rm\,meV/T$ and $g_2\mu_\mathrm{B}=(13.3\pm0.4)\rm\,meV/T$,
corresponding to Landé $g$-factors $g=E_{\rm Zeeman}/(\mu_{\rm B}B)$ of $g_1\approx30$ and $g_2\approx230$, respectively (Fig.~\ref{Fig3}b and Extended Data Fig.~\ref{Fig-SI-peak-extraction}). 
Furthermore, a spatial analysis of the peaks shows that they are consistent across the heterostructure surface and not localized at defects (Extended Data Fig.~\ref{Fig-SI-peak-energies}), pointing to a band{-}structure origin of 
{of these features}. 
Considering the FGT band structure, the observed peaks can be intuitively understood to stem from the flat band edges near the topological gap at the K/K' points, where the orbital magnetic moments {$m^{\rm BO}_z$}, {concentrated} at the band extrema, lead to large $g$-factors (Fig.~\ref{Fig1} and Refs. \onlinecite{li2022manipulation, li2024spin}).
This interpretation also explains the increasing peak intensity and sharpness with increasing magnetic field in Fig.~\ref{Fig3}b, analysed in detail in Extended Data Fig.~\ref{Fig-SI-peak-extraction}: {DFT calculations show opposite orbital moments at the two band edges (Fig.~\ref{Fig1}d), such that the topological gap increases as function of the applied magnetic field.} Concurrently, the band edges at K/K' flatten out further, again because the orbital magnetic moment is concentrated at the band extrema (Fig.~\ref{Fig1}c, d and Refs. \cite{li2022manipulation, li2024spin}).
Extracting the splitting of the two peaks at $B=0$ results in $2\Delta_0=(10.8\pm0.7)\rm\,meV$, which agrees reasonably well with our DFT calculated spin orbit gaps at K/K' {of $\sim52\rm\,meV$} and Refs.~\cite{kim2018large, zhao2021electric}, considering that the gap size decreases when inversion symmetry is broken (Extended Data Fig.~\ref{Fig-SI-trilayer}). 
{Note that the smaller the gap at $B=0$}, the larger the orbital moments $m^{\rm BO}_z\sim\Delta^{-1}$ at K/K' (see Methods section).

{While the model of opposite orbital moments $m^{\rm BO}_z$, concentrated at the upper and lower band edges, respectively, qualitatively explains both the observation of large $g$ factors and of tunneling conductance peaks with increasing intensity and sharpness as the magnetic field is turned up, it disagrees with our experimental observations in that it would predict a $B$-field-induced shift of these conductance peaks in {opposite} directions, whereas we find that the peaks move in the {same} direction (to higher energy)}, albeit with different slopes (Fig.~\ref{Fig3}c). {We therefore conclude} that, in addition to the band orbital effect, there must be a second Zeeman effect at play which shifts both {conductance} peaks to higher energies, as schematically shown in the inset in Fig.~\ref{Fig3}c.
Considering the band structure of FGT, such an effect is indeed expected, because the bands forming the topological gap exhibit the same spin polarization (Fig.~\ref{Fig1}c, Extended Data Fig.~\ref{Fig-SI-exchange_calc}).
However, including {the corresponding} spin Zeeman effect, the {combined $g$-factor of BOM and the band's spins} at K/K' is only $g^{\rm BO}_z = |2 + {\langle L_{} \rangle}/{\langle S_{} \rangle}|\lesssim10 $ (Ref. \onlinecite{jo2025weak} and Extended Data Fig.~\ref{Fig-SI-g-factor}), {where $\langle L_{} \rangle \sim 3\,\hbar$  and $\langle S_{} \rangle \sim 0.5\,\hbar$ are taken from our DFT calculations}, corresponding to shifts of $\sim 0.6\rm\,meV/T$.
Thus, the expected Zeeman effect purely based on FGT's {spin and orbital band magnetism} is much too small to explain the experimental observations {in Fig.~\ref{Fig3}}.

At this point, we recall that the {local Fe moments} at the graphene/FGT interface are canted  {(Fig.~\ref{Fig1}f)}, which is not included in the  DFT calculations. 
The reason for the canting is the electric field perpendicular to the interface between the two vdW layers, which gives rise to a Rashba spin-orbit coupling and thus Dzyaloshinskii–Moriya interaction (DMI), {which acts on the localized moments of primarily Fe-I $3d_{x^2-y^2/xy}$ orbitals} \cite{kim2022fe3gete2, zhao2023strong, srivastava2024unusual}.
The Rashba coefficient at the graphene/FGT interface was theoretically estimated as $\alpha_{\rm R}\approx 246\rm\,meV$\AA, giving rise to a DMI strength of $|\vec{D}|=0.089\rm\,meV$ per unit cell \cite{srivastava2024unusual}, which we expect to be a lower bound {for our experiments, because of an enhancement of the interface dipole by the tip compared to experiments discussed in the literature --- see the next section}.
While at intermediate temperatures $10\rm\,K< T<220\rm\,K$, this effect leads to skyrmion formation \cite{srivastava2024unusual}, at our lower experimental temperatures we most likely stabilize a conical phase \cite{muhlbauer2009skyrmion, li2023tuning}, as schematically shown in Fig.~\ref{Fig1}f, g. 

The un-canting of {local moments} by the application of {an out-of-plane} magnetic field is expected to ultimately result in the {moments} being aligned out-of-plane.
{However, to fully align the moments fields much larger than $0.089\rm\,meV/\mu_{\rm B}\gtrsim1.5\rm\,T$ (Ref. \cite{srivastava2024unusual}) must be applied, i.e. on the order of $10\rm\,T$ which are not accessible in our experiment. }
Also, it cannot explain the observed large $g$-factor of the Zeeman effect in Fig.~\ref{Fig3}c, because when the {moments} are canted away from the surface normal and thus are no longer aligned with the $B$ field, {their} Zeeman effect is weakened instead of enhanced. 
Instead, we propose that the winding of the charge carrier wave functions {of the spin-polarized bands that form the topological gap,} as they traverse the non-collinear magnetic texture at the vdW interface, gives rise to chiral orbital moments {(COM)}, which are described in detail in the literature \cite{shindou2001orbital, tatara2003quantum, bulaevskii2008electronic, dos2016chirality, lux2018engineering, yin2018giant, niu2019mixed}. 
In our system, since the bands {in question} have identical spin polarization, we expect any {local-moment} canting-induced COM to affect both bands in the same way, i.e., result in a Zeeman effect which shifts both bands to higher energy as we increase the magnetic field (inset Fig.~\ref{Fig3}c).
Crucially, the canting, and consequently the COMs, are the result of the Rashba effect and the resulting DMI at the graphene/FGT interface. 
Thus, the resulting {giant} Zeeman effect should be tunable by variation of the electric field across the interface, which we demonstrate in the following.

\subsection{Electric field tunability of the Zeeman effect}
From simple electrostatic considerations, we estimate the electric field at the graphene/FGT interface as 
\begin{eqnarray}
{{E}_{\rm FGT}\equiv\vec E_{\rm FGT}\cdot \hat z=(\Phi_{\rm gr}-\Phi_{\rm FGT})/ed\,\approx 1.41\, \mathrm{GV/m}},\label{eq:E-field}
\end{eqnarray}
with the interlayer distance between graphene and FGT {$d= 3.34\rm\,$\AA} (Fig.~\ref{Fig1}f), the elementary charge $e$, the work function of FGT $\Phi_{\rm FGT}\approx4.08\rm\,eV$ \cite{ko2022hybridized}, and $\Phi_{\rm gr}=\Phi_{\rm gr}^0\approx 4.55\rm\,eV$ \cite{yu2009tuning} the work function of charge-neutral graphene, respectively. 
However, when the heterostructure is initially assembled, the work function difference between graphene and FGT gives rise to a charge transfer {(electrons flow from the low-work function FGT to the high-work function graphene)}, resulting in {an intrinsically interface-driven $n$ doping of the} graphene  {and a corresponding shift of the Dirac point energy $E_{\rm D}^0$ relative to the Fermi level,}
and thus the graphene work function becomes $\Phi_{\rm gr}=\Phi_{\rm gr}^0+E_{\rm D}^0$, where $E_{\rm D}^0$ is {\it a priori} unknown.

In addition to the {interface-driven} doping, the presence of the STM tip {with its own distinct work function}  (${\Phi_{\rm tip}\approx4.{\color{red}7}\rm\,eV}$, see Methods section) should affect $E_{\rm D}$ due to tip gating {(clearly, also the applied bias voltage will contribute to a polarity-dependent gating; however, for bias voltages within the inelastic gap of graphene this effect is so small that it can be neglected).} 
{In Fig.~\ref{Fig4}a, tunnelling spectra recorded at different setpoints and therefore different tip heights $h$ are plotted. Fitting the linear graphene spectrum outside of the inelastic gap to extract the setpoint-dependent Dirac point energy $E_\mathrm{D}$ from the crossing points of the linear fits (Fig.~\ref{Fig4}a and Extended Data Table \ref{Tab-SI-exp}), we observe a shift of $E_\mathrm{D}$ to more positive values as $h$ decreases. This corresponds to the graphene becoming less electron-doped as the tip is brought closer to the sample surface.
In contrast, additional FGT features that are located outside the inelastic gap of graphene show no sign of shifting as function of the setpoint current, demonstrating that mainly the graphene doping level is changed by the tip gating (Fig.~\ref{Fig4}a)}.

{In order to determine the intrinsic doping level of graphene at the interface to FGT, w}e model the experimental tip-sample geometry by a capacitor circuit (lower inset in Fig.~\ref{Fig4}b and Methods section), which we fit to the experiment.
Extrapolating this model to $\delta h\rightarrow-\infty$, i.e., {to the absence of the tunnelling tip}, we find a graphene doping level of {$E_{\rm D}^0=-(579\pm171)\rm\,meV$} (Fig.~\ref{Fig4}b).
The corresponding {intrinsic} electric field {at the graphene/FGT interface is} {${E}_{\rm FGT}=(-0.3\pm 0.5)\rm\,GV/m$}, {i.e., lower and of opposite sign compared to} charge-neutral graphene and FGT.
We note that a non-zero field is the prerequisite for the anomalous Hall effects and giant magnetoresistances at the graphene/FGT interface reported in the literature \cite{srivastava2024unusual, huang2025giant}.
In comparison, in the presence of the STM tip the interfacial electric field is significantly larger, in the range {${E}_{\rm FGT}=1.29 \,\,\text{to}\,\, 1.32\rm\,GV/m$} for tunnelling setpoints $0.1-1\rm\,nA$ (see Extended Data Table 1) and thus also the DMI is expected to be larger.

To demonstrate that the variation of the interfacial electric field leads to a variation of the Rashba splitting and DMI, 
we study the peak positions inside the inelastic gap as function of tunnelling setpoint while keeping $B=2\rm\,T$ fixed (Fig.~\ref{Fig4}c).
Doing so, we find an increase in the intensity of both peaks as we bring the tip closer to the sample surface.
Such behaviour is a known effect for tunnelling into sharp energy levels originating from states with finite in-plane momenta, such as flat bands and localized states at K/K' \cite{kim2013visualizing, luican2019negative, yin2020imaging, zhao2021electric}, which again is consistent with our DFT calculations.
More interestingly though, extracting the {precise} peak positions as function of the tunnelling setpoint once more reveals systematic energy shifts (Fig.~\ref{Fig4}d and Extended Data Fig.~\ref{Fig-SI-peak-extraction2}).
Linear fits to the data (solid lines in Fig.~\ref{Fig4}d) are described by $eV_{\rm s}=(11.8\pm2.8)\mathrm{\,meV\,(GV/\,m)^{-1}} E_{\rm FGT}-(28.1\pm2.8)\rm\,meV$ and $eV_{\rm s}=(26.4\pm8.1)\mathrm{\,meV\,(GV/\,m)^{-1}} E_{\rm FGT}-(13.5\pm10.6)\rm\,meV$, respectively.
Remarkably, the peak positions show a qualitatively similar asymmetric splitting as function of ${E}$ field at fixed $B$ field, as they do for the increasing $B$ field at fixed ${E}$ field (Extended Data Fig.~\ref{Fig-SI-peak-extrapolation}).
\\

\subsection{Coupling between band and chiral orbital moments}
Our observed DMI-induced Zeeman effect indicates a strong inversion symmetry breaking at the graphene/FGT interface, evidenced by measurements on bulk FGT crystals, where no significant Zeeman effect is observed when the inversion symmetry is only weakly broken (Extended Data Fig.~\ref{Fig-SI-bulk}a and Ref. \cite{zhao2021electric}).
In general agreement with this interpretation, such symmetry breaking has recently been proposed as the origin of the anomalous Hall effect in graphene/FGT  \cite{garcia2025origins}, which can be understood as the manifestation of the Zeeman effect in transport.
{Our observations imply that the band-structure-induced and chirality-induced moments are directly coupled beyond a mere superposition, which may explain the enhancement of the BOM by more than one order of magnitude compared to the theoretical moments alone.} 
Currently, the exact interaction mechanism between the chiral and band orbital moments is yet to be understood, but considering that both effects are closely related to the Berry curvature, it does not come as a surprise that they are coupled.
Estimating the $g$-factors of the two contributions according to the inset in Fig.~\ref{Fig3}c results in $g_\mathrm{z}^\mathrm{BO}\approx100$ and $g_\mathrm{z}^\mathrm{CO}\approx130$, respectively, signifying their {similar strength} (see Methods section).
In this context, we note that the {spin} Berry curvature vanishes for the band structures shown in Fig.~\ref{Fig1}c,d.
However, spin canting likely leads to a finite spin Berry curvature, with similar localization at the K/K' points as the Berry curvature \cite{li2024spin}, which would couple the spin canting directly to the topological gap.
Similar effects were recently reported in magnetic Kagome compounds, where the orbital moments of gapped Dirac cones interact with local moments \cite{yin2018giant, sales2019electronic, yin2019negative, xing2020localized, ren2022plethora, li2022manipulation, li2024spin}.
In the absence of spin canting, these bulk compounds can show spin-orbit-driven orbital Zeeman effects \cite{yin2019negative, li2024spin}, which are initially approximately a linear function of the applied $B$ field but saturate as the Berry curvatures decrease with increasing gap sizes.
In the presence of spin canting, however, additional orbital moments with $g\gg2$ {were reported}, due an interplay of the spin-orbit gap and spin chirality \cite{hanke2016role, yin2018giant, grytsiuk2020topological, li2022manipulation}.
{However,} in contrast to the Kagome compounds, where the Zeeman effects typically saturate at fields of $B\lesssim2\rm\,T$, we do not observe any sign of a saturation {at the graphene/FGT interface} up to our maximum accessible field of $3\rm\,T$, which is consistent with the spin un-canting at the graphene/FGT occuring only at much larger fields $B\sim 10\rm\,T$.

In conclusion, despite the lack of strong bonding in graphene/FGT vdW heterostructures, interface dipoles give rise to emergent magnetoelectronic properties, which are localized {at the 2D interface} and result in a rich interplay of orbital magnetism {and local moments}.
{The required interfacial fields are readily achievable in state-of-the-art electronic devices \cite{taychatanapat2010electronic, uwanno2018electrically, chen2020gate}, highlighting the technological potential of our findings.
Thus, the present} vdW interphase provides a flexible platform to engineer magnetic properties, allowing to control them via gate-tuning the Rashba effect and DMI, which may find applications in spintronics or orbitronics.

\section{Methods}
\subsection{Density functional theory}
The orbital magnetization in three-dimensional $k$-space is given by \cite{hanke2016role}
\begin{equation} \label{OMmodern}
\begin{split}
\textbf{M}{^{\mathrm{BO}}} = &\frac{e}{2 \hbar}\mathrm{Im}\sum_{n}\int_{\textrm{BZ}} \frac{d\textbf{k}}{\left( 2\pi\right)^{3}} f_{n\textbf{k}} \langle \partial_{\textbf{k}} u_{n\textbf{k}} \vert \times  \left( \hat{H}_{\textbf{k}} + \varepsilon_{n\textbf{k}} - 2\mu \right)  \\& \vert \partial_{\textbf{k}} u_{n\textbf{k}} \rangle ,
\end{split}
\end{equation}
where $\vert u_{n\textbf{k}} \rangle$ is the Bloch-like eigenstate associated with the $n^\mathrm{th}$ eigenvalue $\varepsilon_{n\mathbf{k}}$ of the Hamiltonian $H_\mathbf{k}$ of the system, $f_{n\textbf{k}}$ is the zero-temperature Fermi occupation factor and $\mu$ the chemical potential. 
Note that the vector product is taken between the two derivatives $\partial_{\textbf{k}} = \frac{\partial}{\partial \textbf{k}}$ {in the bra and ket, respectively}.\\

Unlike transport, which probes all occupied states and is therefore sensitive to the orbital magnetization, STM/S probes the either filled or empty states at a given energy and is thus sensitive to the orbital magnetic moment (OMM) of the band, which is given by
\begin{align}
m_n(\mathbf{k})^{\mathrm{BO}} &= \frac{e}{2 \hbar}\mathrm{Im}\langle \partial_{\textbf{k}} u_{n\textbf{k}} \vert \times  \left( \hat{H}_{\textbf{k}} - \varepsilon_{n\textbf{k}} \right) \vert \partial_{\textbf{k}} u_{n\textbf{k}} \rangle .
\label{eq:mz}
\end{align}
To calculate this quantity using \textit{ab initio} methods, we performed the following procedure:
The electronic structure of bulk \FGT was computed with the density functional theory code FLEUR \cite{FLEUR} using the full-potential linearized augmented plane wave method \cite{wimmer1981full}.
We chose the Perdew-Burke-Ernzerhof exchange correlation functional within the generalized gradient approximation \cite{perdew1996generalized}. The FGT lattice parameters are $a=3.99\rm\,$\AA\ and $c=16.33\rm\,$\AA\ in a hexagonal close-packed lattice \cite{FGT_Springer}.
We performed a full self-consistent calculation to find the charge density including the effect of spin-orbit coupling within FLEUR, with a resulting magnetocrystalline anisotropy of $1.06\rm\,meV$ per Fe atom.
The plane-wave cutoffs for the basis functions, for the charge density and for the exchange correlation functional were set to $5a_0^{-1}$, $15a_0^{-1}$, and $12.5a_0^{-1}$, respectively, where $a_0$ is the Bohr radius. For Fe and Ge, the maximum in the angular momentum expansion was set to $l_\mathrm{max}=8$ and the muffin-tin radii $r_{\rm MT}$ for both  species was set to $2.12a_0$; for Te, $l_\mathrm{max}=10$ and $r_{\rm MT}=2.74a_0$. 
Despite previous work on FGT indicating the effects of strong electronic correlations \cite{zhang2018emergence, zhao2021kondo}, in the present scenario we do not see indications of such correlations, e.g. as the conduction band minimum is moved through the Fermi energy, and thus did not consider them theoretically.
Using the Wannier90 package \cite{wang2006ab, pizzi2020wannier90}, from the DFT-calculated Bloch wave functions $\vert u_{n\textbf{k}} \rangle$, we constructed maximally localised Wannier functions (MLWFs)  
and determine the Bloch-like basis $\vert u_{n\textbf{k}}^{\textrm{W}} \rangle$ \cite{freimuth2008maximally, lopez2012wannier}. 
A mesh of 8 × 8 × 8 $k$-points was used with 192 Bloch states to obtain 96 MLWFs. Initial projections were chosen to be $d$ states for Fe and $p$ states for Ge and Te atoms. The maximum frozen window was set to $2.344\rm\,eV$ above the Fermi energy. The Hamiltonian, spin, and orbital angular momentum (OAM) operators were evaluated in the Bloch basis and transformed into the MLWF basis. The OAM was integrated inside the muffin-tins. Further details of the calculations can be found in Ref.~\onlinecite{saunderson2022hidden}. \\

In the Wannier description, the orbital magnetization can be quantified as proposed by Lopez \textit{et al.} \cite{lopez2012wannier}:
First, the Bloch-like basis $\vert u_{n\textbf{k}}^{\textrm{W}} \rangle$ is transformed into the so called Hamiltonian gauge, using the transformation
\begin{align}
\vert u_{n\textbf{k}}^{\textrm{H}} \rangle = \sum_{m}^{J} \vert u_{m\textbf{k}}^{\textrm{W}} \rangle U_{mn} \left( \textbf{k} \right) .
\end{align}
This is necessary in order to compute the magnetization according to the so-called "modern theory of orbital magnetization" expression Eq.~\ref{OMmodern}.
This expression allowed us to decompose the orbital magnetization into its local
\begin{align}
\textbf{M}{^{\mathrm{BO}}_{\textrm{LC}}} = \frac{e}{2 \hbar}\mathrm{Im}\sum_{n} \int_{\textrm{BZ}} \frac{d\textbf{k}}{\left( 2\pi\right)^{3}} f_{n\textbf{k}} \langle \partial_{\textbf{k}} u_{n\textbf{k}}^{\textrm{H}} \vert \times  \left( \hat{H}_{\textbf{k}} - \mu \right) \vert \partial_{\textbf{k}} u_{n\textbf{k}}^{\textrm{H}} \rangle , \label{eq:local}
\end{align}
and itinerant contributions
\begin{align}
\textbf{M}{^{\mathrm{BO}}_{\textrm{IC}}} = \frac{e}{2 \hbar}\mathrm{Im}\sum_{n} \int_{\textrm{BZ}} \frac{d\textbf{k}}{\left( 2\pi\right)^{3}} f_{n\textbf{k}} \langle \partial_{\textbf{k}} u_{n\textbf{k}}^{\textrm{H}} \vert \times  \left( \varepsilon_{n\textbf{k}} - \mu \right) \vert \partial_{\textbf{k}} u_{n\textbf{k}}^{\textrm{H}} \rangle . \label{eq:itinerant}
\end{align}
We computed the band-resolved contributions to the magnetization
\begin{align}
{m}{^{\mathrm{BO}}_{n, \textrm{LC}}} = \frac{e}{2 \hbar}\mathrm{Im} \langle \partial_{\textbf{k}} u_{n\textbf{k}}^{\textrm{H}} \vert \times  \left( \hat{H}_{\textbf{k}} - \mu \right) \vert \partial_{\textbf{k}} u_{n\textbf{k}}^{\textrm{H}} \rangle , \label{eq:local_m}
\end{align}
and
\begin{align}
{m}{^{\mathrm{BO}}_{n, \textrm{IC}}} = \frac{e}{2 \hbar}\mathrm{Im} \langle \partial_{\textbf{k}} u_{n\textbf{k}}^{\textrm{H}} \vert \times  \left( \varepsilon_{n\textbf{k}} - \mu \right) \vert \partial_{\textbf{k}} u_{n\textbf{k}}^{\textrm{H}} \rangle . \label{eq:itinerant_m}
\end{align}
similar to Ref. \onlinecite{lopez2012wannier}, where it was shown that Eqs.~\ref{eq:local} and \ref{eq:itinerant} can be recast into a gauge-invariant formulation. 
A more detailed theoretical description will be presented in a forthcoming publication.
Finally the orbital magnetic moments in Eq.~\ref{eq:mz} were determined from the $k$-resolved local and itinerant contributions as
$m_n(\mathbf{k})^{\mathrm{BO}}={m}^{\mathrm{BO}}_{n, \textrm{LC}}(\mathbf{k})-{m}^{\mathrm{BO}}_{n, \textrm{IC}}(\mathbf{k})$.

{We note at this point that an earlier proposed model of the relevant bands at K/K' \cite{kim2018large}, when inserted into Eq.~\ref{eq:mz}, results in orbital moments which directly contradict our DFT calculations, indicating that such a model does not seem suitable to capture the full orbital physics.} 

\subsection{Sample fabrication}
Graphite and FGT crystals were exfoliated onto SiO$_2$ and assembled by a standard dry-transfer technique with a microdome polyvinylchlorid/polydimethylsiloxane (PVC/PDMS) stamp \cite{waka2020pvc}.
In the resulting heterostructure, the encapsulation of FGT by the larger graphene and graphite flake prevents the former's degradation.
Ti/Au contacts were fabricated using maskless optical lithography, and samples were subsequently cleaned in an ambient-condition atomic force microscope by repeated scanning in contact mode over night.
Extended Data Fig.~\ref{Fig-SI-hts} shows an AFM image, recorded before the STM measurements, of the heterostructure displayed in Fig.~\ref{Fig1}e. 
After introduction into the STM ultra-high vacuum chamber, the sample was annealed at 240$^{\circ}\rm C$ for $15\rm\,mins$ to desorb residues and water and subsequently transferred to the STM. There, the sample was zero-field cooled to the base temperature of $\sim6.1\rm\,K$, and magnetized by ramping the surface-normal field to $B=2\rm\,T$, prior to STS measurements.

Graphite and \FGT were obtained from commercial sources (graphite from HQ graphene, FGT from 2D semiconductors).
FGT bulk crystals from the same batch were characterized by angle-resolved photoemission spectroscopy (ARPES), showing general agreement with the DFT calculations and literature, and a doping level for which \EF is close to the topological gap (Extended Data Fig.~\ref{Fig-SI-bulk}b).
STM/S spectra taken on the bulk crystals did not exhibit any peak features and virtually no magnetic field dependence (Extended Data Fig.~\ref{Fig-SI-bulk}b), confirming the graphene/FGT interface as the source of the observed behaviour.

\subsection{Scanning tunneling experiments}
All STM/STS measurements were performed in a commercial Sigma Polar instrument at $\sim6.1\rm\,K$. Electro-chemically etched W tips were annealed to orange glow in vacuum and prepared on an Ag(111) surface. 
After introducing the heterostructure sample to the STM scanner, we first approached the tip to the contact leads using visual control and then navigated the tip to the heterostructure by scanning.
On the heterostructure, we performed large-area STM scans to locate the graphene and FGT flakes and to check the cleanliness of the heterostructure surface.
To ensure a clean tunnelling spectrum, we then moved the tip back to the Ti/Au contact and safeguarded that it showed a linear $I(V)$ characteristics by repeated careful indentation.
We then retracted the tip out of tunnelling contact and jumped directly back to the clean surface of the heterostructure, where we performed the measurements, thus preventing any tip alterations that may otherwise occur during intermediate scans.
All spectroscopy measurements were performed after ramping the magnetic field above the saturation field of the FGT flake, to achieve a well-defined magnetization.
STS was performed using standard lock-in techniques at a frequency of $877\rm\,Hz$ and modulation amplitudes of $V_\mathrm{s}^\mathrm{mod}=1 \,\,{\rm to}\,\,3\rm\,meV$.
STM topographs were analyzed using {Gwyddion} \cite{nevcas2012gwyddion}.

{
The trapezoidal shape of the biased tunneling barrier in the WKB-approximation results in a non-linear dependence of the tunneling conductance as the voltage is increased:
\begin{equation}
    dI/dV\propto \rho\ e^{-2\kappa(E) z},
    \label{Eq.R1}
\end{equation}
with the density of states $\rho$, the decay constant
\begin{equation}
    \kappa(E) = \sqrt{\frac{2m}{\hbar^2}(\Phi_\mathrm{eff}-E)+ k_{\parallel}^2},
\end{equation}
the effective barrier height
\begin{equation}
    \Phi_\mathrm{eff}=\frac{\Phi_\mathrm{tip}+\Phi_\mathrm{sample}}{2}
\end{equation} 
for tunneling bias $E=eV_\mathrm{s}\ll\Phi_\mathrm{eff}$, and the momentum in the surface plane $k_{\parallel}$.
As a result, states close to \EF tend to have a smaller $dI/dV$ intensity than those at higher bias.
In graphene, the intensity difference between states at lower and higher tunneling bias is further enhanced by the inelastic tunneling channels that become available at $|eV| > E_\mathrm{ph}$, which significantly increase the tunneling probability \cite{zhang2008giant}.
Inside this inelastic gap, the elastic tunneling current to the graphene is small, such that tunneling from the FGT dominates:
Although the FGT is further away from the tip than the graphene and thus has lower tunneling probability (Eq.~\ref{Eq.R1}), because of its significantly higher density of states compared to the graphene, it dominates the tunneling for $|eV| < E_\mathrm{ph}$:
We estimate the respective elastic tunneling probabilities by evaluating Eqs.~(9-11) at the Fermi energy ($E=0$).
For graphene ($k_{\parallel}=1.7\,\mathrm{\AA}^{-1}$ and $z=h=10\,\mathrm{\AA}$), it results $e^{-2\kappa h}\approx 3 \cdot 10^{-18}$.
For FGT on the other hand ($k_{\parallel}=1.05\,\mathrm{\AA}^{-1}$ and $z=h+d=13.34\,\mathrm{\AA}$), it results $e^{-2\kappa (h+d)}\approx 5 \cdot 10^{-18}$.
Thus, the theoretical elastic tunneling probability to the K-point states in the FGT is larger than in the graphene. In addition, since FGT has a significantly higher density of states than graphene, because the Fe $3d$ bands near K are flatter than the graphene $\pi$-bands, we conclude that the tunneling inside the inelastic gap is dominated by the FGT.
}

{
We further note that, because the graphene is in direct contact to the extended metallic FGT flake, dynamical Coulomb blockade effects are strongly suppressed in our experiments \cite{brun2012dynamical} and thus cannot explain our observed gap features.
This argument is supported by the fact that the observed inelastic gaps are almost identical compared to graphene on insulating substrates \cite{zhang2008giant, decker2011local, qiu2021visualizing}.
}

{The combined observation of a finite peak splitting at zero field followed by an approximately linear shift as function of magnetic field effectively rules out Landau level formation in the graphene as a potential origin of the peak features, since they would follow a square-root dependence $E\propto\sqrt B$ in the absence of lateral confinement \cite{andrei2012electronic}. }
{
We expect lateral confinement to be weak in our heterostructure, because the graphene is coupled to the metallic FGT, which renders it $n$-doped everywhere.
Furthermore, our observations are not consistent with conventional quantum-dot (QD) bound states or whispering gallery modes (WGMs):
Such states typically give rise to a series of peaks in the density of states corresponding to bound states with increasing (angular) momentum quantum numbers \cite{lee2016imaging, ge2023giant}. 
In contrast, our heterostructure exhibits only two well-isolated peak features within the inelastic gap. 
These features instead are phenomenologically much more similar to tunable orbital moments of a gapped graphene Dirac cone \cite{lee2020tunable, ge2021control}. 
In those works, lateral carrier confinement in electrostatically defined potential wells lifts the graphene's inversion symmetry and gives rise to orbital magnetic moments, which couple to external electric and magnetic fields.
In contrast to this work however, in our heterostructure, both, inversion and time-reversal symmetry are already broken, such that we do not require lateral confinement to realize field tunable orbital moments.
Instead, we observe the orbital moments of the gapped FGT nodal line, which dominates the tunneling.
This interpretation is supported by the fact that the states intrinsic to the graphene shift according to the graphene doping level \cite{ge2021control}, whereas the peaks in our experiments do not:
In Fig.~\ref{Fig4}a, we demonstrate that the graphene doping level can be tuned by $\Delta E_\mathrm{D}\sim15\rm\,meV$ by changing the tip-sample distance. 
However, the peaks inside the inelastic gap only shift asymmetrically by $\lesssim1\rm\,meV$, which points to their origin not in the graphene, but in the FGT. 
This interpretation is consistent with our argument above, that tunneling inside the inelastic gap is dominated by the FGT.
Lastly, we rule out the formation of a quantum dot in the FGT itself, because of its metallic nature. 
}

We extracted $E_{\rm D}$ in Fig.~\ref{Fig4}a from the V-shaped Dirac spectrum outside the inelastic gap, by (i) subtracting the energy offset resulting from the inelastic tunnelling process above the phonon threshold $\pm\hbar\omega$, for positive and negative bias voltages separately, and (ii) performing linear fits to the electron and hole sides of the linear Dirac cone spectrum, respectively; the Dirac point energy $E_{\rm D}$ at each setpoint current was then extracted from their intersection.
To determine $\delta h$ in Fig.~\ref{Fig4}, we used the exponential dependence of the tunnelling current on the tip-sample distance $I/I_0=\exp(\frac{{{h_0-h}}}{\lambda})$, where 
{${\delta h}={h_0-h}$} and $\lambda^{-1}\approx(0.455\rm\,$\AA$)^{-1}$ is the decay constant of the tunnelling current for monolayer graphene above the phonon threshold \cite{zhang2008giant}. Then
\begin{eqnarray}
    {\delta h}=\lambda\ln(I/I_0),
\end{eqnarray}
which is independent of {the absolute values of} $h$ and $h_0$.

\subsection{Electrostatic model}
{From the experimental geometry, we estimate the tip and FGT capacitance density with respect to the graphene as
\begin{eqnarray}
        C_{\rm tip}=\frac{\varepsilon_0}{h}&\approx0.88\rm\,\upmu F cm^{-2}\label{eq:tip_capacitance}\\
        C_{\rm FGT}=\frac{\varepsilon_0}{d}&\approx2.651\rm\,\upmu F cm^{-2},
\end{eqnarray}
where $\varepsilon_0$ is the vacuum permittivity and we estimate a graphene/FGT layer spacing of $d=3.34\rm\,$\AA\ and the initial tip height of $h=10\rm\,$\AA\ at setpoint $V_{\rm s}=300\rm\,mV$, $I_0=0.1\rm\,nA$. 
We note that, realistically, the permittivity of the graphene/FGT interface is expected to be higher than $\varepsilon_0$, which would result in an increase of $C_{\rm FGT}$ and thus a decrease of the electric field at the graphene/FGT interface.
The finite screening length of the electric field in the FGT, on the other hand, leads to a decrease of $C_{\rm FGT}$; however, this change is expected to be small because of the metallic character of FGT and the resulting short screening length \cite{black1999electric}.
In any case, both of these effects have only minor influence on the tip-height dependence of the interface field in our study, because the response of the system is dominated by the change of the tip-sample capacitance $C_{\rm tip}$.}

To model the charge distribution across the junction and the heterostructure, we must take into account that the graphene screens the tip electric field from the FGT, similar to a Faraday screen \cite{ambrosetti2019faraday}.
We therefore employed a capacitor model of the junction similar to the one used in Ref.~\onlinecite{lupke2018situ}.
Because of the low density of states of graphene close to the Dirac point, the model needs to  include the graphene quantum capacitance (per unit sample area) \cite{xia2009measurement} 
\begin{eqnarray}
C_{\rm q}= e^2\rho,
\label{eq:Cq}
\end{eqnarray}
where 
\begin{eqnarray}
    \rho=\frac{2 E_{\rm D}}{\pi(\hbar v_{\rm F})^2}
\label{eq:rho}
\end{eqnarray}
is the {spin- and valley-degenerate} graphene density of states at \EF (per unit sample area), and $v_{\rm F}$ the Dirac Fermi velocity. Integrating $\rho$ from the Dirac point to $E_{\rm F}$, one obtains from Eq.~\ref{eq:rho} the number of carriers per unit area in graphene as $n_{\rm gr}=\pi^{-1}(E_{\rm D}/\hbar v_{\rm F})^2$. As usual, the quantum capacitance is implemented as a capacitor in series to the geometric capacitance between tip and graphene, $C_{\rm tip}$ (see inset Fig.~\ref{Fig4}b) \cite{lupke2018situ}.
As a result of the quantum capacitance, the tip electric field is not perfectly screened by the graphene and can penetrate to the FGT underneath, establishing a capacitance $C_{\rm FGT}$ between graphene and FGT, in series with $C_{\rm tip}$. Because it is equivalent to a single capacitor with distance $h+d$, the arrangement of $C_{\rm tip}$ and $C_{\rm FGT}$ in series is consistent with a tunneling path from the tip directly to the FGT. We thus have two parallel tunneling paths (the other  one from the tip to graphene), which suggests putting the two capacitances $C_{\rm q}$ and $C_{\rm FGT}$ in parallel, both connected to the same back contact of the Ti/Au substrate (within the metallic FGT layer, there will be no potential drop, despite its considerable thickness). 
In this model, we neglect the inhomogeneity of the electric field due to the curved tip shape, because we are interested only in the properties right underneath the tip apex and the small graphene-FGT distance compared to the tip radius ($10$ to $100\rm\,nm$) results in effectively parallel electric field lines there.

From the circuit diagram it is clear that
\begin{eqnarray}
    \frac{n_{\rm FGT}e}{C_{\rm FGT}}  = \frac{n_{\rm gr}e}{C_{\rm q}} = \frac{n_{\rm tip}e}{C_{\rm tip}}-V_{\rm ts} 
    \label{eq1}
\end{eqnarray} 
and 
\begin{eqnarray}
    n_{\rm tip}+n_{\rm gr}+n_{\rm FGT}=0, 
    \label{eq2}
\end{eqnarray}
the latter due to charge conservation, where $n_{\rm gr}$, $n_{\rm tip}$, and $n_{\rm FGT}$ are the areal charge densities on graphene and the surfaces of the tip and FGT, respectively. 
We solve the linear equation system by substituting expressions derived from Eq.~\ref{eq1} into Eq.~\ref{eq2}, obtaining 
\begin{eqnarray}
    n_{\rm gr}+\frac{C_{\rm tip}V_{\rm ts}}{e}+\frac{C_{\rm tip}}{C_{\rm q}}n_{\rm gr}+\frac{C_{\rm FGT}}{C_{\rm q}}n_{\rm gr}=0.
\end{eqnarray}
Using the graphene quantum capacitance (Eqs.~\ref{eq:rho} and \ref{eq:Cq}) \cite{xia2009measurement}
\begin{eqnarray}
        C_{\rm q}
        =\frac{2e^2}{\sqrt{\pi(\hbar v_{\rm F})^2}}\sqrt{n_{\rm gr}},
\end{eqnarray}
results in
\begin{eqnarray}
    n_{\rm gr}+ (C_{\rm tip}+C_{\rm FGT}) \frac{\sqrt{\pi(\hbar v_{\rm F})^2}}{2e^2}\sqrt{n_{\rm gr}} + \frac{C_{\rm tip}V_{\rm ts}}{e} = 0.
\end{eqnarray}
Substituting $a\equiv \frac{\sqrt{\pi(\hbar v_{\rm F})^2}}{2e^2}$ gives
\begin{eqnarray}
    n_{\rm gr}+ a(C_{\rm tip}+C_{\rm FGT}) \sqrt{n_{\rm gr}} + \frac{C_{\rm tip}V_{\rm ts}}{e} =0, \label{eq:quadratic}
\end{eqnarray}
which yields 
\begin{align}
    \sqrt{n_{\rm gr}}=&-\frac{a(C_{\rm tip}+C_{\rm FGT})}{2}\nonumber\\
    &+ \sqrt{ \left(\frac{a(C_{\rm tip}+C_{\rm FGT})}{2}\right)^2 - \frac{C_{\rm tip}V_{\rm ts}}{e}}, 
    \label{eq:ngr}
\end{align}
where only the positive solution of the quadratic expression is physical, as $n_{\rm gr}$ must vanish at $V_{\rm ts}=0$. With Eq.~\ref{eq:ngr}, we determined the Dirac point energy as 
\begin{eqnarray}
    E_{\rm D}=\pm2e^2a \sqrt{n_{\rm gr}}\label{eq:E_D}.
\end{eqnarray}
The positive (negative) sign applies to holes (electrons).

The potential difference $V_{\rm ts}$ between the tip and graphene is given by the sum of the applied bias voltage and the contact potential difference between tip and sample surface
\begin{eqnarray}
    V_{\rm ts}=V_{\rm s}+V_{\rm CPD}= V_{\rm s}+(\Phi_{\rm tip}-\Phi_{\rm gr})/e 
\end{eqnarray}
where the $\Phi$ are the respective work functions.
While the work function of charge-neutral graphene is well known (see above), the work function of the tip is more difficult to estimate, and critically depends on the details of the tip apex. 
As a result, for W tips the tip work function typically varies in the range $4.5-5.1\rm\,eV$\cite{zhao2021electric}.
We indent our W tips into Au before measurements and expect Au to stick to the tip as a result of the preparation procedure. 
The work function of bulk Au is $\Phi_{\rm Au}\approx5.1\rm\,eV$, such that we estimate the resulting tip work function to be in the range $\Phi_{\rm tip}=4.6-5.1\rm\,eV$, i.e., greater or equal to that of graphene.
The fit in Fig.~\ref{Fig4}b results in a tip work function of $\Phi_{\rm tip}=(4.66\pm0.03)\rm\,eV$, in excellent agreement with this range of expected values.
Our model further compares well with experiments on graphene/\NbSe heterostructures ($\Phi_{\rm NbSe_2}\approx5.6\rm\,eV$), where a significant hole doping corresponding to $E_{\rm D}\approx+400\rm\,meV$ was reported \cite{Chen2020visualizing}.

\subsection{Angle-resolved photoemission spectroscopy}
Photoemission measurements were performed on cleaved FGT bulk single crystals and were conducted at NanoESCA beamline at Elettra, using the modified Focus GmbH NanoESCA momentum microscope. The sample was cooled using LHe and the temperature was stabilized at $\approx40\rm\,K$. The light was incident at an angle $65^{\circ}$ with respect to the surface normal, along the K-$\Gamma$-K’ direction. Extended Data Fig.~\ref{Fig-SI-bulk}b shows spectra collected at $\hbar\nu = 60 \rm\,eV$ in sweep mode with an energy step size of $0.02\rm\,eV$.

\subsection{Disentangling band-orbital and chiral-orbital Zeeman effects}
We approximate the observed peak shifting behaviour as function of magnetic field by the linear equation system
\begin{eqnarray}
    E_1&=g_1\mu_BB+ E_1^0 &=(g_z^\mathrm{CO}-g_z^\mathrm{BO})\mu_\mathrm{B}B+E_1^0\\
    E_2&=g_2\mu_BB+E_2^0  &=(g_z^\mathrm{CO}+g_z^\mathrm{BO})\mu_\mathrm{B}B+E_2^0,
\end{eqnarray}
where 
$E_i(B)$ are the peak energies, 
with $E_i^0\equiv E_i(B=0)$, and $g_z^\mathrm{CO}$ and $g_z^\mathrm{BO}$ are the chiral- and band-orbital $g$-factors 
, respectively. 
Inserting the fitted slopes from Fig.~\ref{Fig3}c, 
$g_1\mu_\mathrm{B}=1.8\rm\,meV/T$ and $g_2\mu_\mathrm{B}=13.3\rm\,meV/T$, results in
\begin{eqnarray}
    g_z^\mathrm{BO}=\frac{g_2-g_1}{2}\approx100
\end{eqnarray}
and $g_z^\mathrm{CO}\approx130$.

\section{Acknowledgements}
The authors acknowledge Zhoungqui Lyu for assistance in the lab. Furthermore, the authors are grateful to the Helmholtz Nano Facility for its support regarding sample fabrication.
The research was funded by the Deutsche Forschungsgemeinschaft (DFG, German Research Foundation) within the Priority Programme SPP 2244 (project nos.~443416235 and 422707584).
H.B. was supported by the DFG via the project no.~PL 712/5-1.
Y.M., D.G, M.S. and T.S. acknowledge funding by the DFG in the framework of TRR 288 - 422213477 (Project B06), and by the EIC Pathfinder OPEN grant 101129641 ``OBELIX''.
F.L. and F.S.T acknowledge funding from the Bavarian Ministry of Economic Affairs, Regional Development and Energy within Bavaria’s High-Tech Agenda Project ''Bausteine f\"ur das Quantencomputing auf Basis topologischer Materialien mit experimentellen und theoretischen Ans\"atzen'' and Germany’s Excellence Strategy - Cluster of Excellence Matter and Light for Quantum Computing (ML4Q) through an Independence Grant.
F.L. further acknowledges funding from the DFG Emmy Noether Programme (Project No.\ 511561801).
J.M.C. acknowledges funding from the Alexander von Humboldt Foundation.
M.T. acknowledges support from the Heisenberg Program (Grant No. TE 833/2-1) of the DFG.

\vspace{3mm}

\paragraph{Author contributions}
F.L. conceived the project.
T.W. and H.B. fabricated samples.
T.W., K.J., J.M.C, M.T., F.S.T. and F.L. designed the STM experiment.
T.W. and F.L. acquired the STM data, which were analysed by T.W., M.T. and F.L.
M.S., T.G.S., D.G. and Y.M. performed DFT calculations.
F.L. performed the capacitor circuit calculations.
T.W. and S.L. performed auxillary calculations.
H.B. and L.P. designed the ARPES experiment, acquired, and analysed the corresponding data.
L.P., M.T., Y.M., F.S.T. and F.L. supervised the project.
T.W., J.M.C., F.S.T. and F.L. wrote the manuscript, and all authors commented on the manuscript.
\vspace{3mm}

\paragraph{Competing financial interests}
The authors declare no competing financial interests.

\section{References}

\newpage
\onecolumngrid

    \renewcommand{\figurename}{Extended Data Fig.}
    \renewcommand{\tablename}{Extended Data Table}
    \renewcommand{\thefigure}{S\arabic{figure}}
 
\section*{Extended Data}
\renewcommand{\thefigure}{S1}
\begin{figure*}[htbp]
  \centering
  \includegraphics[width=1\textwidth]{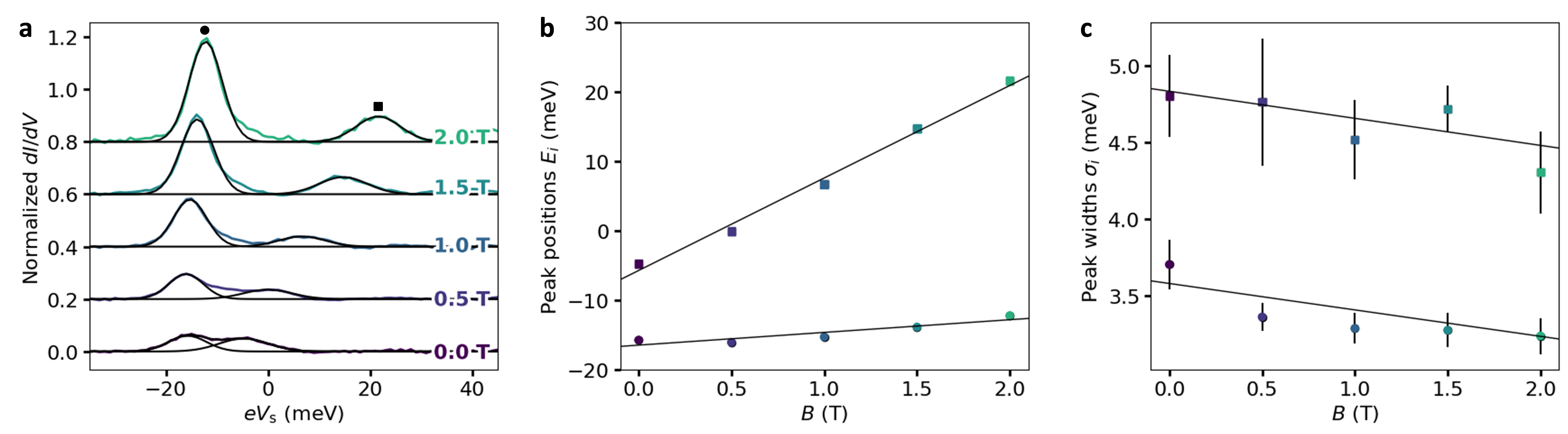}
  \caption{\label{Fig-SI-peak-extraction} \textbf{Fit results for the two \FGT-related peaks in the inelastic tunnelling gap of graphene, as a function of external \textit{B} field.} 
  \textbf{(a)} Differential conductance spectra (arbitrary units) after subtraction of a polynomial background (colored data curves, same data set as in Fig.~\ref{Fig3}b). The displayed  Gaussians $A_ie^{-(eV_{\rm s}-E_i)^2/{2\sigma_i^2}}$ (black curves) were fitted to the data separately for each curve and peak, except at $B=0$, where because of their significant overlap the two peaks were fitted simultaneously. 
  \textbf{(b, c)} Fitted peak energies $E_i$ and widths $\sigma_i$, respectively, with linear fits for each peak.
  Error bars correspond to the fit uncertainties.
  }
\end{figure*}

\renewcommand{\thefigure}{S2}
\begin{figure*}[htbp]
  \centering
  \includegraphics[width=\textwidth]{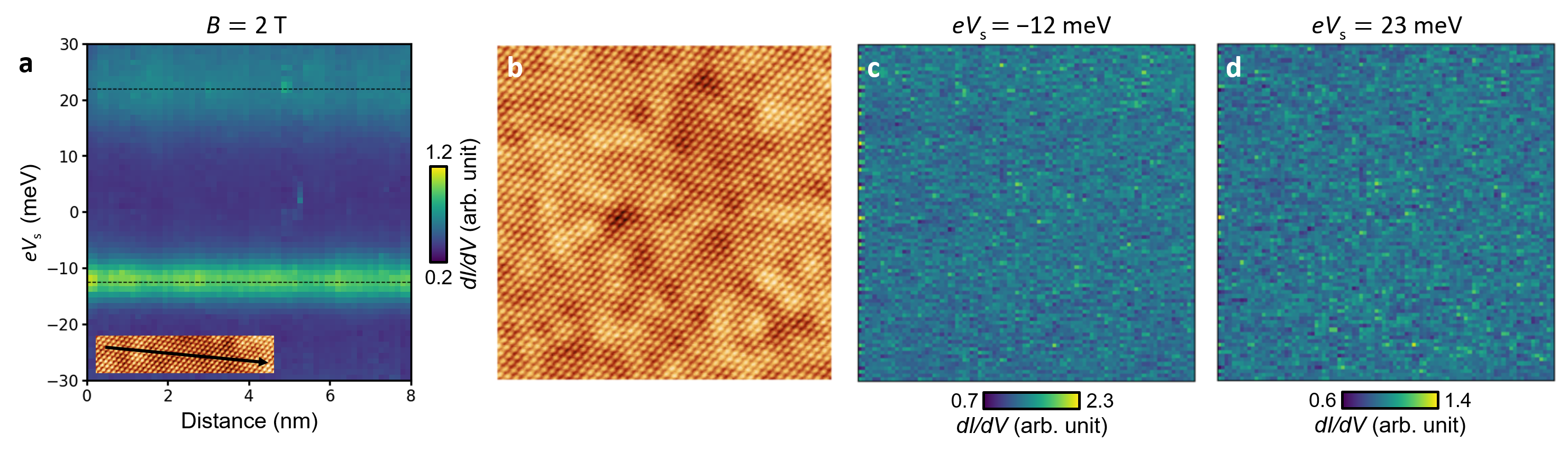}
  \caption{\label{Fig-SI-peak-energies} \textbf{Spatial consistency of the two \FGT-related peaks at \textit{B} = 2\,T.}
  \textbf{(a)} Color map of differential conductance spectra (arbitrary units), recorded at positions along the arrow in the  STM topography in the inset 
  The map reveals constant peak energies and intensities.
  \textbf{(b)} STM topography recorded at another position on the graphene/\FGT heterostructure. Scan size: $10\rm\,nm$.
 \textbf{(c, d)} Color maps of differential conductance spectra (arbitrary units) recorded in the area shown in panel b at the two peak energies marked by dotted lines in panel a. No spatial variations, for example due to scattering at defects, are discernible.
 Tip stabilization parameters for all panels: $V_\mathrm{s}=50\,\rm mV$, $I_\mathrm{t}=1\,\rm nA$.
    }
\end{figure*}

\renewcommand{\thefigure}{S3}
\begin{figure*}[htbp]
  \centering
  \includegraphics[width=0.7\textwidth]{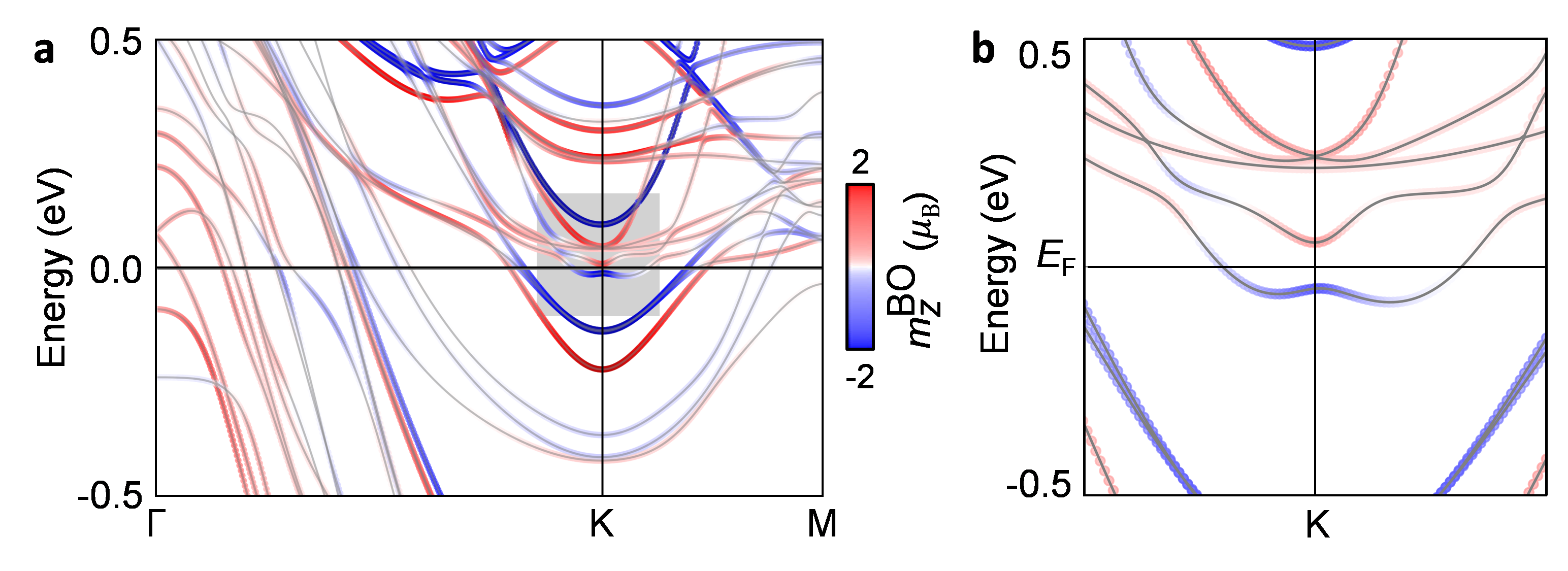}
  \caption{\label{Fig-SI-trilayer} 
  \textbf{DFT-calculated  band structure of trilayer \FGT}. 
  \textbf{(a)} In \FGT trilayers, the inversion symmetry of bulk \FGT (bilayer unit cell) is broken. This leads to additional bands close to \EF, a smaller nodal line gap, and smaller band orbital moments (color coded). Furthermore, the K and K' points are no longer equivalent, although their differences are minimal. We thus only show the K point here. The band structure in this panel should be compared with Fig.~\ref{Fig1}c, which was calculated for bulk \FGT.
  We note that the band crossing at K/K' does not exist in monolayer FGT.
  However, the spin canting, which is strongest in the layer closest to the interface (compare Fig. 1e) may break inversion symmetry for two or more FGT layers and thus can explain a smaller gap at K/K'.
  \textbf{(b)} Zoom into the shaded region in panel a, showing the bands close to the K point. 
  }
\end{figure*}

\renewcommand{\thefigure}{S4}
\begin{figure*}[htbp]
  \centering
  \includegraphics[width=0.8\textwidth]{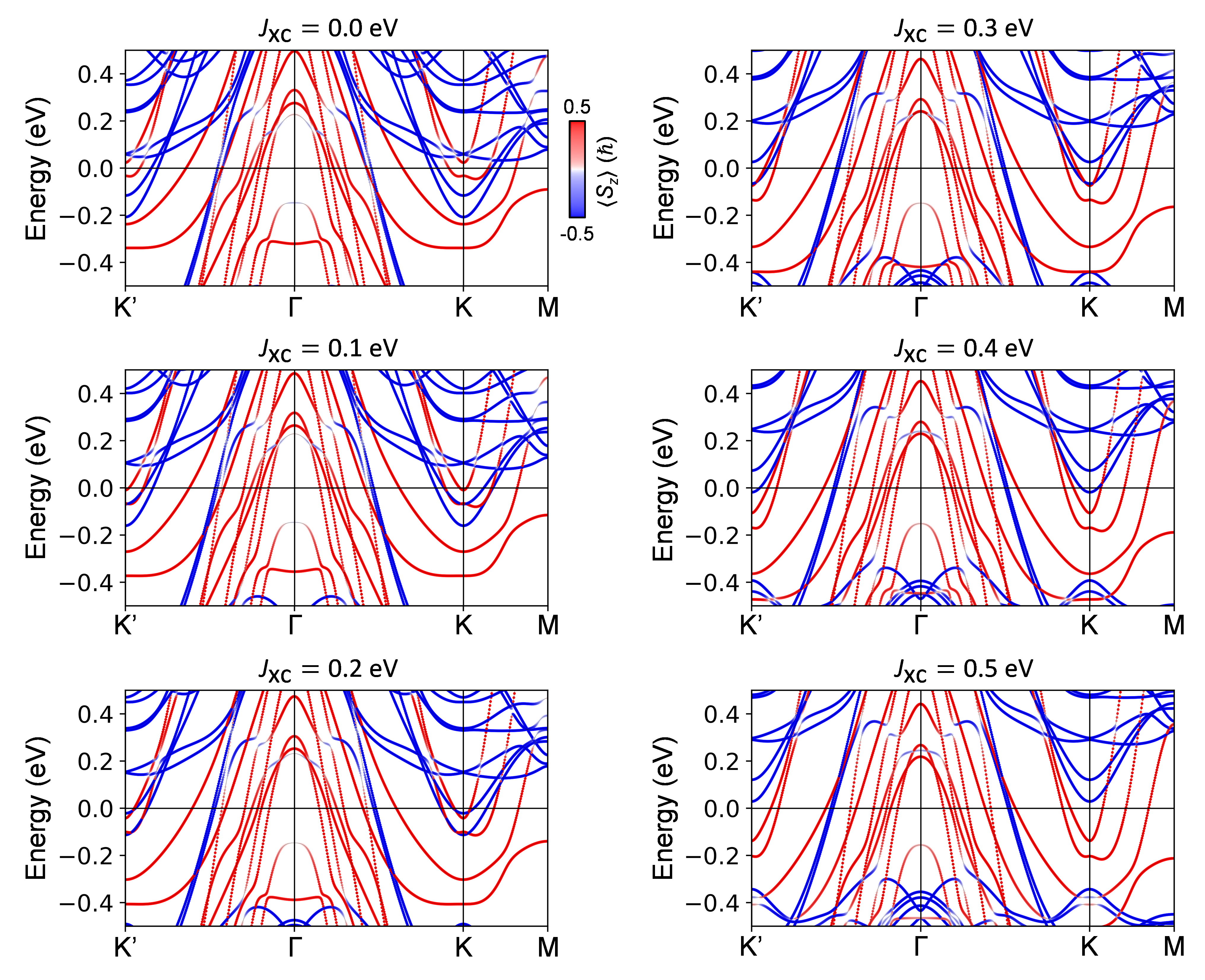}
  \caption{\label{Fig-SI-exchange_calc} \textbf{\FGT band structure as a function of magnetic exchange potential strength.} To model the spin Zeeman effect in an external magnetic field, we took the DFT-calculated single-particle band structure of \FGT (see Fig.~\ref{Fig1}c) and added exchange coupling of varying size, ranging from $J_{\rm xc}=0$ to $0.5\rm\,eV$. The additional exchange coupling is added on top of the FGT Wannier Hamiltonian by using the matrix elements of the spin operator in terms of Wannier functions.
  }
\end{figure*}

\renewcommand{\thefigure}{S5}
\begin{figure*}[htbp]
  \centering
  \includegraphics[width=0.4\textwidth]{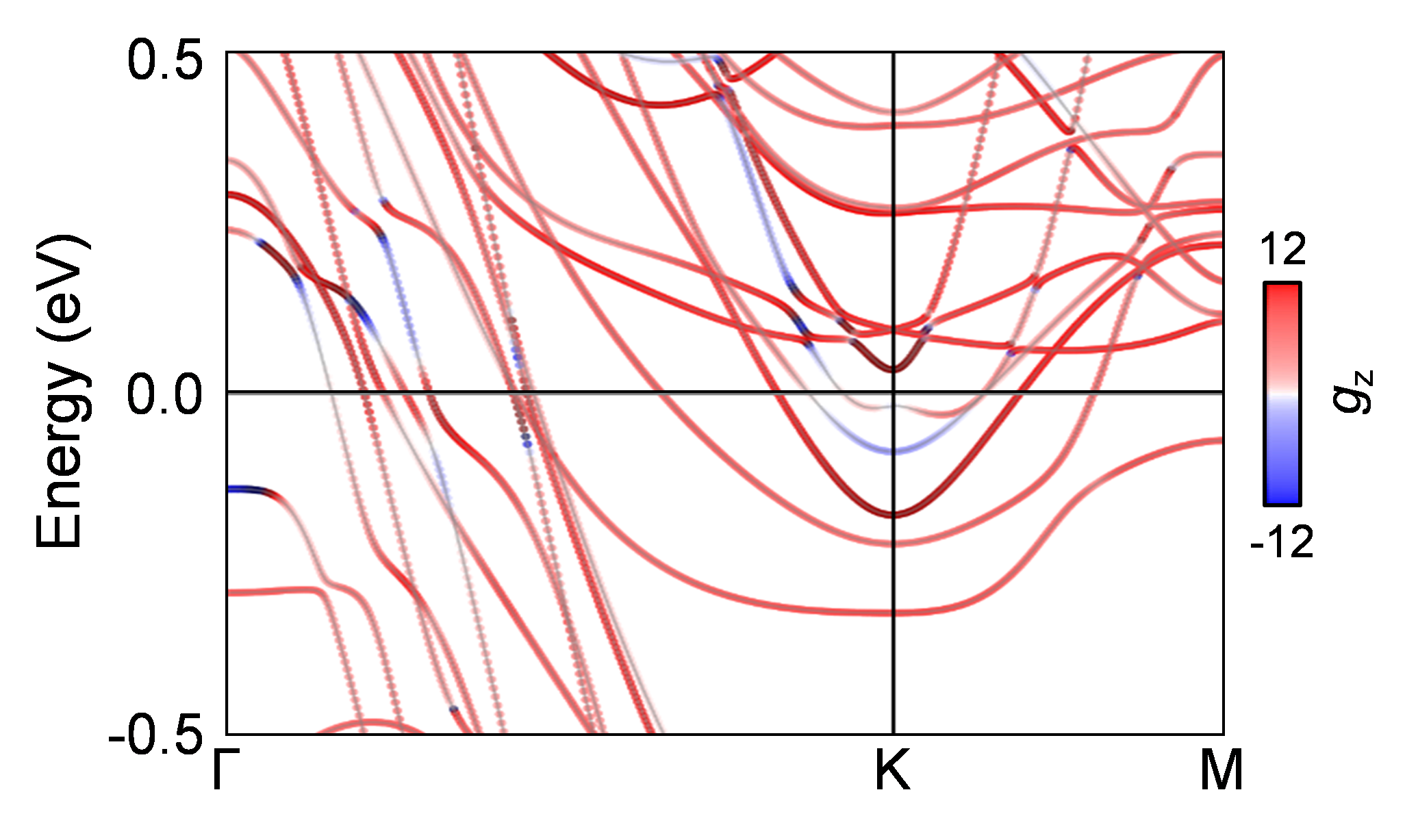}
  \caption{\label{Fig-SI-g-factor} 
  \textbf{Calculated out-of-plane \textit{g}-factor, neglecting the influence of spin canting and chiral orbital moments.} The plot shows the band structure of Fig.~\ref{Fig1}c, with color-coded $g_z^{\rm BO}= 2 + {\langle L_{} \rangle}/{\langle S_{} \rangle}$ resulting from the spin and orbital moments shown as color-coding in Fig.~\ref{Fig1}c and d, respectively.
  }
\end{figure*}

\renewcommand{\thefigure}{S6}
\begin{figure*}[htbp]
  \centering
  \includegraphics[width=\textwidth]{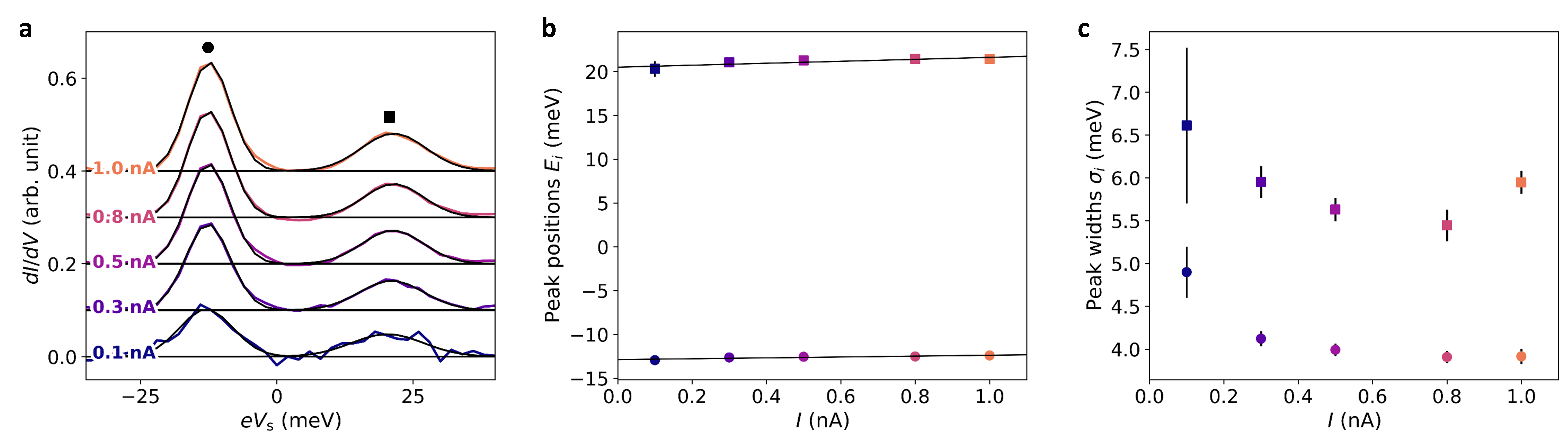}
  \caption{\label{Fig-SI-peak-extraction2} 
  \textbf{Fit results for the two \FGT-related peaks in the inelastic tunnelling gap of graphene, as function of setpoint current.}
  \textbf{(a)} Differential conductance spectra (arbitrary units, measured at $B=2\,\rm T$) after subtraction of a polynomial background (colored data curves, same data set as in Fig.~\ref{Fig4}c). The displayed  Gaussians $A_ie^{-(eV_{\rm s}-E_i)^2/{2\sigma_i^2}}$ (black curves) were fitted to the data separately for each curve and peak. \textbf{(b, c)} Fitted peak energies $E_i$ and widths $\sigma_i$, respectively, with linear fits in panel b for each of the two peaks. 
  Error bars correspond to the fit uncertainties.
  }
\end{figure*}

\renewcommand{\thefigure}{S7}
\begin{figure*}[htbp]
  \centering
  \includegraphics[width=0.7\textwidth]{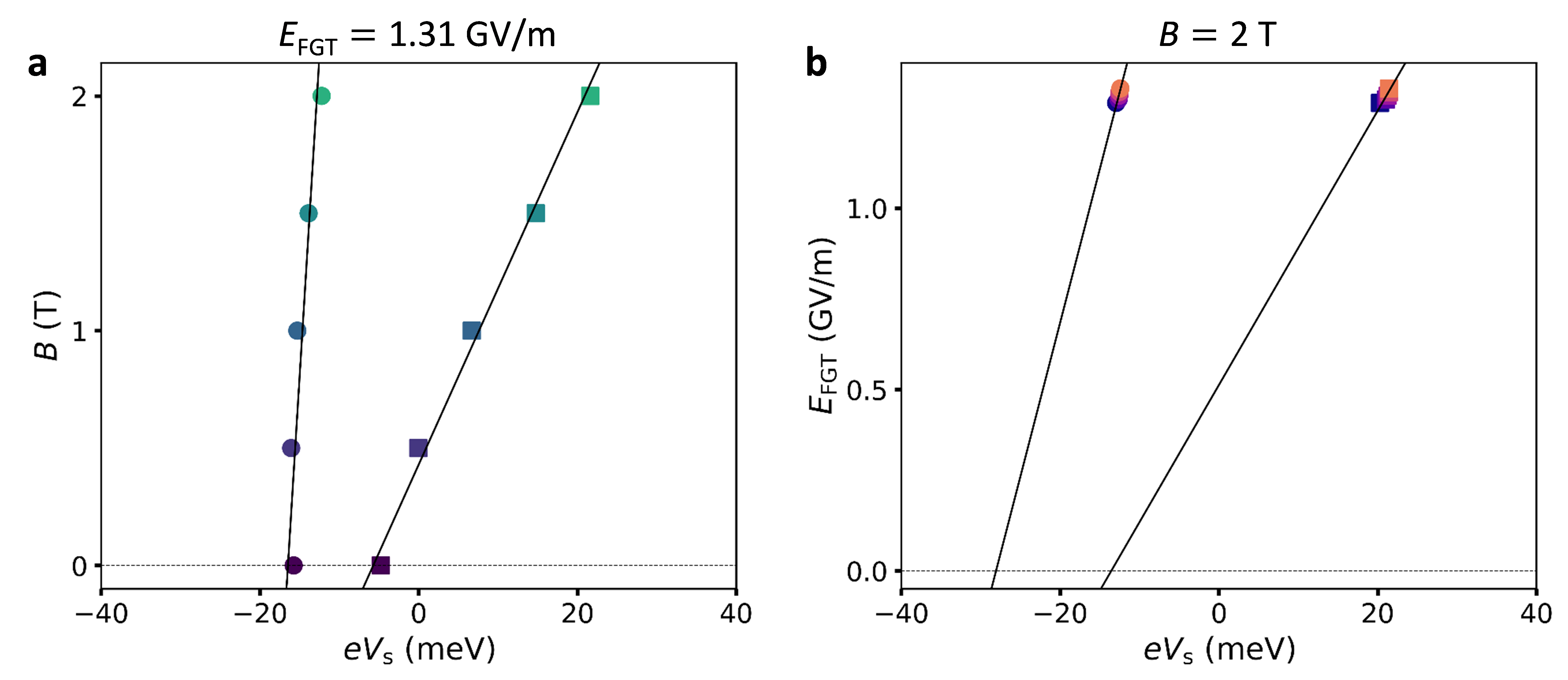}
  \caption{\label{Fig-SI-peak-extrapolation} 
  \textbf{Fitted peak positions for the two \FGT-related peaks in the inelastic tunnelling gap of graphene, as function of magnetic and electric field.} 
  \textbf{(a)} $B$-field-dependent Zeeman effect at $E_\mathrm{FGT}=1.31\rm\,GV/m$. Data points and linear fits (solid lines) are the same as in Fig.~\ref{Fig3}c.  
  \textbf{(b)} ${E_\mathrm{FGT}}$-field-dependent Zeeman effect at $B=2\rm\,T$. Data points and linear fits (solid lines) are the same as in Fig.~\ref{Fig4}d. 
  }
\end{figure*}

\renewcommand{\thefigure}{S8}
\begin{figure*}[htbp]
  \centering
  \includegraphics[width=0.55\textwidth]{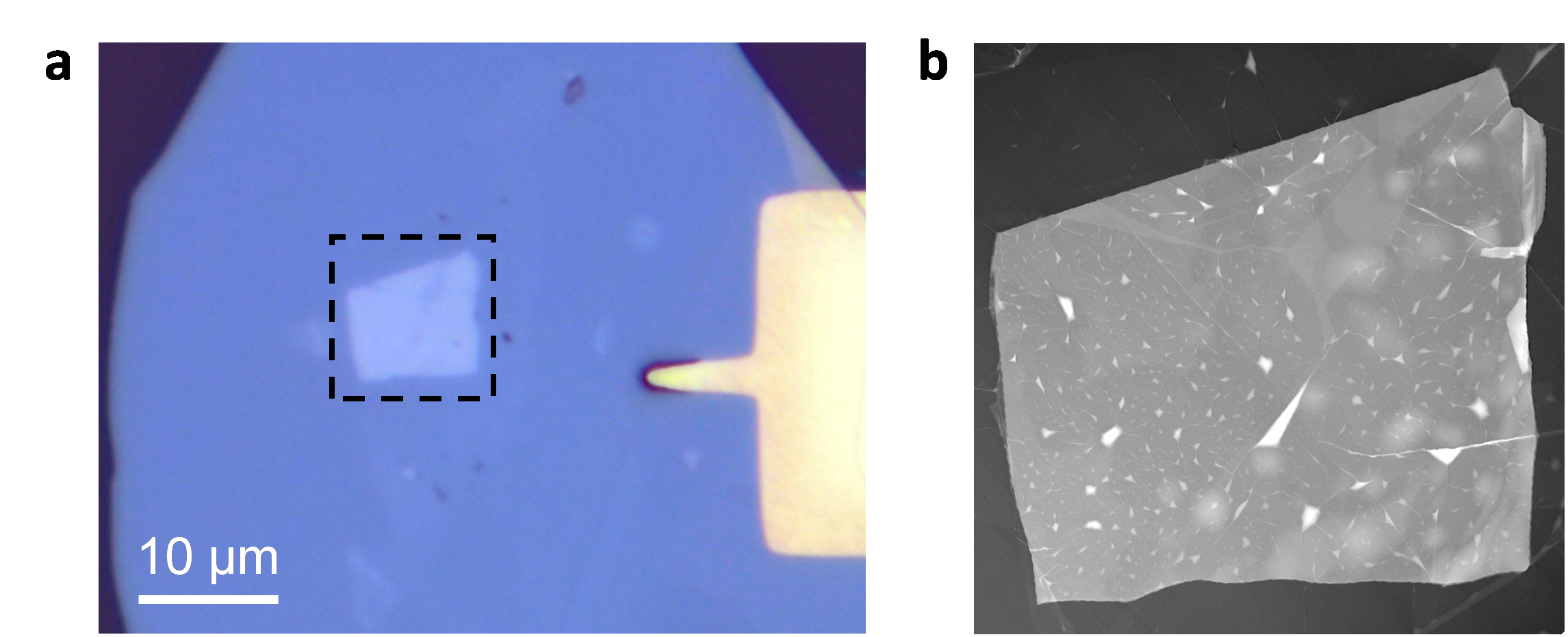}
  \caption{\label{Fig-SI-hts} \textbf{Atomic force microscopy on the the graphene/\FGT heterostructure.}
    \textbf{(a)} Optical micrograph of the heterostructure (same as in Fig.~\ref{Fig1}e), with the black dashed line representing the outline of the AFM measurement in panel b.
    \textbf{(b)} Atomic force micrograph of the graphene-covered \FGT flake. 
    The formation of `dirt pockets' indicates an atomically clean vdW interface in between the pockets, in consistency with the spatially uniform STM measurements that were performed in different spots on the sample.
    }
\end{figure*}

\renewcommand{\thefigure}{S9}
\begin{figure*}[htbp]
  \centering
  \includegraphics[width=0.75\textwidth]{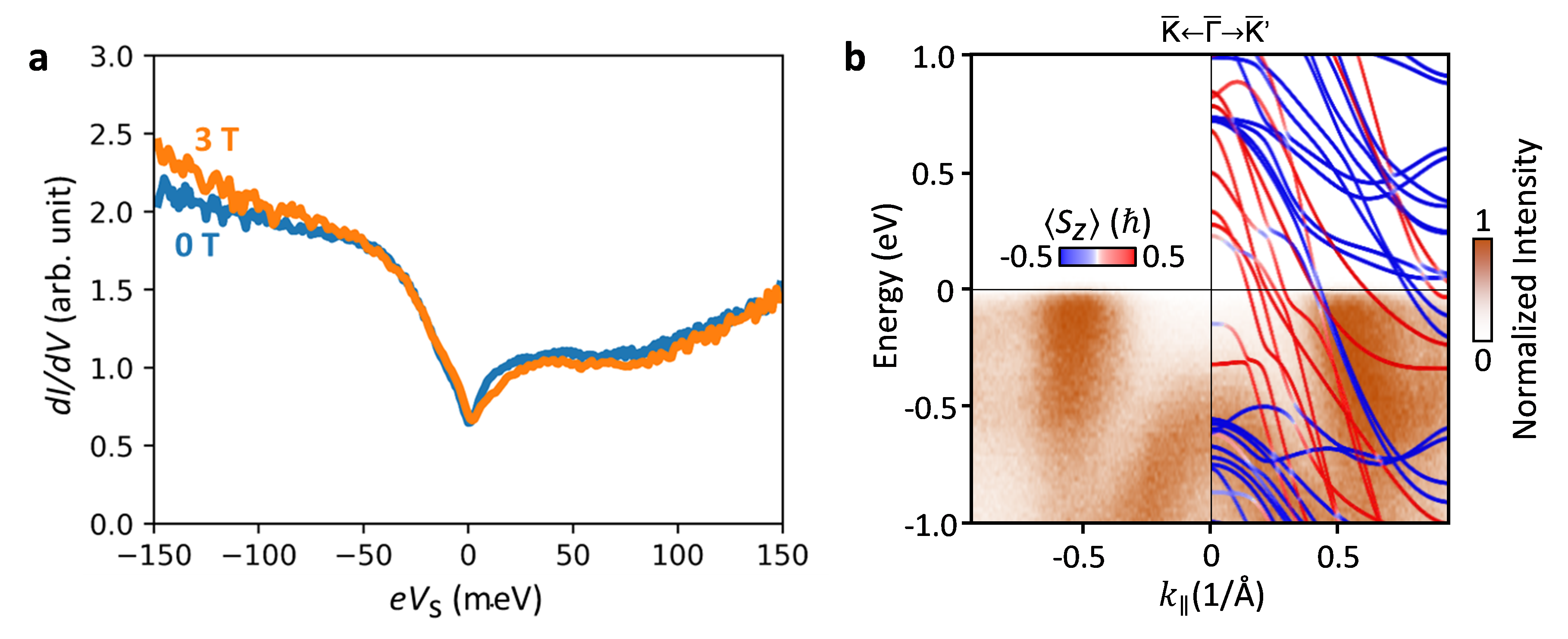}
  \caption{\label{Fig-SI-bulk} 
  \textbf{Characterization of bulk \FGT.}
 \textbf{(a)} Scanning tunneling spectra recorded on a bulk \FGT crystal.
 No significant change is observed as a function of applied magnetic field up to our maximum field of $B=3\rm\,T$. The data is consistent with previous reports in the literature \cite{zhao2021kondo, zhao2021electric} and confirms that breaking of inversion symmetry at the interface is a necessary condition to observe the specific behavior of spectral features reported in the main text.
 \textbf{(b)}
 Angle-resolved photoemission spectroscopy of a bulk \FGT crystal (band maps from $\bar{\Gamma}$ to  $\bar{\rm K}$ on the left and $\bar{\Gamma}$ to  $\bar{\rm K}'$ on the right) are overlayed with the DFT-calculated band structure including SOC (from Fig.~\ref{Fig1}c) in $\bar{\Gamma}{\rm -K}$ direction. Red and blue bands correspond to the majority (minority) spins.  
}
\end{figure*}

\begin{table*}[!htbp]
    \centering
    \begin{tabular}{c|c|c|c|c|c}
        $I_{\rm t}$ (nA) & $\delta h$ (\AA) &  $C_{\rm tip} (\rm\upmu F cm^{-2})$  & $E_{\rm D}$ (meV) & $n_{\rm gr}$ ($10^{10}\rm\,cm^{-2}$) & ${E_\mathrm{FGT}}$ (GV/m) \\\hline
        0.1 & 0 & 0.885  & $-38.7 \pm\, 2.6$ & $5.50 \pm\, 0.74$ & $1.291\pm0.008$ \\
        0.3 & 0.500 & 0.932 & $-33.6 \pm\, 2.6$ & $4.15 \pm\, 0.64$ & $1.306\pm0.008$ \\
        0.5 & 0.732 & 0.955 & $-30.5 \pm\, 2.2$ & $3.42 \pm\, 0.49$ & $1.316\pm0.007$ \\
        0.8 & 0.946 & 0.978 & $-26.5 \pm\, 2.7$ & $2.58 \pm\, 0.53$ & $1.328\pm0.008$ \\
        1.0 & 1.048 & 0.989 & $-27.8 \pm\, 2.4$ & $2.84 \pm\, 0.49$ & $1.324\pm0.007$
    \end{tabular}
    \caption{{{Change in tip height $\delta h$, tip-graphene capacitance $C_\mathrm{tip}$ (Eq.~\ref{eq:tip_capacitance}), Dirac point energy $E_\mathrm{D}$ (Fig.~\ref{Fig4}a), charge carrier concentration (Eq.~\ref{eq:E_D}), and interface electric field $E_\mathrm{FGT}$ (Eq. \ref{eq:E-field}), all as a function of setpoint tunnel current $I_\mathrm{t}$.}}}
    \label{Tab-SI-exp}
\end{table*}

\begin{thebibliography}{75}%
\makeatletter
\providecommand \@ifxundefined [1]{%
 \@ifx{#1\undefined}
}%
\providecommand \@ifnum [1]{%
 \ifnum #1\expandafter \@firstoftwo
 \else \expandafter \@secondoftwo
 \fi
}%
\providecommand \@ifx [1]{%
 \ifx #1\expandafter \@firstoftwo
 \else \expandafter \@secondoftwo
 \fi
}%
\providecommand \natexlab [1]{#1}%
\providecommand \enquote  [1]{``#1''}%
\providecommand \bibnamefont  [1]{#1}%
\providecommand \bibfnamefont [1]{#1}%
\providecommand \citenamefont [1]{#1}%
\providecommand \href@noop [0]{\@secondoftwo}%
\providecommand \href [0]{\begingroup \@sanitize@url \@href}%
\providecommand \@href[1]{\@@startlink{#1}\@@href}%
\providecommand \@@href[1]{\endgroup#1\@@endlink}%
\providecommand \@sanitize@url [0]{\catcode `\\12\catcode `\$12\catcode
  `\&12\catcode `\#12\catcode `\^12\catcode `\_12\catcode `\%12\relax}%
\providecommand \@@startlink[1]{}%
\providecommand \@@endlink[0]{}%
\providecommand \url  [0]{\begingroup\@sanitize@url \@url }%
\providecommand \@url [1]{\endgroup\@href {#1}{\urlprefix }}%
\providecommand \urlprefix  [0]{URL }%
\providecommand \Eprint [0]{\href }%
\providecommand \doibase [0]{https://doi.org/}%
\providecommand \selectlanguage [0]{\@gobble}%
\providecommand \bibinfo  [0]{\@secondoftwo}%
\providecommand \bibfield  [0]{\@secondoftwo}%
\providecommand \translation [1]{[#1]}%
\providecommand \BibitemOpen [0]{}%
\providecommand \bibitemStop [0]{}%
\providecommand \bibitemNoStop [0]{.\EOS\space}%
\providecommand \EOS [0]{\spacefactor3000\relax}%
\providecommand \BibitemShut  [1]{\csname bibitem#1\endcsname}%
\let\auto@bib@innerbib\@empty
%</preamble>
\bibitem [{\citenamefont {Armitage}\ \emph {et~al.}(2018)\citenamefont
  {Armitage}, \citenamefont {Mele},\ and\ \citenamefont
  {Vishwanath}}]{armitage2018weyl}%
  \BibitemOpen
  \bibfield  {author} {\bibinfo {author} {\bibfnamefont {N.}~\bibnamefont
  {Armitage}}, \bibinfo {author} {\bibfnamefont {E.}~\bibnamefont {Mele}},\
  and\ \bibinfo {author} {\bibfnamefont {A.}~\bibnamefont {Vishwanath}},\
  }\bibfield  {title} {\bibinfo {title} {{W}eyl and {D}irac semimetals in
  three-dimensional solids},\ }\href@noop {} {\bibfield  {journal} {\bibinfo
  {journal} {Reviews of Modern Physics}\ }\textbf {\bibinfo {volume} {90}},\
  \bibinfo {pages} {015001} (\bibinfo {year} {2018})}\BibitemShut {NoStop}%
\bibitem [{\citenamefont {Wan}\ \emph {et~al.}(2011)\citenamefont {Wan},
  \citenamefont {Turner}, \citenamefont {Vishwanath},\ and\ \citenamefont
  {Savrasov}}]{wan2011topological}%
  \BibitemOpen
  \bibfield  {author} {\bibinfo {author} {\bibfnamefont {X.}~\bibnamefont
  {Wan}}, \bibinfo {author} {\bibfnamefont {A.~M.}\ \bibnamefont {Turner}},
  \bibinfo {author} {\bibfnamefont {A.}~\bibnamefont {Vishwanath}},\ and\
  \bibinfo {author} {\bibfnamefont {S.~Y.}\ \bibnamefont {Savrasov}},\
  }\bibfield  {title} {\bibinfo {title} {Topological semimetal and fermi-arc
  surface states in the electronic structure of pyrochlore iridates},\
  }\href@noop {} {\bibfield  {journal} {\bibinfo  {journal} {Physical Review
  B}\ }\textbf {\bibinfo {volume} {83}},\ \bibinfo {pages} {205101} (\bibinfo
  {year} {2011})}\BibitemShut {NoStop}%
\bibitem [{\citenamefont {Tokura}\ \emph {et~al.}(2019)\citenamefont {Tokura},
  \citenamefont {Yasuda},\ and\ \citenamefont
  {Tsukazaki}}]{tokura2019magnetic}%
  \BibitemOpen
  \bibfield  {author} {\bibinfo {author} {\bibfnamefont {Y.}~\bibnamefont
  {Tokura}}, \bibinfo {author} {\bibfnamefont {K.}~\bibnamefont {Yasuda}},\
  and\ \bibinfo {author} {\bibfnamefont {A.}~\bibnamefont {Tsukazaki}},\
  }\bibfield  {title} {\bibinfo {title} {Magnetic topological insulators},\
  }\href@noop {} {\bibfield  {journal} {\bibinfo  {journal} {Nature Reviews
  Physics}\ }\textbf {\bibinfo {volume} {1}},\ \bibinfo {pages} {126} (\bibinfo
  {year} {2019})}\BibitemShut {NoStop}%
\bibitem [{\citenamefont {Chen}\ \emph {et~al.}(2013)\citenamefont {Chen},
  \citenamefont {Yang}, \citenamefont {Wang}, \citenamefont {Imai},
  \citenamefont {Ohta}, \citenamefont {Michioka}, \citenamefont {Yoshimura},\
  and\ \citenamefont {Fang}}]{chen2013magnetic}%
  \BibitemOpen
  \bibfield  {author} {\bibinfo {author} {\bibfnamefont {B.}~\bibnamefont
  {Chen}}, \bibinfo {author} {\bibfnamefont {J.}~\bibnamefont {Yang}}, \bibinfo
  {author} {\bibfnamefont {H.}~\bibnamefont {Wang}}, \bibinfo {author}
  {\bibfnamefont {M.}~\bibnamefont {Imai}}, \bibinfo {author} {\bibfnamefont
  {H.}~\bibnamefont {Ohta}}, \bibinfo {author} {\bibfnamefont {C.}~\bibnamefont
  {Michioka}}, \bibinfo {author} {\bibfnamefont {K.}~\bibnamefont
  {Yoshimura}},\ and\ \bibinfo {author} {\bibfnamefont {M.}~\bibnamefont
  {Fang}},\ }\bibfield  {title} {\bibinfo {title} {Magnetic properties of
  layered itinerant electron ferromagnet {Fe$_3$GeTe$_2$}},\ }\href@noop {}
  {\bibfield  {journal} {\bibinfo  {journal} {Journal of the Physical Society
  of Japan}\ }\textbf {\bibinfo {volume} {82}},\ \bibinfo {pages} {124711}
  (\bibinfo {year} {2013})}\BibitemShut {NoStop}%
\bibitem [{\citenamefont {Jang}\ \emph {et~al.}(2020)\citenamefont {Jang},
  \citenamefont {Yoon}, \citenamefont {Jeong}, \citenamefont {Ryee},
  \citenamefont {Kim},\ and\ \citenamefont {Han}}]{jang2020origin}%
  \BibitemOpen
  \bibfield  {author} {\bibinfo {author} {\bibfnamefont {S.~W.}\ \bibnamefont
  {Jang}}, \bibinfo {author} {\bibfnamefont {H.}~\bibnamefont {Yoon}}, \bibinfo
  {author} {\bibfnamefont {M.~Y.}\ \bibnamefont {Jeong}}, \bibinfo {author}
  {\bibfnamefont {S.}~\bibnamefont {Ryee}}, \bibinfo {author} {\bibfnamefont
  {H.-S.}\ \bibnamefont {Kim}},\ and\ \bibinfo {author} {\bibfnamefont {M.~J.}\
  \bibnamefont {Han}},\ }\bibfield  {title} {\bibinfo {title} {Origin of
  ferromagnetism and the effect of doping on {Fe$_3$GeTe$_2$}},\ }\href@noop {}
  {\bibfield  {journal} {\bibinfo  {journal} {Nanoscale}\ }\textbf {\bibinfo
  {volume} {12}},\ \bibinfo {pages} {13501} (\bibinfo {year}
  {2020})}\BibitemShut {NoStop}%
\bibitem [{\citenamefont {Zhu}\ \emph {et~al.}(2016)\citenamefont {Zhu},
  \citenamefont {Janoschek}, \citenamefont {Chaves}, \citenamefont {Cezar},
  \citenamefont {Durakiewicz}, \citenamefont {Ronning}, \citenamefont {Sassa},
  \citenamefont {Mansson}, \citenamefont {Scott}, \citenamefont {Wakeham} \emph
  {et~al.}}]{zhu2016electronic}%
  \BibitemOpen
  \bibfield  {author} {\bibinfo {author} {\bibfnamefont {J.-X.}\ \bibnamefont
  {Zhu}}, \bibinfo {author} {\bibfnamefont {M.}~\bibnamefont {Janoschek}},
  \bibinfo {author} {\bibfnamefont {D.}~\bibnamefont {Chaves}}, \bibinfo
  {author} {\bibfnamefont {J.}~\bibnamefont {Cezar}}, \bibinfo {author}
  {\bibfnamefont {T.}~\bibnamefont {Durakiewicz}}, \bibinfo {author}
  {\bibfnamefont {F.}~\bibnamefont {Ronning}}, \bibinfo {author} {\bibfnamefont
  {Y.}~\bibnamefont {Sassa}}, \bibinfo {author} {\bibfnamefont
  {M.}~\bibnamefont {Mansson}}, \bibinfo {author} {\bibfnamefont
  {B.}~\bibnamefont {Scott}}, \bibinfo {author} {\bibfnamefont
  {N.}~\bibnamefont {Wakeham}}, \emph {et~al.},\ }\bibfield  {title} {\bibinfo
  {title} {{Electronic correlation and magnetism in the ferromagnetic metal
  Fe$_3$GeTe$_2$}},\ }\href@noop {} {\bibfield  {journal} {\bibinfo  {journal}
  {Physical Review B}\ }\textbf {\bibinfo {volume} {93}},\ \bibinfo {pages}
  {144404} (\bibinfo {year} {2016})}\BibitemShut {NoStop}%
\bibitem [{\citenamefont {Zhang}\ \emph {et~al.}(2018)\citenamefont {Zhang},
  \citenamefont {Lu}, \citenamefont {Zhu}, \citenamefont {Tan}, \citenamefont
  {Feng}, \citenamefont {Liu}, \citenamefont {Zhang}, \citenamefont {Chen},
  \citenamefont {Liu}, \citenamefont {Luo} \emph
  {et~al.}}]{zhang2018emergence}%
  \BibitemOpen
  \bibfield  {author} {\bibinfo {author} {\bibfnamefont {Y.}~\bibnamefont
  {Zhang}}, \bibinfo {author} {\bibfnamefont {H.}~\bibnamefont {Lu}}, \bibinfo
  {author} {\bibfnamefont {X.}~\bibnamefont {Zhu}}, \bibinfo {author}
  {\bibfnamefont {S.}~\bibnamefont {Tan}}, \bibinfo {author} {\bibfnamefont
  {W.}~\bibnamefont {Feng}}, \bibinfo {author} {\bibfnamefont {Q.}~\bibnamefont
  {Liu}}, \bibinfo {author} {\bibfnamefont {W.}~\bibnamefont {Zhang}}, \bibinfo
  {author} {\bibfnamefont {Q.}~\bibnamefont {Chen}}, \bibinfo {author}
  {\bibfnamefont {Y.}~\bibnamefont {Liu}}, \bibinfo {author} {\bibfnamefont
  {X.}~\bibnamefont {Luo}}, \emph {et~al.},\ }\bibfield  {title} {\bibinfo
  {title} {Emergence of {K}ondo lattice behavior in a van der {W}aals itinerant
  ferromagnet, {Fe$_3$GeTe$_2$}},\ }\href@noop {} {\bibfield  {journal}
  {\bibinfo  {journal} {Science Advances}\ }\textbf {\bibinfo {volume} {4}},\
  \bibinfo {pages} {eaao6791} (\bibinfo {year} {2018})}\BibitemShut {NoStop}%
\bibitem [{\citenamefont {Kim}\ \emph {et~al.}(2022)\citenamefont {Kim},
  \citenamefont {Ryee},\ and\ \citenamefont {Han}}]{kim2022fe3gete2}%
  \BibitemOpen
  \bibfield  {author} {\bibinfo {author} {\bibfnamefont {T.~J.}\ \bibnamefont
  {Kim}}, \bibinfo {author} {\bibfnamefont {S.}~\bibnamefont {Ryee}},\ and\
  \bibinfo {author} {\bibfnamefont {M.~J.}\ \bibnamefont {Han}},\ }\bibfield
  {title} {\bibinfo {title} {{Fe$_3$GeTe$_2$}: a site-differentiated {H}und
  metal},\ }\href@noop {} {\bibfield  {journal} {\bibinfo  {journal} {npj
  Computational Materials}\ }\textbf {\bibinfo {volume} {8}},\ \bibinfo {pages}
  {245} (\bibinfo {year} {2022})}\BibitemShut {NoStop}%
\bibitem [{\citenamefont {Kim}\ \emph {et~al.}(2018)\citenamefont {Kim},
  \citenamefont {Seo}, \citenamefont {Lee}, \citenamefont {Ko}, \citenamefont
  {Kim}, \citenamefont {Jang}, \citenamefont {Ok}, \citenamefont {Lee},
  \citenamefont {Jo}, \citenamefont {Kang} \emph {et~al.}}]{kim2018large}%
  \BibitemOpen
  \bibfield  {author} {\bibinfo {author} {\bibfnamefont {K.}~\bibnamefont
  {Kim}}, \bibinfo {author} {\bibfnamefont {J.}~\bibnamefont {Seo}}, \bibinfo
  {author} {\bibfnamefont {E.}~\bibnamefont {Lee}}, \bibinfo {author}
  {\bibfnamefont {K.-T.}\ \bibnamefont {Ko}}, \bibinfo {author} {\bibfnamefont
  {B.}~\bibnamefont {Kim}}, \bibinfo {author} {\bibfnamefont {B.~G.}\
  \bibnamefont {Jang}}, \bibinfo {author} {\bibfnamefont {J.~M.}\ \bibnamefont
  {Ok}}, \bibinfo {author} {\bibfnamefont {J.}~\bibnamefont {Lee}}, \bibinfo
  {author} {\bibfnamefont {Y.~J.}\ \bibnamefont {Jo}}, \bibinfo {author}
  {\bibfnamefont {W.}~\bibnamefont {Kang}}, \emph {et~al.},\ }\bibfield
  {title} {\bibinfo {title} {Large anomalous {H}all current induced by
  topological nodal lines in a ferromagnetic van der {W}aals semimetal},\
  }\href@noop {} {\bibfield  {journal} {\bibinfo  {journal} {Nature Materials}\
  }\textbf {\bibinfo {volume} {17}},\ \bibinfo {pages} {794} (\bibinfo {year}
  {2018})}\BibitemShut {NoStop}%
\bibitem [{\citenamefont {Hanke}\ \emph {et~al.}(2016)\citenamefont {Hanke},
  \citenamefont {Freimuth}, \citenamefont {Nandy}, \citenamefont {Zhang},
  \citenamefont {Bl{\"u}gel},\ and\ \citenamefont {Mokrousov}}]{hanke2016role}%
  \BibitemOpen
  \bibfield  {author} {\bibinfo {author} {\bibfnamefont {J.-P.}\ \bibnamefont
  {Hanke}}, \bibinfo {author} {\bibfnamefont {F.}~\bibnamefont {Freimuth}},
  \bibinfo {author} {\bibfnamefont {A.~K.}\ \bibnamefont {Nandy}}, \bibinfo
  {author} {\bibfnamefont {H.}~\bibnamefont {Zhang}}, \bibinfo {author}
  {\bibfnamefont {S.}~\bibnamefont {Bl{\"u}gel}},\ and\ \bibinfo {author}
  {\bibfnamefont {Y.}~\bibnamefont {Mokrousov}},\ }\bibfield  {title} {\bibinfo
  {title} {Role of {B}erry phase theory for describing orbital magnetism: From
  magnetic heterostructures to topological orbital ferromagnets},\ }\href@noop
  {} {\bibfield  {journal} {\bibinfo  {journal} {Physical Review B}\ }\textbf
  {\bibinfo {volume} {94}},\ \bibinfo {pages} {121114} (\bibinfo {year}
  {2016})}\BibitemShut {NoStop}%
\bibitem [{\citenamefont {Zhao}\ \emph
  {et~al.}(2021{\natexlab{a}})\citenamefont {Zhao}, \citenamefont {Zhao},
  \citenamefont {Xi}, \citenamefont {Xu}, \citenamefont {Feng}, \citenamefont
  {Xu}, \citenamefont {Hao}, \citenamefont {Zhou}, \citenamefont {Zhao},
  \citenamefont {Dou} \emph {et~al.}}]{zhao2021electric}%
  \BibitemOpen
  \bibfield  {author} {\bibinfo {author} {\bibfnamefont {M.}~\bibnamefont
  {Zhao}}, \bibinfo {author} {\bibfnamefont {Y.}~\bibnamefont {Zhao}}, \bibinfo
  {author} {\bibfnamefont {Y.}~\bibnamefont {Xi}}, \bibinfo {author}
  {\bibfnamefont {H.}~\bibnamefont {Xu}}, \bibinfo {author} {\bibfnamefont
  {H.}~\bibnamefont {Feng}}, \bibinfo {author} {\bibfnamefont {X.}~\bibnamefont
  {Xu}}, \bibinfo {author} {\bibfnamefont {W.}~\bibnamefont {Hao}}, \bibinfo
  {author} {\bibfnamefont {S.}~\bibnamefont {Zhou}}, \bibinfo {author}
  {\bibfnamefont {J.}~\bibnamefont {Zhao}}, \bibinfo {author} {\bibfnamefont
  {S.~X.}\ \bibnamefont {Dou}}, \emph {et~al.},\ }\bibfield  {title} {\bibinfo
  {title} {Electric-field-driven negative differential conductance in {2D} van
  der {W}aals ferromagnet {Fe$_3$GeTe$_2$}},\ }\href@noop {} {\bibfield
  {journal} {\bibinfo  {journal} {Nano Letters}\ }\textbf {\bibinfo {volume}
  {21}},\ \bibinfo {pages} {9233} (\bibinfo {year}
  {2021}{\natexlab{a}})}\BibitemShut {NoStop}%
\bibitem [{\citenamefont {Zhao}\ \emph
  {et~al.}(2021{\natexlab{b}})\citenamefont {Zhao}, \citenamefont {Chen},
  \citenamefont {Xi}, \citenamefont {Zhao}, \citenamefont {Xu}, \citenamefont
  {Zhang}, \citenamefont {Cheng}, \citenamefont {Feng}, \citenamefont {Zhuang},
  \citenamefont {Pan} \emph {et~al.}}]{zhao2021kondo}%
  \BibitemOpen
  \bibfield  {author} {\bibinfo {author} {\bibfnamefont {M.}~\bibnamefont
  {Zhao}}, \bibinfo {author} {\bibfnamefont {B.-B.}\ \bibnamefont {Chen}},
  \bibinfo {author} {\bibfnamefont {Y.}~\bibnamefont {Xi}}, \bibinfo {author}
  {\bibfnamefont {Y.}~\bibnamefont {Zhao}}, \bibinfo {author} {\bibfnamefont
  {H.}~\bibnamefont {Xu}}, \bibinfo {author} {\bibfnamefont {H.}~\bibnamefont
  {Zhang}}, \bibinfo {author} {\bibfnamefont {N.}~\bibnamefont {Cheng}},
  \bibinfo {author} {\bibfnamefont {H.}~\bibnamefont {Feng}}, \bibinfo {author}
  {\bibfnamefont {J.}~\bibnamefont {Zhuang}}, \bibinfo {author} {\bibfnamefont
  {F.}~\bibnamefont {Pan}}, \emph {et~al.},\ }\bibfield  {title} {\bibinfo
  {title} {{K}ondo holes in the two-dimensional itinerant ising ferromagnet
  {Fe$_3$GeTe$_2$}},\ }\href@noop {} {\bibfield  {journal} {\bibinfo  {journal}
  {Nano Letters}\ }\textbf {\bibinfo {volume} {21}},\ \bibinfo {pages} {6117}
  (\bibinfo {year} {2021}{\natexlab{b}})}\BibitemShut {NoStop}%
\bibitem [{\citenamefont {Srivastava}\ \emph {et~al.}(2024)\citenamefont
  {Srivastava}, \citenamefont {Hassan}, \citenamefont {Lee}, \citenamefont
  {Joe}, \citenamefont {Abbas}, \citenamefont {Ahn}, \citenamefont {Tiwari},
  \citenamefont {Ghosh}, \citenamefont {Yoo}, \citenamefont {Singh} \emph
  {et~al.}}]{srivastava2024unusual}%
  \BibitemOpen
  \bibfield  {author} {\bibinfo {author} {\bibfnamefont {P.~K.}\ \bibnamefont
  {Srivastava}}, \bibinfo {author} {\bibfnamefont {Y.}~\bibnamefont {Hassan}},
  \bibinfo {author} {\bibfnamefont {S.}~\bibnamefont {Lee}}, \bibinfo {author}
  {\bibfnamefont {M.}~\bibnamefont {Joe}}, \bibinfo {author} {\bibfnamefont
  {M.~S.}\ \bibnamefont {Abbas}}, \bibinfo {author} {\bibfnamefont
  {H.}~\bibnamefont {Ahn}}, \bibinfo {author} {\bibfnamefont {A.}~\bibnamefont
  {Tiwari}}, \bibinfo {author} {\bibfnamefont {S.}~\bibnamefont {Ghosh}},
  \bibinfo {author} {\bibfnamefont {W.~J.}\ \bibnamefont {Yoo}}, \bibinfo
  {author} {\bibfnamefont {B.}~\bibnamefont {Singh}}, \emph {et~al.},\
  }\bibfield  {title} {\bibinfo {title} {Unusual {Dzyaloshinskii-Moriya}
  interaction in graphene/{Fe$_3$GeTe$_2$} van der {W}aals heterostructure},\
  }\href@noop {} {\bibfield  {journal} {\bibinfo  {journal} {Small}\ }\textbf
  {\bibinfo {volume} {20}},\ \bibinfo {pages} {2402604} (\bibinfo {year}
  {2024})}\BibitemShut {NoStop}%
\bibitem [{\citenamefont {Deng}\ \emph {et~al.}(2018)\citenamefont {Deng},
  \citenamefont {Yu}, \citenamefont {Song}, \citenamefont {Zhang},
  \citenamefont {Wang}, \citenamefont {Sun}, \citenamefont {Yi}, \citenamefont
  {Wu}, \citenamefont {Wu}, \citenamefont {Zhu} \emph {et~al.}}]{deng2018gate}%
  \BibitemOpen
  \bibfield  {author} {\bibinfo {author} {\bibfnamefont {Y.}~\bibnamefont
  {Deng}}, \bibinfo {author} {\bibfnamefont {Y.}~\bibnamefont {Yu}}, \bibinfo
  {author} {\bibfnamefont {Y.}~\bibnamefont {Song}}, \bibinfo {author}
  {\bibfnamefont {J.}~\bibnamefont {Zhang}}, \bibinfo {author} {\bibfnamefont
  {N.~Z.}\ \bibnamefont {Wang}}, \bibinfo {author} {\bibfnamefont
  {Z.}~\bibnamefont {Sun}}, \bibinfo {author} {\bibfnamefont {Y.}~\bibnamefont
  {Yi}}, \bibinfo {author} {\bibfnamefont {Y.~Z.}\ \bibnamefont {Wu}}, \bibinfo
  {author} {\bibfnamefont {S.}~\bibnamefont {Wu}}, \bibinfo {author}
  {\bibfnamefont {J.}~\bibnamefont {Zhu}}, \emph {et~al.},\ }\bibfield  {title}
  {\bibinfo {title} {Gate-tunable room-temperature ferromagnetism in
  two-dimensional {Fe$_3$GeTe$_2$}},\ }\href@noop {} {\bibfield  {journal}
  {\bibinfo  {journal} {Nature}\ }\textbf {\bibinfo {volume} {563}},\ \bibinfo
  {pages} {94} (\bibinfo {year} {2018})}\BibitemShut {NoStop}%
\bibitem [{\citenamefont {Wang}\ \emph {et~al.}(2019)\citenamefont {Wang},
  \citenamefont {Tang}, \citenamefont {Xia}, \citenamefont {He}, \citenamefont
  {Zhang}, \citenamefont {Liu}, \citenamefont {Wan}, \citenamefont {Fang},
  \citenamefont {Guo}, \citenamefont {Yang} \emph {et~al.}}]{wang2019current}%
  \BibitemOpen
  \bibfield  {author} {\bibinfo {author} {\bibfnamefont {X.}~\bibnamefont
  {Wang}}, \bibinfo {author} {\bibfnamefont {J.}~\bibnamefont {Tang}}, \bibinfo
  {author} {\bibfnamefont {X.}~\bibnamefont {Xia}}, \bibinfo {author}
  {\bibfnamefont {C.}~\bibnamefont {He}}, \bibinfo {author} {\bibfnamefont
  {J.}~\bibnamefont {Zhang}}, \bibinfo {author} {\bibfnamefont
  {Y.}~\bibnamefont {Liu}}, \bibinfo {author} {\bibfnamefont {C.}~\bibnamefont
  {Wan}}, \bibinfo {author} {\bibfnamefont {C.}~\bibnamefont {Fang}}, \bibinfo
  {author} {\bibfnamefont {C.}~\bibnamefont {Guo}}, \bibinfo {author}
  {\bibfnamefont {W.}~\bibnamefont {Yang}}, \emph {et~al.},\ }\bibfield
  {title} {\bibinfo {title} {Current-driven magnetization switching in a van
  der {W}aals ferromagnet {Fe$_3$GeTe$_2$}},\ }\href@noop {} {\bibfield
  {journal} {\bibinfo  {journal} {Science Advances}\ }\textbf {\bibinfo
  {volume} {5}},\ \bibinfo {pages} {eaaw8904} (\bibinfo {year}
  {2019})}\BibitemShut {NoStop}%
\bibitem [{\citenamefont {Park}\ \emph {et~al.}(2019)\citenamefont {Park},
  \citenamefont {Kim}, \citenamefont {Liu}, \citenamefont {Hwang},
  \citenamefont {Kim}, \citenamefont {Kim}, \citenamefont {Kim}, \citenamefont
  {Petrovic}, \citenamefont {Hwang}, \citenamefont {Mo} \emph
  {et~al.}}]{park2019controlling}%
  \BibitemOpen
  \bibfield  {author} {\bibinfo {author} {\bibfnamefont {S.~Y.}\ \bibnamefont
  {Park}}, \bibinfo {author} {\bibfnamefont {D.~S.}\ \bibnamefont {Kim}},
  \bibinfo {author} {\bibfnamefont {Y.}~\bibnamefont {Liu}}, \bibinfo {author}
  {\bibfnamefont {J.}~\bibnamefont {Hwang}}, \bibinfo {author} {\bibfnamefont
  {Y.}~\bibnamefont {Kim}}, \bibinfo {author} {\bibfnamefont {W.}~\bibnamefont
  {Kim}}, \bibinfo {author} {\bibfnamefont {J.-Y.}\ \bibnamefont {Kim}},
  \bibinfo {author} {\bibfnamefont {C.}~\bibnamefont {Petrovic}}, \bibinfo
  {author} {\bibfnamefont {C.}~\bibnamefont {Hwang}}, \bibinfo {author}
  {\bibfnamefont {S.-K.}\ \bibnamefont {Mo}}, \emph {et~al.},\ }\bibfield
  {title} {\bibinfo {title} {{Controlling the magnetic anisotropy of the van
  der {W}aals ferromagnet Fe$_3$GeTe$_2$ through hole doping}},\ }\href@noop {}
  {\bibfield  {journal} {\bibinfo  {journal} {Nano Letters}\ }\textbf {\bibinfo
  {volume} {20}},\ \bibinfo {pages} {95} (\bibinfo {year} {2019})}\BibitemShut
  {NoStop}%
\bibitem [{\citenamefont {Shindou}\ and\ \citenamefont
  {Nagaosa}(2001)}]{shindou2001orbital}%
  \BibitemOpen
  \bibfield  {author} {\bibinfo {author} {\bibfnamefont {R.}~\bibnamefont
  {Shindou}}\ and\ \bibinfo {author} {\bibfnamefont {N.}~\bibnamefont
  {Nagaosa}},\ }\bibfield  {title} {\bibinfo {title} {Orbital ferromagnetism
  and anomalous {H}all effect in antiferromagnets on the distorted fcc
  lattice},\ }\href@noop {} {\bibfield  {journal} {\bibinfo  {journal}
  {Physical Review Letters}\ }\textbf {\bibinfo {volume} {87}},\ \bibinfo
  {pages} {116801} (\bibinfo {year} {2001})}\BibitemShut {NoStop}%
\bibitem [{\citenamefont {Tatara}\ and\ \citenamefont
  {Garcia}(2003)}]{tatara2003quantum}%
  \BibitemOpen
  \bibfield  {author} {\bibinfo {author} {\bibfnamefont {G.}~\bibnamefont
  {Tatara}}\ and\ \bibinfo {author} {\bibfnamefont {N.}~\bibnamefont
  {Garcia}},\ }\bibfield  {title} {\bibinfo {title} {Quantum toys for quantum
  computing: Persistent currents controlled by the spin {J}osephson effect},\
  }\href@noop {} {\bibfield  {journal} {\bibinfo  {journal} {Physical Review
  Letters}\ }\textbf {\bibinfo {volume} {91}},\ \bibinfo {pages} {076806}
  (\bibinfo {year} {2003})}\BibitemShut {NoStop}%
\bibitem [{\citenamefont {Bulaevskii}\ \emph {et~al.}(2008)\citenamefont
  {Bulaevskii}, \citenamefont {Batista}, \citenamefont {Mostovoy},\ and\
  \citenamefont {Khomskii}}]{bulaevskii2008electronic}%
  \BibitemOpen
  \bibfield  {author} {\bibinfo {author} {\bibfnamefont {L.}~\bibnamefont
  {Bulaevskii}}, \bibinfo {author} {\bibfnamefont {C.}~\bibnamefont {Batista}},
  \bibinfo {author} {\bibfnamefont {M.}~\bibnamefont {Mostovoy}},\ and\
  \bibinfo {author} {\bibfnamefont {D.}~\bibnamefont {Khomskii}},\ }\bibfield
  {title} {\bibinfo {title} {Electronic orbital currents and polarization in
  {M}ott insulators},\ }\href@noop {} {\bibfield  {journal} {\bibinfo
  {journal} {Physical Review B}\ }\textbf {\bibinfo {volume} {78}},\ \bibinfo
  {pages} {024402} (\bibinfo {year} {2008})}\BibitemShut {NoStop}%
\bibitem [{\citenamefont {dos Santos~Dias}\ \emph {et~al.}(2016)\citenamefont
  {dos Santos~Dias}, \citenamefont {Bouaziz}, \citenamefont {Bouhassoune},
  \citenamefont {Bl{\"u}gel},\ and\ \citenamefont {Lounis}}]{dos2016chirality}%
  \BibitemOpen
  \bibfield  {author} {\bibinfo {author} {\bibfnamefont {M.}~\bibnamefont {dos
  Santos~Dias}}, \bibinfo {author} {\bibfnamefont {J.}~\bibnamefont {Bouaziz}},
  \bibinfo {author} {\bibfnamefont {M.}~\bibnamefont {Bouhassoune}}, \bibinfo
  {author} {\bibfnamefont {S.}~\bibnamefont {Bl{\"u}gel}},\ and\ \bibinfo
  {author} {\bibfnamefont {S.}~\bibnamefont {Lounis}},\ }\bibfield  {title}
  {\bibinfo {title} {Chirality-driven orbital magnetic moments as a new probe
  for topological magnetic structures},\ }\href@noop {} {\bibfield  {journal}
  {\bibinfo  {journal} {Nature Communications}\ }\textbf {\bibinfo {volume}
  {7}},\ \bibinfo {pages} {13613} (\bibinfo {year} {2016})}\BibitemShut
  {NoStop}%
\bibitem [{\citenamefont {Momma}\ and\ \citenamefont
  {Izumi}(2011)}]{momma2011vesta}%
  \BibitemOpen
  \bibfield  {author} {\bibinfo {author} {\bibfnamefont {K.}~\bibnamefont
  {Momma}}\ and\ \bibinfo {author} {\bibfnamefont {F.}~\bibnamefont {Izumi}},\
  }\bibfield  {title} {\bibinfo {title} {{VESTA 3} for three-dimensional
  visualization of crystal, volumetric and morphology data},\ }\href@noop {}
  {\bibfield  {journal} {\bibinfo  {journal} {Applied Crystallography}\
  }\textbf {\bibinfo {volume} {44}},\ \bibinfo {pages} {1272} (\bibinfo {year}
  {2011})}\BibitemShut {NoStop}%
\bibitem [{\citenamefont {Zhao}\ \emph {et~al.}(2023)\citenamefont {Zhao},
  \citenamefont {Karpiak}, \citenamefont {Hoque}, \citenamefont {Dhagat},\ and\
  \citenamefont {Dash}}]{zhao2023strong}%
  \BibitemOpen
  \bibfield  {author} {\bibinfo {author} {\bibfnamefont {B.}~\bibnamefont
  {Zhao}}, \bibinfo {author} {\bibfnamefont {B.}~\bibnamefont {Karpiak}},
  \bibinfo {author} {\bibfnamefont {A.~M.}\ \bibnamefont {Hoque}}, \bibinfo
  {author} {\bibfnamefont {P.}~\bibnamefont {Dhagat}},\ and\ \bibinfo {author}
  {\bibfnamefont {S.~P.}\ \bibnamefont {Dash}},\ }\bibfield  {title} {\bibinfo
  {title} {Strong perpendicular anisotropic ferromagnet
  {Fe$_3$GeTe$_2$/graphene} van der {W}aals heterostructure},\ }\href@noop {}
  {\bibfield  {journal} {\bibinfo  {journal} {Journal of Physics D: Applied
  Physics}\ }\textbf {\bibinfo {volume} {56}},\ \bibinfo {pages} {094001}
  (\bibinfo {year} {2023})}\BibitemShut {NoStop}%
\bibitem [{\citenamefont {Yin}\ \emph {et~al.}(2018)\citenamefont {Yin},
  \citenamefont {Zhang}, \citenamefont {Li}, \citenamefont {Jiang},
  \citenamefont {Chang}, \citenamefont {Zhang}, \citenamefont {Lian},
  \citenamefont {Xiang}, \citenamefont {Belopolski}, \citenamefont {Zheng}
  \emph {et~al.}}]{yin2018giant}%
  \BibitemOpen
  \bibfield  {author} {\bibinfo {author} {\bibfnamefont {J.-X.}\ \bibnamefont
  {Yin}}, \bibinfo {author} {\bibfnamefont {S.~S.}\ \bibnamefont {Zhang}},
  \bibinfo {author} {\bibfnamefont {H.}~\bibnamefont {Li}}, \bibinfo {author}
  {\bibfnamefont {K.}~\bibnamefont {Jiang}}, \bibinfo {author} {\bibfnamefont
  {G.}~\bibnamefont {Chang}}, \bibinfo {author} {\bibfnamefont
  {B.}~\bibnamefont {Zhang}}, \bibinfo {author} {\bibfnamefont
  {B.}~\bibnamefont {Lian}}, \bibinfo {author} {\bibfnamefont {C.}~\bibnamefont
  {Xiang}}, \bibinfo {author} {\bibfnamefont {I.}~\bibnamefont {Belopolski}},
  \bibinfo {author} {\bibfnamefont {H.}~\bibnamefont {Zheng}}, \emph {et~al.},\
  }\bibfield  {title} {\bibinfo {title} {Giant and anisotropic many-body
  spin--orbit tunability in a strongly correlated {K}agome magnet},\
  }\href@noop {} {\bibfield  {journal} {\bibinfo  {journal} {Nature}\ }\textbf
  {\bibinfo {volume} {562}},\ \bibinfo {pages} {91} (\bibinfo {year}
  {2018})}\BibitemShut {NoStop}%
\bibitem [{\citenamefont {Lux}\ \emph {et~al.}(2018)\citenamefont {Lux},
  \citenamefont {Freimuth}, \citenamefont {Bl{\"u}gel},\ and\ \citenamefont
  {Mokrousov}}]{lux2018engineering}%
  \BibitemOpen
  \bibfield  {author} {\bibinfo {author} {\bibfnamefont {F.~R.}\ \bibnamefont
  {Lux}}, \bibinfo {author} {\bibfnamefont {F.}~\bibnamefont {Freimuth}},
  \bibinfo {author} {\bibfnamefont {S.}~\bibnamefont {Bl{\"u}gel}},\ and\
  \bibinfo {author} {\bibfnamefont {Y.}~\bibnamefont {Mokrousov}},\ }\bibfield
  {title} {\bibinfo {title} {Engineering chiral and topological orbital
  magnetism of domain walls and skyrmions},\ }\href@noop {} {\bibfield
  {journal} {\bibinfo  {journal} {Communications Physics}\ }\textbf {\bibinfo
  {volume} {1}},\ \bibinfo {pages} {60} (\bibinfo {year} {2018})}\BibitemShut
  {NoStop}%
\bibitem [{\citenamefont {Grytsiuk}\ \emph {et~al.}(2020)\citenamefont
  {Grytsiuk}, \citenamefont {Hanke}, \citenamefont {Hoffmann}, \citenamefont
  {Bouaziz}, \citenamefont {Gomonay}, \citenamefont {Bihlmayer}, \citenamefont
  {Lounis}, \citenamefont {Mokrousov},\ and\ \citenamefont
  {Bl{\"u}gel}}]{grytsiuk2020topological}%
  \BibitemOpen
  \bibfield  {author} {\bibinfo {author} {\bibfnamefont {S.}~\bibnamefont
  {Grytsiuk}}, \bibinfo {author} {\bibfnamefont {J.-P.}\ \bibnamefont {Hanke}},
  \bibinfo {author} {\bibfnamefont {M.}~\bibnamefont {Hoffmann}}, \bibinfo
  {author} {\bibfnamefont {J.}~\bibnamefont {Bouaziz}}, \bibinfo {author}
  {\bibfnamefont {O.}~\bibnamefont {Gomonay}}, \bibinfo {author} {\bibfnamefont
  {G.}~\bibnamefont {Bihlmayer}}, \bibinfo {author} {\bibfnamefont
  {S.}~\bibnamefont {Lounis}}, \bibinfo {author} {\bibfnamefont
  {Y.}~\bibnamefont {Mokrousov}},\ and\ \bibinfo {author} {\bibfnamefont
  {S.}~\bibnamefont {Bl{\"u}gel}},\ }\bibfield  {title} {\bibinfo {title}
  {Topological--chiral magnetic interactions driven by emergent orbital
  magnetism},\ }\href@noop {} {\bibfield  {journal} {\bibinfo  {journal}
  {Nature Communications}\ }\textbf {\bibinfo {volume} {11}},\ \bibinfo {pages}
  {511} (\bibinfo {year} {2020})}\BibitemShut {NoStop}%
\bibitem [{\citenamefont {Fei}\ \emph {et~al.}(2018)\citenamefont {Fei},
  \citenamefont {Huang}, \citenamefont {Malinowski}, \citenamefont {Wang},
  \citenamefont {Song}, \citenamefont {Sanchez}, \citenamefont {Yao},
  \citenamefont {Xiao}, \citenamefont {Zhu}, \citenamefont {May} \emph
  {et~al.}}]{fei2018two}%
  \BibitemOpen
  \bibfield  {author} {\bibinfo {author} {\bibfnamefont {Z.}~\bibnamefont
  {Fei}}, \bibinfo {author} {\bibfnamefont {B.}~\bibnamefont {Huang}}, \bibinfo
  {author} {\bibfnamefont {P.}~\bibnamefont {Malinowski}}, \bibinfo {author}
  {\bibfnamefont {W.}~\bibnamefont {Wang}}, \bibinfo {author} {\bibfnamefont
  {T.}~\bibnamefont {Song}}, \bibinfo {author} {\bibfnamefont {J.}~\bibnamefont
  {Sanchez}}, \bibinfo {author} {\bibfnamefont {W.}~\bibnamefont {Yao}},
  \bibinfo {author} {\bibfnamefont {D.}~\bibnamefont {Xiao}}, \bibinfo {author}
  {\bibfnamefont {X.}~\bibnamefont {Zhu}}, \bibinfo {author} {\bibfnamefont
  {A.~F.}\ \bibnamefont {May}}, \emph {et~al.},\ }\bibfield  {title} {\bibinfo
  {title} {Two-dimensional itinerant ferromagnetism in atomically thin
  {Fe$_3$GeTe$_2$}},\ }\href@noop {} {\bibfield  {journal} {\bibinfo  {journal}
  {Nature Materials}\ }\textbf {\bibinfo {volume} {17}},\ \bibinfo {pages}
  {778} (\bibinfo {year} {2018})}\BibitemShut {NoStop}%
\bibitem [{\citenamefont {Decker}\ \emph {et~al.}(2011)\citenamefont {Decker},
  \citenamefont {Wang}, \citenamefont {Brar}, \citenamefont {Regan},
  \citenamefont {Tsai}, \citenamefont {Wu}, \citenamefont {Gannett},
  \citenamefont {Zettl},\ and\ \citenamefont {Crommie}}]{decker2011local}%
  \BibitemOpen
  \bibfield  {author} {\bibinfo {author} {\bibfnamefont {R.}~\bibnamefont
  {Decker}}, \bibinfo {author} {\bibfnamefont {Y.}~\bibnamefont {Wang}},
  \bibinfo {author} {\bibfnamefont {V.~W.}\ \bibnamefont {Brar}}, \bibinfo
  {author} {\bibfnamefont {W.}~\bibnamefont {Regan}}, \bibinfo {author}
  {\bibfnamefont {H.-Z.}\ \bibnamefont {Tsai}}, \bibinfo {author}
  {\bibfnamefont {Q.}~\bibnamefont {Wu}}, \bibinfo {author} {\bibfnamefont
  {W.}~\bibnamefont {Gannett}}, \bibinfo {author} {\bibfnamefont
  {A.}~\bibnamefont {Zettl}},\ and\ \bibinfo {author} {\bibfnamefont {M.~F.}\
  \bibnamefont {Crommie}},\ }\bibfield  {title} {\bibinfo {title} {Local
  electronic properties of graphene on a {BN} substrate via scanning tunneling
  microscopy},\ }\href@noop {} {\bibfield  {journal} {\bibinfo  {journal} {Nano
  Letters}\ }\textbf {\bibinfo {volume} {11}},\ \bibinfo {pages} {2291}
  (\bibinfo {year} {2011})}\BibitemShut {NoStop}%
\bibitem [{\citenamefont {Zhang}\ \emph {et~al.}(2008)\citenamefont {Zhang},
  \citenamefont {Brar}, \citenamefont {Wang}, \citenamefont {Girit},
  \citenamefont {Yayon}, \citenamefont {Panlasigui}, \citenamefont {Zettl},\
  and\ \citenamefont {Crommie}}]{zhang2008giant}%
  \BibitemOpen
  \bibfield  {author} {\bibinfo {author} {\bibfnamefont {Y.}~\bibnamefont
  {Zhang}}, \bibinfo {author} {\bibfnamefont {V.~W.}\ \bibnamefont {Brar}},
  \bibinfo {author} {\bibfnamefont {F.}~\bibnamefont {Wang}}, \bibinfo {author}
  {\bibfnamefont {C.}~\bibnamefont {Girit}}, \bibinfo {author} {\bibfnamefont
  {Y.}~\bibnamefont {Yayon}}, \bibinfo {author} {\bibfnamefont
  {M.}~\bibnamefont {Panlasigui}}, \bibinfo {author} {\bibfnamefont
  {A.}~\bibnamefont {Zettl}},\ and\ \bibinfo {author} {\bibfnamefont {M.~F.}\
  \bibnamefont {Crommie}},\ }\bibfield  {title} {\bibinfo {title} {Giant
  phonon-induced conductance in scanning tunnelling spectroscopy of
  gate-tunable graphene},\ }\href@noop {} {\bibfield  {journal} {\bibinfo
  {journal} {Nature Physics}\ }\textbf {\bibinfo {volume} {4}},\ \bibinfo
  {pages} {627} (\bibinfo {year} {2008})}\BibitemShut {NoStop}%
\bibitem [{\citenamefont {Wehling}\ \emph {et~al.}(2008)\citenamefont
  {Wehling}, \citenamefont {Grigorenko}, \citenamefont {Lichtenstein},\ and\
  \citenamefont {Balatsky}}]{wehling2008phonon}%
  \BibitemOpen
  \bibfield  {author} {\bibinfo {author} {\bibfnamefont {T.}~\bibnamefont
  {Wehling}}, \bibinfo {author} {\bibfnamefont {I.}~\bibnamefont {Grigorenko}},
  \bibinfo {author} {\bibfnamefont {A.}~\bibnamefont {Lichtenstein}},\ and\
  \bibinfo {author} {\bibfnamefont {A.}~\bibnamefont {Balatsky}},\ }\bibfield
  {title} {\bibinfo {title} {Phonon-mediated tunneling into graphene},\
  }\href@noop {} {\bibfield  {journal} {\bibinfo  {journal} {Physical Review
  Letters}\ }\textbf {\bibinfo {volume} {101}},\ \bibinfo {pages} {216803}
  (\bibinfo {year} {2008})}\BibitemShut {NoStop}%
\bibitem [{\citenamefont {Natterer}\ \emph {et~al.}(2015)\citenamefont
  {Natterer}, \citenamefont {Zhao}, \citenamefont {Wyrick}, \citenamefont
  {Chan}, \citenamefont {Ruan}, \citenamefont {Chou}, \citenamefont {Watanabe},
  \citenamefont {Taniguchi}, \citenamefont {Zhitenev},\ and\ \citenamefont
  {Stroscio}}]{natterer2015strong}%
  \BibitemOpen
  \bibfield  {author} {\bibinfo {author} {\bibfnamefont {F.~D.}\ \bibnamefont
  {Natterer}}, \bibinfo {author} {\bibfnamefont {Y.}~\bibnamefont {Zhao}},
  \bibinfo {author} {\bibfnamefont {J.}~\bibnamefont {Wyrick}}, \bibinfo
  {author} {\bibfnamefont {Y.-H.}\ \bibnamefont {Chan}}, \bibinfo {author}
  {\bibfnamefont {W.-Y.}\ \bibnamefont {Ruan}}, \bibinfo {author}
  {\bibfnamefont {M.-Y.}\ \bibnamefont {Chou}}, \bibinfo {author}
  {\bibfnamefont {K.}~\bibnamefont {Watanabe}}, \bibinfo {author}
  {\bibfnamefont {T.}~\bibnamefont {Taniguchi}}, \bibinfo {author}
  {\bibfnamefont {N.~B.}\ \bibnamefont {Zhitenev}},\ and\ \bibinfo {author}
  {\bibfnamefont {J.~A.}\ \bibnamefont {Stroscio}},\ }\bibfield  {title}
  {\bibinfo {title} {Strong asymmetric charge carrier dependence in inelastic
  electron tunneling spectroscopy of graphene phonons},\ }\href@noop {}
  {\bibfield  {journal} {\bibinfo  {journal} {Physical Review Letters}\
  }\textbf {\bibinfo {volume} {114}},\ \bibinfo {pages} {245502} (\bibinfo
  {year} {2015})}\BibitemShut {NoStop}%
\bibitem [{\citenamefont {Qiu}\ \emph {et~al.}(2021)\citenamefont {Qiu},
  \citenamefont {Holwill}, \citenamefont {Olsen}, \citenamefont {Lyu},
  \citenamefont {Li}, \citenamefont {Fang}, \citenamefont {Yang}, \citenamefont
  {Kashchenko}, \citenamefont {Novoselov},\ and\ \citenamefont
  {Lu}}]{qiu2021visualizing}%
  \BibitemOpen
  \bibfield  {author} {\bibinfo {author} {\bibfnamefont {Z.}~\bibnamefont
  {Qiu}}, \bibinfo {author} {\bibfnamefont {M.}~\bibnamefont {Holwill}},
  \bibinfo {author} {\bibfnamefont {T.}~\bibnamefont {Olsen}}, \bibinfo
  {author} {\bibfnamefont {P.}~\bibnamefont {Lyu}}, \bibinfo {author}
  {\bibfnamefont {J.}~\bibnamefont {Li}}, \bibinfo {author} {\bibfnamefont
  {H.}~\bibnamefont {Fang}}, \bibinfo {author} {\bibfnamefont {H.}~\bibnamefont
  {Yang}}, \bibinfo {author} {\bibfnamefont {M.}~\bibnamefont {Kashchenko}},
  \bibinfo {author} {\bibfnamefont {K.~S.}\ \bibnamefont {Novoselov}},\ and\
  \bibinfo {author} {\bibfnamefont {J.}~\bibnamefont {Lu}},\ }\bibfield
  {title} {\bibinfo {title} {Visualizing atomic structure and magnetism of {2D}
  magnetic insulators via tunneling through graphene},\ }\href@noop {}
  {\bibfield  {journal} {\bibinfo  {journal} {Nature Communications}\ }\textbf
  {\bibinfo {volume} {12}},\ \bibinfo {pages} {1} (\bibinfo {year}
  {2021})}\BibitemShut {NoStop}%
\bibitem [{\citenamefont {Sun}\ \emph {et~al.}(2023)\citenamefont {Sun},
  \citenamefont {Rademaker}, \citenamefont {Mauro}, \citenamefont {Scarfato},
  \citenamefont {P{\'a}sztor}, \citenamefont {Guti{\'e}rrez-Lezama},
  \citenamefont {Wang}, \citenamefont {Martinez-Castro}, \citenamefont
  {Morpurgo},\ and\ \citenamefont {Renner}}]{sun2023determining}%
  \BibitemOpen
  \bibfield  {author} {\bibinfo {author} {\bibfnamefont {L.}~\bibnamefont
  {Sun}}, \bibinfo {author} {\bibfnamefont {L.}~\bibnamefont {Rademaker}},
  \bibinfo {author} {\bibfnamefont {D.}~\bibnamefont {Mauro}}, \bibinfo
  {author} {\bibfnamefont {A.}~\bibnamefont {Scarfato}}, \bibinfo {author}
  {\bibfnamefont {{\'A}.}~\bibnamefont {P{\'a}sztor}}, \bibinfo {author}
  {\bibfnamefont {I.}~\bibnamefont {Guti{\'e}rrez-Lezama}}, \bibinfo {author}
  {\bibfnamefont {Z.}~\bibnamefont {Wang}}, \bibinfo {author} {\bibfnamefont
  {J.}~\bibnamefont {Martinez-Castro}}, \bibinfo {author} {\bibfnamefont
  {A.~F.}\ \bibnamefont {Morpurgo}},\ and\ \bibinfo {author} {\bibfnamefont
  {C.}~\bibnamefont {Renner}},\ }\bibfield  {title} {\bibinfo {title}
  {Determining spin-orbit coupling in graphene by quasiparticle interference
  imaging},\ }\href@noop {} {\bibfield  {journal} {\bibinfo  {journal} {Nature
  Communications}\ }\textbf {\bibinfo {volume} {14}},\ \bibinfo {pages} {3771}
  (\bibinfo {year} {2023})}\BibitemShut {NoStop}%
\bibitem [{\citenamefont {Li}\ \emph {et~al.}(2022)\citenamefont {Li},
  \citenamefont {Zhao}, \citenamefont {Jiang}, \citenamefont {Wang},
  \citenamefont {Yin}, \citenamefont {Zhao}, \citenamefont {Liu}, \citenamefont
  {Wang}, \citenamefont {Lei},\ and\ \citenamefont
  {Zeljkovic}}]{li2022manipulation}%
  \BibitemOpen
  \bibfield  {author} {\bibinfo {author} {\bibfnamefont {H.}~\bibnamefont
  {Li}}, \bibinfo {author} {\bibfnamefont {H.}~\bibnamefont {Zhao}}, \bibinfo
  {author} {\bibfnamefont {K.}~\bibnamefont {Jiang}}, \bibinfo {author}
  {\bibfnamefont {Q.}~\bibnamefont {Wang}}, \bibinfo {author} {\bibfnamefont
  {Q.}~\bibnamefont {Yin}}, \bibinfo {author} {\bibfnamefont {N.-N.}\
  \bibnamefont {Zhao}}, \bibinfo {author} {\bibfnamefont {K.}~\bibnamefont
  {Liu}}, \bibinfo {author} {\bibfnamefont {Z.}~\bibnamefont {Wang}}, \bibinfo
  {author} {\bibfnamefont {H.}~\bibnamefont {Lei}},\ and\ \bibinfo {author}
  {\bibfnamefont {I.}~\bibnamefont {Zeljkovic}},\ }\bibfield  {title} {\bibinfo
  {title} {Manipulation of {D}irac band curvature and momentum-dependent
  $g$-factor in a {K}agome magnet},\ }\href@noop {} {\bibfield  {journal}
  {\bibinfo  {journal} {Nature Physics}\ }\textbf {\bibinfo {volume} {18}},\
  \bibinfo {pages} {644} (\bibinfo {year} {2022})}\BibitemShut {NoStop}%
\bibitem [{\citenamefont {Li}\ \emph {et~al.}(2024)\citenamefont {Li},
  \citenamefont {Cheng}, \citenamefont {Pokharel}, \citenamefont {Eck},
  \citenamefont {Bigi}, \citenamefont {Mazzola}, \citenamefont {Sangiovanni},
  \citenamefont {Wilson}, \citenamefont {Di~Sante}, \citenamefont {Wang} \emph
  {et~al.}}]{li2024spin}%
  \BibitemOpen
  \bibfield  {author} {\bibinfo {author} {\bibfnamefont {H.}~\bibnamefont
  {Li}}, \bibinfo {author} {\bibfnamefont {S.}~\bibnamefont {Cheng}}, \bibinfo
  {author} {\bibfnamefont {G.}~\bibnamefont {Pokharel}}, \bibinfo {author}
  {\bibfnamefont {P.}~\bibnamefont {Eck}}, \bibinfo {author} {\bibfnamefont
  {C.}~\bibnamefont {Bigi}}, \bibinfo {author} {\bibfnamefont {F.}~\bibnamefont
  {Mazzola}}, \bibinfo {author} {\bibfnamefont {G.}~\bibnamefont
  {Sangiovanni}}, \bibinfo {author} {\bibfnamefont {S.~D.}\ \bibnamefont
  {Wilson}}, \bibinfo {author} {\bibfnamefont {D.}~\bibnamefont {Di~Sante}},
  \bibinfo {author} {\bibfnamefont {Z.}~\bibnamefont {Wang}}, \emph {et~al.},\
  }\bibfield  {title} {\bibinfo {title} {{Spin {B}erry curvature-enhanced
  orbital {Z}eeman effect in a {K}agome metal}},\ }\href@noop {} {\bibfield
  {journal} {\bibinfo  {journal} {Nature Physics}\ }\textbf {\bibinfo {volume}
  {20}},\ \bibinfo {pages} {1103} (\bibinfo {year} {2024})}\BibitemShut
  {NoStop}%
\bibitem [{\citenamefont {Jo}\ \emph {et~al.}(2025)\citenamefont {Jo},
  \citenamefont {Go}, \citenamefont {Mokrousov}, \citenamefont {Oppeneer},
  \citenamefont {Cheong},\ and\ \citenamefont {Lee}}]{jo2025weak}%
  \BibitemOpen
  \bibfield  {author} {\bibinfo {author} {\bibfnamefont {D.}~\bibnamefont
  {Jo}}, \bibinfo {author} {\bibfnamefont {D.}~\bibnamefont {Go}}, \bibinfo
  {author} {\bibfnamefont {Y.}~\bibnamefont {Mokrousov}}, \bibinfo {author}
  {\bibfnamefont {P.~M.}\ \bibnamefont {Oppeneer}}, \bibinfo {author}
  {\bibfnamefont {S.-W.}\ \bibnamefont {Cheong}},\ and\ \bibinfo {author}
  {\bibfnamefont {H.-W.}\ \bibnamefont {Lee}},\ }\bibfield  {title} {\bibinfo
  {title} {Weak ferromagnetism in altermagnets from alternating $g$-{T}ensor
  anisotropy},\ }\href@noop {} {\bibfield  {journal} {\bibinfo  {journal}
  {Physical Review Letters}\ }\textbf {\bibinfo {volume} {134}},\ \bibinfo
  {pages} {196703} (\bibinfo {year} {2025})}\BibitemShut {NoStop}%
\bibitem [{\citenamefont {M\"uhlbauer}\ \emph {et~al.}(2009)\citenamefont
  {M\"uhlbauer}, \citenamefont {Binz}, \citenamefont {Jonietz}, \citenamefont
  {Pfleiderer}, \citenamefont {Rosch}, \citenamefont {Neubauer}, \citenamefont
  {Georgii},\ and\ \citenamefont {B\"oni}}]{muhlbauer2009skyrmion}%
  \BibitemOpen
  \bibfield  {author} {\bibinfo {author} {\bibfnamefont {S.}~\bibnamefont
  {M\"uhlbauer}}, \bibinfo {author} {\bibfnamefont {B.}~\bibnamefont {Binz}},
  \bibinfo {author} {\bibfnamefont {F.}~\bibnamefont {Jonietz}}, \bibinfo
  {author} {\bibfnamefont {C.}~\bibnamefont {Pfleiderer}}, \bibinfo {author}
  {\bibfnamefont {A.}~\bibnamefont {Rosch}}, \bibinfo {author} {\bibfnamefont
  {A.}~\bibnamefont {Neubauer}}, \bibinfo {author} {\bibfnamefont
  {R.}~\bibnamefont {Georgii}},\ and\ \bibinfo {author} {\bibfnamefont
  {P.}~\bibnamefont {B\"oni}},\ }\bibfield  {title} {\bibinfo {title} {Skyrmion
  lattice in a chiral magnet},\ }\href@noop {} {\bibfield  {journal} {\bibinfo
  {journal} {Science}\ }\textbf {\bibinfo {volume} {323}},\ \bibinfo {pages}
  {915} (\bibinfo {year} {2009})}\BibitemShut {NoStop}%
\bibitem [{\citenamefont {Li}\ \emph {et~al.}(2023)\citenamefont {Li},
  \citenamefont {Haldar}, \citenamefont {Drevelow},\ and\ \citenamefont
  {Heinze}}]{li2023tuning}%
  \BibitemOpen
  \bibfield  {author} {\bibinfo {author} {\bibfnamefont {D.}~\bibnamefont
  {Li}}, \bibinfo {author} {\bibfnamefont {S.}~\bibnamefont {Haldar}}, \bibinfo
  {author} {\bibfnamefont {T.}~\bibnamefont {Drevelow}},\ and\ \bibinfo
  {author} {\bibfnamefont {S.}~\bibnamefont {Heinze}},\ }\bibfield  {title}
  {\bibinfo {title} {Tuning the magnetic interactions in van der {W}aals
  {Fe$_3$GeTe$_2$} heterostructures: {A} comparative study of ab initio
  methods},\ }\href@noop {} {\bibfield  {journal} {\bibinfo  {journal}
  {Physical Review B}\ }\textbf {\bibinfo {volume} {107}},\ \bibinfo {pages}
  {104428} (\bibinfo {year} {2023})}\BibitemShut {NoStop}%
\bibitem [{\citenamefont {Niu}\ \emph {et~al.}(2019)\citenamefont {Niu},
  \citenamefont {Hanke}, \citenamefont {Buhl}, \citenamefont {Zhang},
  \citenamefont {Plucinski}, \citenamefont {Wortmann}, \citenamefont
  {Bl{\"u}gel}, \citenamefont {Bihlmayer},\ and\ \citenamefont
  {Mokrousov}}]{niu2019mixed}%
  \BibitemOpen
  \bibfield  {author} {\bibinfo {author} {\bibfnamefont {C.}~\bibnamefont
  {Niu}}, \bibinfo {author} {\bibfnamefont {J.-P.}\ \bibnamefont {Hanke}},
  \bibinfo {author} {\bibfnamefont {P.~M.}\ \bibnamefont {Buhl}}, \bibinfo
  {author} {\bibfnamefont {H.}~\bibnamefont {Zhang}}, \bibinfo {author}
  {\bibfnamefont {L.}~\bibnamefont {Plucinski}}, \bibinfo {author}
  {\bibfnamefont {D.}~\bibnamefont {Wortmann}}, \bibinfo {author}
  {\bibfnamefont {S.}~\bibnamefont {Bl{\"u}gel}}, \bibinfo {author}
  {\bibfnamefont {G.}~\bibnamefont {Bihlmayer}},\ and\ \bibinfo {author}
  {\bibfnamefont {Y.}~\bibnamefont {Mokrousov}},\ }\bibfield  {title} {\bibinfo
  {title} {Mixed topological semimetals driven by orbital complexity in
  two-dimensional ferromagnets},\ }\href@noop {} {\bibfield  {journal}
  {\bibinfo  {journal} {Nature Communications}\ }\textbf {\bibinfo {volume}
  {10}},\ \bibinfo {pages} {3179} (\bibinfo {year} {2019})}\BibitemShut
  {NoStop}%
\bibitem [{\citenamefont {Ko}(2022)}]{ko2022hybridized}%
  \BibitemOpen
  \bibfield  {author} {\bibinfo {author} {\bibfnamefont {E.}~\bibnamefont
  {Ko}},\ }\bibfield  {title} {\bibinfo {title} {Hybridized bands and
  stacking-dependent band edges in ferromagnetic {Fe$_3$GeTe$_2$/CrGeTe$_3$}
  moir{\'e} heterobilayer},\ }\href@noop {} {\bibfield  {journal} {\bibinfo
  {journal} {Scientific reports}\ }\textbf {\bibinfo {volume} {12}},\ \bibinfo
  {pages} {1} (\bibinfo {year} {2022})}\BibitemShut {NoStop}%
\bibitem [{\citenamefont {Yu}\ \emph {et~al.}(2009)\citenamefont {Yu},
  \citenamefont {Zhao}, \citenamefont {Ryu}, \citenamefont {Brus},
  \citenamefont {Kim},\ and\ \citenamefont {Kim}}]{yu2009tuning}%
  \BibitemOpen
  \bibfield  {author} {\bibinfo {author} {\bibfnamefont {Y.-J.}\ \bibnamefont
  {Yu}}, \bibinfo {author} {\bibfnamefont {Y.}~\bibnamefont {Zhao}}, \bibinfo
  {author} {\bibfnamefont {S.}~\bibnamefont {Ryu}}, \bibinfo {author}
  {\bibfnamefont {L.~E.}\ \bibnamefont {Brus}}, \bibinfo {author}
  {\bibfnamefont {K.~S.}\ \bibnamefont {Kim}},\ and\ \bibinfo {author}
  {\bibfnamefont {P.}~\bibnamefont {Kim}},\ }\bibfield  {title} {\bibinfo
  {title} {Tuning the graphene work function by electric field effect},\
  }\href@noop {} {\bibfield  {journal} {\bibinfo  {journal} {Nano Letters}\
  }\textbf {\bibinfo {volume} {9}},\ \bibinfo {pages} {3430} (\bibinfo {year}
  {2009})}\BibitemShut {NoStop}%
\bibitem [{\citenamefont {Huang}\ \emph {et~al.}(2025)\citenamefont {Huang},
  \citenamefont {Zhu}, \citenamefont {Zhao}, \citenamefont {Watanabe},
  \citenamefont {Taniguchi}, \citenamefont {Xiao}, \citenamefont {Wang},
  \citenamefont {Mei}, \citenamefont {Huang}, \citenamefont {Zhang} \emph
  {et~al.}}]{huang2025giant}%
  \BibitemOpen
  \bibfield  {author} {\bibinfo {author} {\bibfnamefont {S.}~\bibnamefont
  {Huang}}, \bibinfo {author} {\bibfnamefont {L.}~\bibnamefont {Zhu}}, \bibinfo
  {author} {\bibfnamefont {Y.}~\bibnamefont {Zhao}}, \bibinfo {author}
  {\bibfnamefont {K.}~\bibnamefont {Watanabe}}, \bibinfo {author}
  {\bibfnamefont {T.}~\bibnamefont {Taniguchi}}, \bibinfo {author}
  {\bibfnamefont {J.}~\bibnamefont {Xiao}}, \bibinfo {author} {\bibfnamefont
  {L.}~\bibnamefont {Wang}}, \bibinfo {author} {\bibfnamefont {J.}~\bibnamefont
  {Mei}}, \bibinfo {author} {\bibfnamefont {H.}~\bibnamefont {Huang}}, \bibinfo
  {author} {\bibfnamefont {F.}~\bibnamefont {Zhang}}, \emph {et~al.},\
  }\bibfield  {title} {\bibinfo {title} {Giant magnetoresistance induced by
  spin-dependent orbital coupling in {Fe$_3$GeTe$_2$}/graphene
  heterostructures},\ }\href@noop {} {\bibfield  {journal} {\bibinfo  {journal}
  {Nature Communications}\ }\textbf {\bibinfo {volume} {16}},\ \bibinfo {pages}
  {2866} (\bibinfo {year} {2025})}\BibitemShut {NoStop}%
\bibitem [{\citenamefont {Kim}\ \emph {et~al.}(2013)\citenamefont {Kim},
  \citenamefont {Kim}, \citenamefont {Walter}, \citenamefont {Seyller},
  \citenamefont {Yeom}, \citenamefont {Rotenberg},\ and\ \citenamefont
  {Bostwick}}]{kim2013visualizing}%
  \BibitemOpen
  \bibfield  {author} {\bibinfo {author} {\bibfnamefont {K.~S.}\ \bibnamefont
  {Kim}}, \bibinfo {author} {\bibfnamefont {T.-H.}\ \bibnamefont {Kim}},
  \bibinfo {author} {\bibfnamefont {A.~L.}\ \bibnamefont {Walter}}, \bibinfo
  {author} {\bibfnamefont {T.}~\bibnamefont {Seyller}}, \bibinfo {author}
  {\bibfnamefont {H.~W.}\ \bibnamefont {Yeom}}, \bibinfo {author}
  {\bibfnamefont {E.}~\bibnamefont {Rotenberg}},\ and\ \bibinfo {author}
  {\bibfnamefont {A.}~\bibnamefont {Bostwick}},\ }\bibfield  {title} {\bibinfo
  {title} {Visualizing atomic-scale negative differential resistance in bilayer
  graphene},\ }\href@noop {} {\bibfield  {journal} {\bibinfo  {journal}
  {Physical Review Letters}\ }\textbf {\bibinfo {volume} {110}},\ \bibinfo
  {pages} {036804} (\bibinfo {year} {2013})}\BibitemShut {NoStop}%
\bibitem [{\citenamefont {Luican-Mayer}\ \emph {et~al.}(2019)\citenamefont
  {Luican-Mayer}, \citenamefont {Zhang}, \citenamefont {DiLullo}, \citenamefont
  {Li}, \citenamefont {Fisher}, \citenamefont {Ulloa},\ and\ \citenamefont
  {Hla}}]{luican2019negative}%
  \BibitemOpen
  \bibfield  {author} {\bibinfo {author} {\bibfnamefont {A.}~\bibnamefont
  {Luican-Mayer}}, \bibinfo {author} {\bibfnamefont {Y.}~\bibnamefont {Zhang}},
  \bibinfo {author} {\bibfnamefont {A.}~\bibnamefont {DiLullo}}, \bibinfo
  {author} {\bibfnamefont {Y.}~\bibnamefont {Li}}, \bibinfo {author}
  {\bibfnamefont {B.}~\bibnamefont {Fisher}}, \bibinfo {author} {\bibfnamefont
  {S.~E.}\ \bibnamefont {Ulloa}},\ and\ \bibinfo {author} {\bibfnamefont
  {S.-W.}\ \bibnamefont {Hla}},\ }\bibfield  {title} {\bibinfo {title}
  {Negative differential resistance observed on the charge density wave of a
  transition metal dichalcogenide},\ }\href@noop {} {\bibfield  {journal}
  {\bibinfo  {journal} {Nanoscale}\ }\textbf {\bibinfo {volume} {11}},\
  \bibinfo {pages} {22351} (\bibinfo {year} {2019})}\BibitemShut {NoStop}%
\bibitem [{\citenamefont {Yin}\ \emph {et~al.}(2020)\citenamefont {Yin},
  \citenamefont {Yang}, \citenamefont {Zhang}, \citenamefont {Wu},
  \citenamefont {Fu}, \citenamefont {Tong}, \citenamefont {Yang}, \citenamefont
  {Tian}, \citenamefont {Zhang},\ and\ \citenamefont {Qin}}]{yin2020imaging}%
  \BibitemOpen
  \bibfield  {author} {\bibinfo {author} {\bibfnamefont {L.-J.}\ \bibnamefont
  {Yin}}, \bibinfo {author} {\bibfnamefont {L.-Z.}\ \bibnamefont {Yang}},
  \bibinfo {author} {\bibfnamefont {L.}~\bibnamefont {Zhang}}, \bibinfo
  {author} {\bibfnamefont {Q.}~\bibnamefont {Wu}}, \bibinfo {author}
  {\bibfnamefont {X.}~\bibnamefont {Fu}}, \bibinfo {author} {\bibfnamefont
  {L.-H.}\ \bibnamefont {Tong}}, \bibinfo {author} {\bibfnamefont
  {G.}~\bibnamefont {Yang}}, \bibinfo {author} {\bibfnamefont {Y.}~\bibnamefont
  {Tian}}, \bibinfo {author} {\bibfnamefont {L.}~\bibnamefont {Zhang}},\ and\
  \bibinfo {author} {\bibfnamefont {Z.}~\bibnamefont {Qin}},\ }\bibfield
  {title} {\bibinfo {title} {Imaging of nearly flat band induced atomic-scale
  negative differential conductivity in {ABC}-stacked trilayer graphene},\
  }\href@noop {} {\bibfield  {journal} {\bibinfo  {journal} {Physical Review
  B}\ }\textbf {\bibinfo {volume} {102}},\ \bibinfo {pages} {241403} (\bibinfo
  {year} {2020})}\BibitemShut {NoStop}%
\bibitem [{\citenamefont {Garc{\'\i}a-D{\'\i}ez}\ \emph
  {et~al.}(2025)\citenamefont {Garc{\'\i}a-D{\'\i}ez}, \citenamefont
  {Beidenkopf}, \citenamefont {Robredo},\ and\ \citenamefont
  {Vergniory}}]{garcia2025origins}%
  \BibitemOpen
  \bibfield  {author} {\bibinfo {author} {\bibfnamefont {M.}~\bibnamefont
  {Garc{\'\i}a-D{\'\i}ez}}, \bibinfo {author} {\bibfnamefont {H.}~\bibnamefont
  {Beidenkopf}}, \bibinfo {author} {\bibfnamefont {I.}~\bibnamefont
  {Robredo}},\ and\ \bibinfo {author} {\bibfnamefont {M.~G.}\ \bibnamefont
  {Vergniory}},\ }\bibfield  {title} {\bibinfo {title} {Origins of the
  anomalous {H}all conductivity in the symmetry enforced {Fe$_3$GeTe$_2$}
  nodal-line ferromagnet},\ }\href@noop {} {\bibfield  {journal} {\bibinfo
  {journal} {Journal of Physics: Materials}\ }\textbf {\bibinfo {volume} {8}},\
  \bibinfo {pages} {035012} (\bibinfo {year} {2025})}\BibitemShut {NoStop}%
\bibitem [{\citenamefont {Sales}\ \emph {et~al.}(2019)\citenamefont {Sales},
  \citenamefont {Yan}, \citenamefont {Meier}, \citenamefont {Christianson},
  \citenamefont {Okamoto},\ and\ \citenamefont
  {McGuire}}]{sales2019electronic}%
  \BibitemOpen
  \bibfield  {author} {\bibinfo {author} {\bibfnamefont {B.~C.}\ \bibnamefont
  {Sales}}, \bibinfo {author} {\bibfnamefont {J.}~\bibnamefont {Yan}}, \bibinfo
  {author} {\bibfnamefont {W.~R.}\ \bibnamefont {Meier}}, \bibinfo {author}
  {\bibfnamefont {A.~D.}\ \bibnamefont {Christianson}}, \bibinfo {author}
  {\bibfnamefont {S.}~\bibnamefont {Okamoto}},\ and\ \bibinfo {author}
  {\bibfnamefont {M.~A.}\ \bibnamefont {McGuire}},\ }\bibfield  {title}
  {\bibinfo {title} {Electronic, magnetic, and thermodynamic properties of the
  {K}agome layer compound {FeSn}},\ }\href@noop {} {\bibfield  {journal}
  {\bibinfo  {journal} {Physical Review Materials}\ }\textbf {\bibinfo {volume}
  {3}},\ \bibinfo {pages} {114203} (\bibinfo {year} {2019})}\BibitemShut
  {NoStop}%
\bibitem [{\citenamefont {Yin}\ \emph {et~al.}(2019)\citenamefont {Yin},
  \citenamefont {Zhang}, \citenamefont {Chang}, \citenamefont {Wang},
  \citenamefont {Tsirkin}, \citenamefont {Guguchia}, \citenamefont {Lian},
  \citenamefont {Zhou}, \citenamefont {Jiang}, \citenamefont {Belopolski} \emph
  {et~al.}}]{yin2019negative}%
  \BibitemOpen
  \bibfield  {author} {\bibinfo {author} {\bibfnamefont {J.-X.}\ \bibnamefont
  {Yin}}, \bibinfo {author} {\bibfnamefont {S.~S.}\ \bibnamefont {Zhang}},
  \bibinfo {author} {\bibfnamefont {G.}~\bibnamefont {Chang}}, \bibinfo
  {author} {\bibfnamefont {Q.}~\bibnamefont {Wang}}, \bibinfo {author}
  {\bibfnamefont {S.~S.}\ \bibnamefont {Tsirkin}}, \bibinfo {author}
  {\bibfnamefont {Z.}~\bibnamefont {Guguchia}}, \bibinfo {author}
  {\bibfnamefont {B.}~\bibnamefont {Lian}}, \bibinfo {author} {\bibfnamefont
  {H.}~\bibnamefont {Zhou}}, \bibinfo {author} {\bibfnamefont {K.}~\bibnamefont
  {Jiang}}, \bibinfo {author} {\bibfnamefont {I.}~\bibnamefont {Belopolski}},
  \emph {et~al.},\ }\bibfield  {title} {\bibinfo {title} {Negative flat band
  magnetism in a spin--orbit-coupled correlated {K}agome magnet},\ }\href@noop
  {} {\bibfield  {journal} {\bibinfo  {journal} {Nature Physics}\ }\textbf
  {\bibinfo {volume} {15}},\ \bibinfo {pages} {443} (\bibinfo {year}
  {2019})}\BibitemShut {NoStop}%
\bibitem [{\citenamefont {Xing}\ \emph {et~al.}(2020)\citenamefont {Xing},
  \citenamefont {Shen}, \citenamefont {Chen}, \citenamefont {Huang},
  \citenamefont {Gao}, \citenamefont {Zheng}, \citenamefont {Zhang},
  \citenamefont {Li}, \citenamefont {Hu}, \citenamefont {Qian} \emph
  {et~al.}}]{xing2020localized}%
  \BibitemOpen
  \bibfield  {author} {\bibinfo {author} {\bibfnamefont {Y.}~\bibnamefont
  {Xing}}, \bibinfo {author} {\bibfnamefont {J.}~\bibnamefont {Shen}}, \bibinfo
  {author} {\bibfnamefont {H.}~\bibnamefont {Chen}}, \bibinfo {author}
  {\bibfnamefont {L.}~\bibnamefont {Huang}}, \bibinfo {author} {\bibfnamefont
  {Y.}~\bibnamefont {Gao}}, \bibinfo {author} {\bibfnamefont {Q.}~\bibnamefont
  {Zheng}}, \bibinfo {author} {\bibfnamefont {Y.-Y.}\ \bibnamefont {Zhang}},
  \bibinfo {author} {\bibfnamefont {G.}~\bibnamefont {Li}}, \bibinfo {author}
  {\bibfnamefont {B.}~\bibnamefont {Hu}}, \bibinfo {author} {\bibfnamefont
  {G.}~\bibnamefont {Qian}}, \emph {et~al.},\ }\bibfield  {title} {\bibinfo
  {title} {Localized spin-orbit polaron in magnetic {W}eyl semimetal
  {Co$_3$Sn$_2$S$_2$}},\ }\href@noop {} {\bibfield  {journal} {\bibinfo
  {journal} {Nature Communications}\ }\textbf {\bibinfo {volume} {11}},\
  \bibinfo {pages} {5613} (\bibinfo {year} {2020})}\BibitemShut {NoStop}%
\bibitem [{\citenamefont {Ren}\ \emph {et~al.}(2022)\citenamefont {Ren},
  \citenamefont {Li}, \citenamefont {Sharma}, \citenamefont {Bhattarai},
  \citenamefont {Zhao}, \citenamefont {Rachmilowitz}, \citenamefont {Bahrami},
  \citenamefont {Tafti}, \citenamefont {Fang}, \citenamefont {Ghimire} \emph
  {et~al.}}]{ren2022plethora}%
  \BibitemOpen
  \bibfield  {author} {\bibinfo {author} {\bibfnamefont {Z.}~\bibnamefont
  {Ren}}, \bibinfo {author} {\bibfnamefont {H.}~\bibnamefont {Li}}, \bibinfo
  {author} {\bibfnamefont {S.}~\bibnamefont {Sharma}}, \bibinfo {author}
  {\bibfnamefont {D.}~\bibnamefont {Bhattarai}}, \bibinfo {author}
  {\bibfnamefont {H.}~\bibnamefont {Zhao}}, \bibinfo {author} {\bibfnamefont
  {B.}~\bibnamefont {Rachmilowitz}}, \bibinfo {author} {\bibfnamefont
  {F.}~\bibnamefont {Bahrami}}, \bibinfo {author} {\bibfnamefont
  {F.}~\bibnamefont {Tafti}}, \bibinfo {author} {\bibfnamefont
  {S.}~\bibnamefont {Fang}}, \bibinfo {author} {\bibfnamefont {M.~P.}\
  \bibnamefont {Ghimire}}, \emph {et~al.},\ }\bibfield  {title} {\bibinfo
  {title} {Plethora of tunable {W}eyl fermions in {K}agome magnet
  {Fe$_3$Sn$_2$} thin films},\ }\href@noop {} {\bibfield  {journal} {\bibinfo
  {journal} {npj Quantum Materials}\ }\textbf {\bibinfo {volume} {7}},\
  \bibinfo {pages} {109} (\bibinfo {year} {2022})}\BibitemShut {NoStop}%
\bibitem [{\citenamefont {Taychatanapat}\ and\ \citenamefont
  {Jarillo-Herrero}(2010)}]{taychatanapat2010electronic}%
  \BibitemOpen
  \bibfield  {author} {\bibinfo {author} {\bibfnamefont {T.}~\bibnamefont
  {Taychatanapat}}\ and\ \bibinfo {author} {\bibfnamefont {P.}~\bibnamefont
  {Jarillo-Herrero}},\ }\bibfield  {title} {\bibinfo {title} {Electronic
  transport in dual-gated bilayer graphene at large displacement fields},\
  }\href@noop {} {\bibfield  {journal} {\bibinfo  {journal} {Physical Review
  Letters}\ }\textbf {\bibinfo {volume} {105}},\ \bibinfo {pages} {166601}
  (\bibinfo {year} {2010})}\BibitemShut {NoStop}%
\bibitem [{\citenamefont {Uwanno}\ \emph {et~al.}(2018)\citenamefont {Uwanno},
  \citenamefont {Taniguchi}, \citenamefont {Watanabe},\ and\ \citenamefont
  {Nagashio}}]{uwanno2018electrically}%
  \BibitemOpen
  \bibfield  {author} {\bibinfo {author} {\bibfnamefont {T.}~\bibnamefont
  {Uwanno}}, \bibinfo {author} {\bibfnamefont {T.}~\bibnamefont {Taniguchi}},
  \bibinfo {author} {\bibfnamefont {K.}~\bibnamefont {Watanabe}},\ and\
  \bibinfo {author} {\bibfnamefont {K.}~\bibnamefont {Nagashio}},\ }\bibfield
  {title} {\bibinfo {title} {Electrically inert h-bn/bilayer graphene interface
  in all-two-dimensional heterostructure field effect transistors},\
  }\href@noop {} {\bibfield  {journal} {\bibinfo  {journal} {ACS applied
  materials \& interfaces}\ }\textbf {\bibinfo {volume} {10}},\ \bibinfo
  {pages} {28780} (\bibinfo {year} {2018})}\BibitemShut {NoStop}%
\bibitem [{\citenamefont {Chen}\ \emph
  {et~al.}(2020{\natexlab{a}})\citenamefont {Chen}, \citenamefont {Zhou},
  \citenamefont {Liu}, \citenamefont {Qiao}, \citenamefont {Oezyilmaz},\ and\
  \citenamefont {Martin}}]{chen2020gate}%
  \BibitemOpen
  \bibfield  {author} {\bibinfo {author} {\bibfnamefont {H.}~\bibnamefont
  {Chen}}, \bibinfo {author} {\bibfnamefont {P.}~\bibnamefont {Zhou}}, \bibinfo
  {author} {\bibfnamefont {J.}~\bibnamefont {Liu}}, \bibinfo {author}
  {\bibfnamefont {J.}~\bibnamefont {Qiao}}, \bibinfo {author} {\bibfnamefont
  {B.}~\bibnamefont {Oezyilmaz}},\ and\ \bibinfo {author} {\bibfnamefont
  {J.}~\bibnamefont {Martin}},\ }\bibfield  {title} {\bibinfo {title} {Gate
  controlled valley polarizer in bilayer graphene},\ }\href@noop {} {\bibfield
  {journal} {\bibinfo  {journal} {Nature Communications}\ }\textbf {\bibinfo
  {volume} {11}},\ \bibinfo {pages} {1202} (\bibinfo {year}
  {2020}{\natexlab{a}})}\BibitemShut {NoStop}%
\bibitem [{FLE()}]{FLEUR}%
  \BibitemOpen
  \href@noop {} {}\bibinfo {note} {For the program description, see
  "\url{https://www.flapw.de}"\:(accessed 2025/10/28)}\BibitemShut {NoStop}%
\bibitem [{\citenamefont {Wimmer}\ \emph {et~al.}(1981)\citenamefont {Wimmer},
  \citenamefont {Krakauer}, \citenamefont {Weinert},\ and\ \citenamefont
  {Freeman}}]{wimmer1981full}%
  \BibitemOpen
  \bibfield  {author} {\bibinfo {author} {\bibfnamefont {E.}~\bibnamefont
  {Wimmer}}, \bibinfo {author} {\bibfnamefont {H.}~\bibnamefont {Krakauer}},
  \bibinfo {author} {\bibfnamefont {M.}~\bibnamefont {Weinert}},\ and\ \bibinfo
  {author} {\bibfnamefont {A.~J.}\ \bibnamefont {Freeman}},\ }\bibfield
  {title} {\bibinfo {title} {Full-potential self-consistent
  linearized-augmented-plane-wave method for calculating the electronic
  structure of molecules and surfaces: {O$_2$} molecule},\ }\href@noop {}
  {\bibfield  {journal} {\bibinfo  {journal} {Physical Review B}\ }\textbf
  {\bibinfo {volume} {24}},\ \bibinfo {pages} {864} (\bibinfo {year}
  {1981})}\BibitemShut {NoStop}%
\bibitem [{\citenamefont {Perdew}\ \emph {et~al.}(1996)\citenamefont {Perdew},
  \citenamefont {Burke},\ and\ \citenamefont
  {Ernzerhof}}]{perdew1996generalized}%
  \BibitemOpen
  \bibfield  {author} {\bibinfo {author} {\bibfnamefont {J.~P.}\ \bibnamefont
  {Perdew}}, \bibinfo {author} {\bibfnamefont {K.}~\bibnamefont {Burke}},\ and\
  \bibinfo {author} {\bibfnamefont {M.}~\bibnamefont {Ernzerhof}},\ }\bibfield
  {title} {\bibinfo {title} {Generalized gradient approximation made simple},\
  }\href@noop {} {\bibfield  {journal} {\bibinfo  {journal} {Physical Review
  Letters}\ }\textbf {\bibinfo {volume} {77}},\ \bibinfo {pages} {3865}
  (\bibinfo {year} {1996})}\BibitemShut {NoStop}%
\bibitem [{FGT()}]{FGT_Springer}%
  \BibitemOpen
  \href@noop {} {}\bibinfo {note}
  {\url{https://materials.springer.com/isp/crystallographic/docs/sd_1420956}\:(accessed
  2025/10/28)}\BibitemShut {NoStop}%
\bibitem [{\citenamefont {Wang}\ \emph {et~al.}(2006)\citenamefont {Wang},
  \citenamefont {Yates}, \citenamefont {Souza},\ and\ \citenamefont
  {Vanderbilt}}]{wang2006ab}%
  \BibitemOpen
  \bibfield  {author} {\bibinfo {author} {\bibfnamefont {X.}~\bibnamefont
  {Wang}}, \bibinfo {author} {\bibfnamefont {J.~R.}\ \bibnamefont {Yates}},
  \bibinfo {author} {\bibfnamefont {I.}~\bibnamefont {Souza}},\ and\ \bibinfo
  {author} {\bibfnamefont {D.}~\bibnamefont {Vanderbilt}},\ }\bibfield  {title}
  {\bibinfo {title} {Ab initio calculation of the anomalous {H}all conductivity
  by {W}annier interpolation},\ }\href@noop {} {\bibfield  {journal} {\bibinfo
  {journal} {Physical Review B}\ }\textbf {\bibinfo {volume} {74}},\ \bibinfo
  {pages} {195118} (\bibinfo {year} {2006})}\BibitemShut {NoStop}%
\bibitem [{\citenamefont {Pizzi}\ \emph {et~al.}(2020)\citenamefont {Pizzi},
  \citenamefont {Vitale}, \citenamefont {Arita}, \citenamefont {Bl{\"u}gel},
  \citenamefont {Freimuth}, \citenamefont {G{\'e}ranton}, \citenamefont
  {Gibertini}, \citenamefont {Gresch}, \citenamefont {Johnson}, \citenamefont
  {Koretsune} \emph {et~al.}}]{pizzi2020wannier90}%
  \BibitemOpen
  \bibfield  {author} {\bibinfo {author} {\bibfnamefont {G.}~\bibnamefont
  {Pizzi}}, \bibinfo {author} {\bibfnamefont {V.}~\bibnamefont {Vitale}},
  \bibinfo {author} {\bibfnamefont {R.}~\bibnamefont {Arita}}, \bibinfo
  {author} {\bibfnamefont {S.}~\bibnamefont {Bl{\"u}gel}}, \bibinfo {author}
  {\bibfnamefont {F.}~\bibnamefont {Freimuth}}, \bibinfo {author}
  {\bibfnamefont {G.}~\bibnamefont {G{\'e}ranton}}, \bibinfo {author}
  {\bibfnamefont {M.}~\bibnamefont {Gibertini}}, \bibinfo {author}
  {\bibfnamefont {D.}~\bibnamefont {Gresch}}, \bibinfo {author} {\bibfnamefont
  {C.}~\bibnamefont {Johnson}}, \bibinfo {author} {\bibfnamefont
  {T.}~\bibnamefont {Koretsune}}, \emph {et~al.},\ }\bibfield  {title}
  {\bibinfo {title} {{Wannier90} as a community code: new features and
  applications},\ }\href@noop {} {\bibfield  {journal} {\bibinfo  {journal}
  {Journal of Physics: Condensed Matter}\ }\textbf {\bibinfo {volume} {32}},\
  \bibinfo {pages} {165902} (\bibinfo {year} {2020})}\BibitemShut {NoStop}%
\bibitem [{\citenamefont {Freimuth}\ \emph {et~al.}(2008)\citenamefont
  {Freimuth}, \citenamefont {Mokrousov}, \citenamefont {Wortmann},
  \citenamefont {Heinze},\ and\ \citenamefont
  {Bl{\"u}gel}}]{freimuth2008maximally}%
  \BibitemOpen
  \bibfield  {author} {\bibinfo {author} {\bibfnamefont {F.}~\bibnamefont
  {Freimuth}}, \bibinfo {author} {\bibfnamefont {Y.}~\bibnamefont {Mokrousov}},
  \bibinfo {author} {\bibfnamefont {D.}~\bibnamefont {Wortmann}}, \bibinfo
  {author} {\bibfnamefont {S.}~\bibnamefont {Heinze}},\ and\ \bibinfo {author}
  {\bibfnamefont {S.}~\bibnamefont {Bl{\"u}gel}},\ }\bibfield  {title}
  {\bibinfo {title} {Maximally localized {W}annier functions within the {FLAPW}
  formalism},\ }\href@noop {} {\bibfield  {journal} {\bibinfo  {journal}
  {Physical Review B}\ }\textbf {\bibinfo {volume} {78}},\ \bibinfo {pages}
  {035120} (\bibinfo {year} {2008})}\BibitemShut {NoStop}%
\bibitem [{\citenamefont {Lopez}\ \emph {et~al.}(2012)\citenamefont {Lopez},
  \citenamefont {Vanderbilt}, \citenamefont {Thonhauser},\ and\ \citenamefont
  {Souza}}]{lopez2012wannier}%
  \BibitemOpen
  \bibfield  {author} {\bibinfo {author} {\bibfnamefont {M.}~\bibnamefont
  {Lopez}}, \bibinfo {author} {\bibfnamefont {D.}~\bibnamefont {Vanderbilt}},
  \bibinfo {author} {\bibfnamefont {T.}~\bibnamefont {Thonhauser}},\ and\
  \bibinfo {author} {\bibfnamefont {I.}~\bibnamefont {Souza}},\ }\bibfield
  {title} {\bibinfo {title} {{W}annier-based calculation of the orbital
  magnetization in crystals},\ }\href@noop {} {\bibfield  {journal} {\bibinfo
  {journal} {Physical Review B}\ }\textbf {\bibinfo {volume} {85}},\ \bibinfo
  {pages} {014435} (\bibinfo {year} {2012})}\BibitemShut {NoStop}%
\bibitem [{\citenamefont {Saunderson}\ \emph {et~al.}(2022)\citenamefont
  {Saunderson}, \citenamefont {Go}, \citenamefont {Bl{\"u}gel}, \citenamefont
  {Kl{\"a}ui},\ and\ \citenamefont {Mokrousov}}]{saunderson2022hidden}%
  \BibitemOpen
  \bibfield  {author} {\bibinfo {author} {\bibfnamefont {T.~G.}\ \bibnamefont
  {Saunderson}}, \bibinfo {author} {\bibfnamefont {D.}~\bibnamefont {Go}},
  \bibinfo {author} {\bibfnamefont {S.}~\bibnamefont {Bl{\"u}gel}}, \bibinfo
  {author} {\bibfnamefont {M.}~\bibnamefont {Kl{\"a}ui}},\ and\ \bibinfo
  {author} {\bibfnamefont {Y.}~\bibnamefont {Mokrousov}},\ }\bibfield  {title}
  {\bibinfo {title} {Hidden interplay of current-induced spin and orbital
  torques in bulk {Fe$_3$GeTe$_2$}},\ }\href@noop {} {\bibfield  {journal}
  {\bibinfo  {journal} {Physical Review Research}\ }\textbf {\bibinfo {volume}
  {4}},\ \bibinfo {pages} {L042022} (\bibinfo {year} {2022})}\BibitemShut
  {NoStop}%
\bibitem [{\citenamefont {Wakafuji}\ \emph {et~al.}(2020)\citenamefont
  {Wakafuji}, \citenamefont {Moriya}, \citenamefont {Masubuchi}, \citenamefont
  {Watanabe}, \citenamefont {Taniguchi},\ and\ \citenamefont
  {Machida}}]{waka2020pvc}%
  \BibitemOpen
  \bibfield  {author} {\bibinfo {author} {\bibfnamefont {Y.}~\bibnamefont
  {Wakafuji}}, \bibinfo {author} {\bibfnamefont {R.}~\bibnamefont {Moriya}},
  \bibinfo {author} {\bibfnamefont {S.}~\bibnamefont {Masubuchi}}, \bibinfo
  {author} {\bibfnamefont {K.}~\bibnamefont {Watanabe}}, \bibinfo {author}
  {\bibfnamefont {T.}~\bibnamefont {Taniguchi}},\ and\ \bibinfo {author}
  {\bibfnamefont {T.}~\bibnamefont {Machida}},\ }\bibfield  {title} {\bibinfo
  {title} {{3D} manipulation of {2D} materials using microdome polymer},\
  }\href@noop {} {\bibfield  {journal} {\bibinfo  {journal} {Nano Letters}\
  }\textbf {\bibinfo {volume} {20}},\ \bibinfo {pages} {2486} (\bibinfo {year}
  {2020})}\BibitemShut {NoStop}%
\bibitem [{\citenamefont {Ne{\v{c}}as}\ and\ \citenamefont
  {Klapetek}(2012)}]{nevcas2012gwyddion}%
  \BibitemOpen
  \bibfield  {author} {\bibinfo {author} {\bibfnamefont {D.}~\bibnamefont
  {Ne{\v{c}}as}}\ and\ \bibinfo {author} {\bibfnamefont {P.}~\bibnamefont
  {Klapetek}},\ }\bibfield  {title} {\bibinfo {title} {Gwyddion: an open-source
  software for {SPM} data analysis},\ }\href@noop {} {\bibfield  {journal}
  {\bibinfo  {journal} {Central European Journal of Physics}\ }\textbf
  {\bibinfo {volume} {10}},\ \bibinfo {pages} {181} (\bibinfo {year}
  {2012})}\BibitemShut {NoStop}%
\bibitem [{\citenamefont {Brun}\ \emph {et~al.}(2012)\citenamefont {Brun},
  \citenamefont {M{\"u}ller}, \citenamefont {Hong}, \citenamefont {Patthey},
  \citenamefont {Flindt},\ and\ \citenamefont {Schneider}}]{brun2012dynamical}%
  \BibitemOpen
  \bibfield  {author} {\bibinfo {author} {\bibfnamefont {C.}~\bibnamefont
  {Brun}}, \bibinfo {author} {\bibfnamefont {K.~H.}\ \bibnamefont
  {M{\"u}ller}}, \bibinfo {author} {\bibfnamefont {I.-P.}\ \bibnamefont
  {Hong}}, \bibinfo {author} {\bibfnamefont {F.}~\bibnamefont {Patthey}},
  \bibinfo {author} {\bibfnamefont {C.}~\bibnamefont {Flindt}},\ and\ \bibinfo
  {author} {\bibfnamefont {W.-D.}\ \bibnamefont {Schneider}},\ }\bibfield
  {title} {\bibinfo {title} {Dynamical coulomb blockade observed in nanosized
  electrical contacts},\ }\href@noop {} {\bibfield  {journal} {\bibinfo
  {journal} {Physical Review Letters}\ }\textbf {\bibinfo {volume} {108}},\
  \bibinfo {pages} {126802} (\bibinfo {year} {2012})}\BibitemShut {NoStop}%
\bibitem [{\citenamefont {Andrei}\ \emph {et~al.}(2012)\citenamefont {Andrei},
  \citenamefont {Li},\ and\ \citenamefont {Du}}]{andrei2012electronic}%
  \BibitemOpen
  \bibfield  {author} {\bibinfo {author} {\bibfnamefont {E.~Y.}\ \bibnamefont
  {Andrei}}, \bibinfo {author} {\bibfnamefont {G.}~\bibnamefont {Li}},\ and\
  \bibinfo {author} {\bibfnamefont {X.}~\bibnamefont {Du}},\ }\bibfield
  {title} {\bibinfo {title} {Electronic properties of graphene: a perspective
  from scanning tunneling microscopy and magnetotransport},\ }\href@noop {}
  {\bibfield  {journal} {\bibinfo  {journal} {Reports on Progress in Physics}\
  }\textbf {\bibinfo {volume} {75}},\ \bibinfo {pages} {056501} (\bibinfo
  {year} {2012})}\BibitemShut {NoStop}%
\bibitem [{\citenamefont {Lee}\ \emph {et~al.}(2016)\citenamefont {Lee},
  \citenamefont {Wong}, \citenamefont {Velasco~Jr}, \citenamefont
  {Rodriguez-Nieva}, \citenamefont {Kahn}, \citenamefont {Tsai}, \citenamefont
  {Taniguchi}, \citenamefont {Watanabe}, \citenamefont {Zettl}, \citenamefont
  {Wang} \emph {et~al.}}]{lee2016imaging}%
  \BibitemOpen
  \bibfield  {author} {\bibinfo {author} {\bibfnamefont {J.}~\bibnamefont
  {Lee}}, \bibinfo {author} {\bibfnamefont {D.}~\bibnamefont {Wong}}, \bibinfo
  {author} {\bibfnamefont {J.}~\bibnamefont {Velasco~Jr}}, \bibinfo {author}
  {\bibfnamefont {J.~F.}\ \bibnamefont {Rodriguez-Nieva}}, \bibinfo {author}
  {\bibfnamefont {S.}~\bibnamefont {Kahn}}, \bibinfo {author} {\bibfnamefont
  {H.-Z.}\ \bibnamefont {Tsai}}, \bibinfo {author} {\bibfnamefont
  {T.}~\bibnamefont {Taniguchi}}, \bibinfo {author} {\bibfnamefont
  {K.}~\bibnamefont {Watanabe}}, \bibinfo {author} {\bibfnamefont
  {A.}~\bibnamefont {Zettl}}, \bibinfo {author} {\bibfnamefont
  {F.}~\bibnamefont {Wang}}, \emph {et~al.},\ }\bibfield  {title} {\bibinfo
  {title} {Imaging electrostatically confined dirac fermions in graphene
  quantum dots},\ }\href@noop {} {\bibfield  {journal} {\bibinfo  {journal}
  {Nature Physics}\ }\textbf {\bibinfo {volume} {12}},\ \bibinfo {pages} {1032}
  (\bibinfo {year} {2016})}\BibitemShut {NoStop}%
\bibitem [{\citenamefont {Ge}\ \emph {et~al.}(2023)\citenamefont {Ge},
  \citenamefont {Slizovskiy}, \citenamefont {Polizogopoulos}, \citenamefont
  {Joshi}, \citenamefont {Taniguchi}, \citenamefont {Watanabe}, \citenamefont
  {Lederman}, \citenamefont {Fal’ko},\ and\ \citenamefont
  {Velasco~Jr}}]{ge2023giant}%
  \BibitemOpen
  \bibfield  {author} {\bibinfo {author} {\bibfnamefont {Z.}~\bibnamefont
  {Ge}}, \bibinfo {author} {\bibfnamefont {S.}~\bibnamefont {Slizovskiy}},
  \bibinfo {author} {\bibfnamefont {P.}~\bibnamefont {Polizogopoulos}},
  \bibinfo {author} {\bibfnamefont {T.}~\bibnamefont {Joshi}}, \bibinfo
  {author} {\bibfnamefont {T.}~\bibnamefont {Taniguchi}}, \bibinfo {author}
  {\bibfnamefont {K.}~\bibnamefont {Watanabe}}, \bibinfo {author}
  {\bibfnamefont {D.}~\bibnamefont {Lederman}}, \bibinfo {author}
  {\bibfnamefont {V.~I.}\ \bibnamefont {Fal’ko}},\ and\ \bibinfo {author}
  {\bibfnamefont {J.}~\bibnamefont {Velasco~Jr}},\ }\bibfield  {title}
  {\bibinfo {title} {Giant orbital magnetic moments and paramagnetic shift in
  artificial relativistic atoms and molecules},\ }\href@noop {} {\bibfield
  {journal} {\bibinfo  {journal} {Nature Nanotechnology}\ }\textbf {\bibinfo
  {volume} {18}},\ \bibinfo {pages} {250} (\bibinfo {year} {2023})}\BibitemShut
  {NoStop}%
\bibitem [{\citenamefont {Lee}\ \emph {et~al.}(2020)\citenamefont {Lee},
  \citenamefont {Knothe}, \citenamefont {Overweg}, \citenamefont {Eich},
  \citenamefont {Gold}, \citenamefont {Kurzmann}, \citenamefont {Klasovika},
  \citenamefont {Taniguchi}, \citenamefont {Wantanabe}, \citenamefont
  {Fal’ko} \emph {et~al.}}]{lee2020tunable}%
  \BibitemOpen
  \bibfield  {author} {\bibinfo {author} {\bibfnamefont {Y.}~\bibnamefont
  {Lee}}, \bibinfo {author} {\bibfnamefont {A.}~\bibnamefont {Knothe}},
  \bibinfo {author} {\bibfnamefont {H.}~\bibnamefont {Overweg}}, \bibinfo
  {author} {\bibfnamefont {M.}~\bibnamefont {Eich}}, \bibinfo {author}
  {\bibfnamefont {C.}~\bibnamefont {Gold}}, \bibinfo {author} {\bibfnamefont
  {A.}~\bibnamefont {Kurzmann}}, \bibinfo {author} {\bibfnamefont
  {V.}~\bibnamefont {Klasovika}}, \bibinfo {author} {\bibfnamefont
  {T.}~\bibnamefont {Taniguchi}}, \bibinfo {author} {\bibfnamefont
  {K.}~\bibnamefont {Wantanabe}}, \bibinfo {author} {\bibfnamefont
  {V.}~\bibnamefont {Fal’ko}}, \emph {et~al.},\ }\bibfield  {title} {\bibinfo
  {title} {Tunable valley splitting due to topological orbital magnetic moment
  in bilayer graphene quantum point contacts},\ }\href@noop {} {\bibfield
  {journal} {\bibinfo  {journal} {Physical review letters}\ }\textbf {\bibinfo
  {volume} {124}},\ \bibinfo {pages} {126802} (\bibinfo {year}
  {2020})}\BibitemShut {NoStop}%
\bibitem [{\citenamefont {Ge}\ \emph {et~al.}(2021)\citenamefont {Ge},
  \citenamefont {Slizovskiy}, \citenamefont {Joucken}, \citenamefont {Quezada},
  \citenamefont {Taniguchi}, \citenamefont {Watanabe}, \citenamefont
  {Fal’ko},\ and\ \citenamefont {Velasco~Jr}}]{ge2021control}%
  \BibitemOpen
  \bibfield  {author} {\bibinfo {author} {\bibfnamefont {Z.}~\bibnamefont
  {Ge}}, \bibinfo {author} {\bibfnamefont {S.}~\bibnamefont {Slizovskiy}},
  \bibinfo {author} {\bibfnamefont {F.}~\bibnamefont {Joucken}}, \bibinfo
  {author} {\bibfnamefont {E.~A.}\ \bibnamefont {Quezada}}, \bibinfo {author}
  {\bibfnamefont {T.}~\bibnamefont {Taniguchi}}, \bibinfo {author}
  {\bibfnamefont {K.}~\bibnamefont {Watanabe}}, \bibinfo {author}
  {\bibfnamefont {V.~I.}\ \bibnamefont {Fal’ko}},\ and\ \bibinfo {author}
  {\bibfnamefont {J.}~\bibnamefont {Velasco~Jr}},\ }\bibfield  {title}
  {\bibinfo {title} {Control of giant topological magnetic moment and valley
  splitting in trilayer graphene},\ }\href@noop {} {\bibfield  {journal}
  {\bibinfo  {journal} {Physical Review Letters}\ }\textbf {\bibinfo {volume}
  {127}},\ \bibinfo {pages} {136402} (\bibinfo {year} {2021})}\BibitemShut
  {NoStop}%
\bibitem [{\citenamefont {Black}\ and\ \citenamefont
  {Welser}(1999)}]{black1999electric}%
  \BibitemOpen
  \bibfield  {author} {\bibinfo {author} {\bibfnamefont {C.~T.}\ \bibnamefont
  {Black}}\ and\ \bibinfo {author} {\bibfnamefont {J.~J.}\ \bibnamefont
  {Welser}},\ }\bibfield  {title} {\bibinfo {title} {Electric-field penetration
  into metals: consequences for high-dielectric-constant capacitors},\
  }\href@noop {} {\bibfield  {journal} {\bibinfo  {journal} {IEEE Transactions
  on Electron Devices}\ }\textbf {\bibinfo {volume} {46}},\ \bibinfo {pages}
  {776} (\bibinfo {year} {1999})}\BibitemShut {NoStop}%
\bibitem [{\citenamefont {Ambrosetti}\ and\ \citenamefont
  {Silvestrelli}(2019)}]{ambrosetti2019faraday}%
  \BibitemOpen
  \bibfield  {author} {\bibinfo {author} {\bibfnamefont {A.}~\bibnamefont
  {Ambrosetti}}\ and\ \bibinfo {author} {\bibfnamefont {P.~L.}\ \bibnamefont
  {Silvestrelli}},\ }\bibfield  {title} {\bibinfo {title} {Faraday-like
  screening by two-dimensional nanomaterials: A scale-dependent tunable
  effect},\ }\href@noop {} {\bibfield  {journal} {\bibinfo  {journal} {The
  Journal of Physical Chemistry Letters}\ }\textbf {\bibinfo {volume} {10}},\
  \bibinfo {pages} {2044} (\bibinfo {year} {2019})}\BibitemShut {NoStop}%
\bibitem [{\citenamefont {L{\"u}pke}\ \emph {et~al.}(2018)\citenamefont
  {L{\"u}pke}, \citenamefont {Just}, \citenamefont {Eschbach}, \citenamefont
  {Heider}, \citenamefont {M{\l}y{\'n}czak}, \citenamefont {Lanius},
  \citenamefont {Sch{\"u}ffelgen}, \citenamefont {Rosenbach}, \citenamefont
  {von~den Driesch}, \citenamefont {Cherepanov} \emph
  {et~al.}}]{lupke2018situ}%
  \BibitemOpen
  \bibfield  {author} {\bibinfo {author} {\bibfnamefont {F.}~\bibnamefont
  {L{\"u}pke}}, \bibinfo {author} {\bibfnamefont {S.}~\bibnamefont {Just}},
  \bibinfo {author} {\bibfnamefont {M.}~\bibnamefont {Eschbach}}, \bibinfo
  {author} {\bibfnamefont {T.}~\bibnamefont {Heider}}, \bibinfo {author}
  {\bibfnamefont {E.}~\bibnamefont {M{\l}y{\'n}czak}}, \bibinfo {author}
  {\bibfnamefont {M.}~\bibnamefont {Lanius}}, \bibinfo {author} {\bibfnamefont
  {P.}~\bibnamefont {Sch{\"u}ffelgen}}, \bibinfo {author} {\bibfnamefont
  {D.}~\bibnamefont {Rosenbach}}, \bibinfo {author} {\bibfnamefont
  {N.}~\bibnamefont {von~den Driesch}}, \bibinfo {author} {\bibfnamefont
  {V.}~\bibnamefont {Cherepanov}}, \emph {et~al.},\ }\bibfield  {title}
  {\bibinfo {title} {In situ disentangling surface state transport channels of
  a topological insulator thin film by gating},\ }\href@noop {} {\bibfield
  {journal} {\bibinfo  {journal} {npj Quantum Materials}\ }\textbf {\bibinfo
  {volume} {3}},\ \bibinfo {pages} {46} (\bibinfo {year} {2018})}\BibitemShut
  {NoStop}%
\bibitem [{\citenamefont {Xia}\ \emph {et~al.}(2009)\citenamefont {Xia},
  \citenamefont {Chen}, \citenamefont {Li},\ and\ \citenamefont
  {Tao}}]{xia2009measurement}%
  \BibitemOpen
  \bibfield  {author} {\bibinfo {author} {\bibfnamefont {J.}~\bibnamefont
  {Xia}}, \bibinfo {author} {\bibfnamefont {F.}~\bibnamefont {Chen}}, \bibinfo
  {author} {\bibfnamefont {J.}~\bibnamefont {Li}},\ and\ \bibinfo {author}
  {\bibfnamefont {N.}~\bibnamefont {Tao}},\ }\bibfield  {title} {\bibinfo
  {title} {Measurement of the quantum capacitance of graphene},\ }\href@noop {}
  {\bibfield  {journal} {\bibinfo  {journal} {Nature Nanotechnology}\ }\textbf
  {\bibinfo {volume} {4}},\ \bibinfo {pages} {505} (\bibinfo {year}
  {2009})}\BibitemShut {NoStop}%
\bibitem [{\citenamefont {Chen}\ \emph
  {et~al.}(2020{\natexlab{b}})\citenamefont {Chen}, \citenamefont {Wu},
  \citenamefont {Xu}, \citenamefont {Cong}, \citenamefont {Li}, \citenamefont
  {Feng}, \citenamefont {Zhang}, \citenamefont {Zou}, \citenamefont {Shang},
  \citenamefont {Yang}, \citenamefont {Loh}, \citenamefont {Huang},\ and\
  \citenamefont {Yu}}]{Chen2020visualizing}%
  \BibitemOpen
  \bibfield  {author} {\bibinfo {author} {\bibfnamefont {Y.}~\bibnamefont
  {Chen}}, \bibinfo {author} {\bibfnamefont {L.}~\bibnamefont {Wu}}, \bibinfo
  {author} {\bibfnamefont {H.}~\bibnamefont {Xu}}, \bibinfo {author}
  {\bibfnamefont {C.}~\bibnamefont {Cong}}, \bibinfo {author} {\bibfnamefont
  {S.}~\bibnamefont {Li}}, \bibinfo {author} {\bibfnamefont {S.}~\bibnamefont
  {Feng}}, \bibinfo {author} {\bibfnamefont {H.}~\bibnamefont {Zhang}},
  \bibinfo {author} {\bibfnamefont {C.}~\bibnamefont {Zou}}, \bibinfo {author}
  {\bibfnamefont {J.}~\bibnamefont {Shang}}, \bibinfo {author} {\bibfnamefont
  {S.~A.}\ \bibnamefont {Yang}}, \bibinfo {author} {\bibfnamefont {K.~P.}\
  \bibnamefont {Loh}}, \bibinfo {author} {\bibfnamefont {W.}~\bibnamefont
  {Huang}},\ and\ \bibinfo {author} {\bibfnamefont {T.}~\bibnamefont {Yu}},\
  }\bibfield  {title} {\bibinfo {title} {Visualizing the anomalous charge
  density wave states in graphene/{NbSe$_2$} heterostructures},\ }\href@noop {}
  {\bibfield  {journal} {\bibinfo  {journal} {Advanced Materials}\ }\textbf
  {\bibinfo {volume} {32}},\ \bibinfo {pages} {2003746} (\bibinfo {year}
  {2020}{\natexlab{b}})}\BibitemShut {NoStop}%
\end{thebibliography}
\end{document}